\documentclass{article}

\usepackage{amssymb,amsfonts,amsmath,stmaryrd}
\usepackage{cite,enumerate,float,indentfirst}
\usepackage{color}
\usepackage{ytableau}
\usepackage[vcentermath]{youngtab}
\usepackage{pict2e}

\def\be{\begin{eqnarray}}
\def\ee{\end{eqnarray}}
\def\nn{\nonumber}

\def\p{\partial}

\def\Tr{{\rm Tr}\,}

\newcommand{\beq}{\begin{equation}}
\newcommand{\eeq}{\end{equation}}
\newcommand{\beqa}{\begin{eqnarray}}
\newcommand{\eeqa}{\end{eqnarray}}

\definecolor{red}{rgb}{1,0,0}
\definecolor{orange}{rgb}{1,0.5,0}
\definecolor{violet}{rgb}{0.7,0,1}

\def\ba{\bf}
\def\bb{}

\newcommand{\ttop}[1]{
  q^{\hat{D}_#1}
}

\hoffset= - 1.0in         % switch off for draft style
\begin{document}

\title{\vspace{1.5cm}\bf
Commutative families in DIM algebra,\\
integrable many-body systems and $q,t$ matrix models
}

\author{
A. Mironov$^{b,c,d,}$\footnote{mironov@lpi.ru,mironov@itep.ru},
A. Morozov$^{a,c,d,}$\footnote{morozov@itep.ru},
A. Popolitov$^{a,c,d,}$\footnote{popolit@gmail.com}
}

\date{ }

\maketitle

\vspace{-6.5cm}

\begin{center}
\hfill FIAN/TD-07/24\\
\hfill IITP/TH-13/24\\
\hfill ITEP/TH-15/24\\
\hfill MIPT/TH-11/24
\end{center}

\vspace{4.5cm}

\begin{center}
$^a$ {\small {\it MIPT, Dolgoprudny, 141701, Russia}}\\
$^b$ {\small {\it Lebedev Physics Institute, Moscow 119991, Russia}}\\
$^c$ {\small {\it NRC ``Kurchatov Institute", 123182, Moscow, Russia}}\\
$^d$ {\small {\it Institute for Information Transmission Problems, Moscow 127994, Russia}}
%$^e$ {\small{\it Institute for Theoretical and Mathematical Physics, Lomonosov Moscow State University, Moscow 119991, Russia}}
\end{center}

\vspace{.1cm}

\begin{abstract}
We extend our consideration of commutative subalgebras (rays) in different representations
of the $W_{1+\infty}$ algebra to the elliptic Hall algebra (or, equivalently, to the Ding-Iohara-Miki (DIM) algebra  $U_{q,t}(\widehat{\widehat{\mathfrak{gl}}}_1)$).
Its advantage is that it possesses the Miki automorphism,
which makes all commutative rays equivalent.
Integrable systems associated with these rays become
finite-difference and, apart from the trigonometric Ruijsenaars system not too much familiar.
We concentrate on the simplest many-body and Fock representations, and derive explicit formulas
for all generators of the elliptic Hall algebra $e_{n,m}$.
In the one-body representation, they differ just by normalization from $z^nq^{m\hat D}$
of the $W_{1+\infty}$ Lie algebra, and, in the $N$-body case, they are
non-trivially generalized to monomials of the Cherednik operators with action restricted to symmetric polynomials.
In the Fock representation, the resulting operators are expressed
through auxiliary polynomials of $n$ variables, which define weights in the residues formulas.
We also discuss $q,t$-deformation of matrix models associated with constructed commutative subalgebras.
\end{abstract}

\bigskip

\section{Introduction}

The problem of constructing commutative subalgebras of finite- and infinite-dimensional algebras is a standard problem of theoretical physics related to many important issues, from describing irreducible representations associated with physical systems/particles and their charges to integrable systems, both many-body integrable systems and infinite integrable hierarchies.
Original integrable Hamiltonians were made from Casimir operators lying in the universal enveloping algebras of the Lie algebras \cite{HCas,BR}, and were a part of conceptually complicated construction.
The recent progress \cite{MMCal,MMMP1} shifted the accent and dealt with these  Hamiltonians as with generators of the algebra itself.
Commutative sets are then provided by {\it rays}, and the entire algebra describes an infinite set of integrable systems.
In this paper, we further lift this identification to the level of Ding-Iohara-Miki (DIM) algebra \cite{DI,Miki} (see also \cite{AKMZ}),
where Hamiltonians can be substituted by the well-defined generators,
and {\it rays} get related by Miki rotations \cite{Miki1,Miki}.
This gives rise to {\bf a conceptually new interpretation of the DIM algebra as an ensemble of interrelated integrable systems}.

We now describe  the steps of this development in slightly more details.

In \cite{MMCal,MMMP1}, we searched for commutative subalgebras of the $W_{1+\infty}$ algebra. These commutative subalgebras turned out to be related with Hamiltonians of rational Calogero and trigonometric Calogero-Sutherland systems at the free fermion point and their generalizations, and, at the same time, they generate the hypergeometric and skew hypergeometric $\tau$-functions of the Toda/KP hierarchy \cite{GKM2,OS,Ch1,Ch2}, which are partition functions of some matrix models \cite{Ch1,Ch2}. These commutative subalgebras are enumerated by a coprime pair of integers (p,r).

These results were later extended \cite{MMMP1,MMMP2} to the affine Yangian algebra, where the commutative subalgebras were associated with Hamiltonians of rational Calogero and trigonometric Calogero-Sutherland systems at arbitrary coupling and their generalizations, and, at the same time, they generated partition functions of $\beta$-ensembles \cite{MOP1,MOP2}. However, these commutative subalgebras in the Yangian case are enumerated only by a pair (1,r).

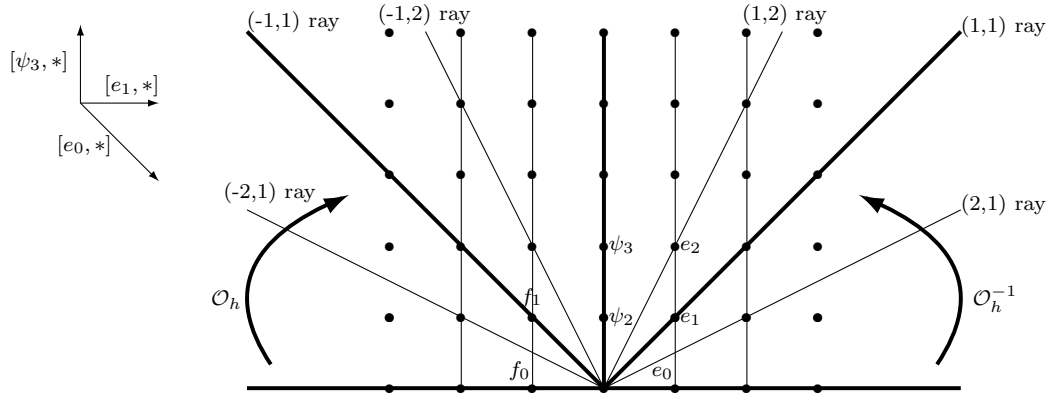
\begin{figure}[h]
\setlength{\unitlength}{.9pt}
\begin{picture}(350,210)(-285,-30)
{\footnotesize
\put(-30,0){\line(0,1){150}}
\put(-60,0){\line(0,1){150}}
\put(30,0){\line(0,1){150}}
\put(60,0){\line(0,1){150}}

%\put(-2.5,-2.5){\mbox{$\bullet$}}

\put(-32.5,147.5){\mbox{$\bullet$}}
\put(-32.5,117.5){\mbox{$\bullet$}}
\put(-32.5,87.5){\mbox{$\bullet$}}
\put(-32.5,57.5){\mbox{$\bullet$}}
\put(-32.5,27.5){\mbox{$\bullet$}}
\put(-32.5,-2.5){\mbox{$\bullet$}}

\put(-62.5,147.5){\mbox{$\bullet$}}
\put(-62.5,117.5){\mbox{$\bullet$}}
\put(-62.5,87.5){\mbox{$\bullet$}}
\put(-62.5,57.5){\mbox{$\bullet$}}
\put(-62.5,27.5){\mbox{$\bullet$}}
\put(-62.5,-2.5){\mbox{$\bullet$}}

\put(-92.5,147.5){\mbox{$\bullet$}}
\put(-92.5,117.5){\mbox{$\bullet$}}
\put(-92.5,87.5){\mbox{$\bullet$}}
\put(-92.5,57.5){\mbox{$\bullet$}}
\put(-92.5,27.5){\mbox{$\bullet$}}
\put(-92.5,-2.5){\mbox{$\bullet$}}
\put(-92.5,27.5){\mbox{$\bullet$}}

\put(-2.5,147.5){\mbox{$\bullet$}}
\put(-2.5,117.5){\mbox{$\bullet$}}
\put(-2.5,87.5){\mbox{$\bullet$}}
\put(-2.5,57.5){\mbox{$\bullet$}}
\put(-2.5,27.5){\mbox{$\bullet$}}
\put(-2.5,-2.5){\mbox{$\bullet$}}
\put(-2.5,27.5){\mbox{$\bullet$}}

\put(27.5,147.5){\mbox{$\bullet$}}
\put(27.5,117.5){\mbox{$\bullet$}}
\put(27.5,87.5){\mbox{$\bullet$}}
\put(27.5,57.5){\mbox{$\bullet$}}
\put(27.5,27.5){\mbox{$\bullet$}}
\put(27.5,-2.5){\mbox{$\bullet$}}
\put(27.5,27.5){\mbox{$\bullet$}}

\put(57.5,147.5){\mbox{$\bullet$}}
\put(57.5,117.5){\mbox{$\bullet$}}
\put(57.5,87.5){\mbox{$\bullet$}}
\put(57.5,57.5){\mbox{$\bullet$}}
\put(57.5,27.5){\mbox{$\bullet$}}
\put(57.5,-2.5){\mbox{$\bullet$}}
\put(57.5,27.5){\mbox{$\bullet$}}

\put(87.5,147.5){\mbox{$\bullet$}}
\put(87.5,117.5){\mbox{$\bullet$}}
\put(87.5,87.5){\mbox{$\bullet$}}
\put(87.5,57.5){\mbox{$\bullet$}}
\put(87.5,27.5){\mbox{$\bullet$}}
\put(87.5,-2.5){\mbox{$\bullet$}}
\put(87.5,27.5){\mbox{$\bullet$}}

\put(0,0){\line(-2,1){150}}
\put(0,0){\line(2,1){150}}
\put(0,0){\line(-1,2){75}}
\put(0,0){\line(1,2){75}}

\put(150,150){\mbox{(1,1) ray}}
\put(150,75){\mbox{(2,1) ray}}
\put(-150,152){\mbox{(-1,1) ray}}
\put(-160,80){\mbox{(-2,1) ray}}
\put(58,155){\mbox{(1,2) ray}}
\put(-95,155){\mbox{(-1,2) ray}}

\put(-40,5){\mbox{$f_0$}}
%\put(-70,5){\mbox{$f_2$}}
\put(20,5){\mbox{$e_0$}}
%\put(50,5){\mbox{$e_2$}}

\put(-35,36){\mbox{$f_1$}}
\put(32,28){\mbox{$e_1$}}
\put(2,28){\mbox{$\psi_2$}}
\put(2,58){\mbox{$\psi_3$}}
\put(32,58){\mbox{$e_2$}}
%\put(62,58){\mbox{$e_{(2,2)}$}}

\put(-220,120){\line(0,1){30}}
\put(-220,150){\vector(0,1){3}}
\put(-250,135){\mbox{$[\psi_3,\ast]$}}
\put(-220,120){\line(1,0){30}}
\put(-190,120){\vector(1,0){3}}
\put(-210,125){\mbox{$[e_1,\ast]$}}
\put(-220,120){\line(1,-1){30}}
\put(-190,90){\vector(1,-1){3}}
\put(-230,100){\mbox{$[e_0,\ast]$}}

\linethickness{1.5pt}
\put(0,0){\line(-1,1){150}}
\put(0,0){\line(1,1){150}}
\put(0,0){\line(0,1){150}}
\put(-150,0){\line(1,0){300}}

\qbezier(-140,10)(-170,55)(-110,80)
\put(-110,80){\vector(2,1){3}}
\qbezier(140,10)(170,55)(110,80)
\put(110,80){\vector(-2,1){3}}
\put(-165,35){\mbox{${\cal O}_h$}}
\put(155,35){\mbox{${\cal O}_h^{-1}$}}

}
\end{picture}
\caption{\footnotesize $2d$ integer half-plane lattice of generators of the affine Yangian/$W_{1+\infty}$ algebra. Each ray (1,r) gives rise to commutative subalgebras in the both algebras, and each ray (p,r) gives rise to a commutative subalgebra of the $W_{1+\infty}$ algebra. Depicted at the left upper corner is the action of commutators with $e_0$, $e_1$ and $\psi_3=6\hat W_0$.}
\label{YWfig}
\end{figure}

Basically, the scheme of constructing the commutative subalgebras of these algebras works as follows. The generators of the both of them occupy the integer half-plane lattice, see Fig.\ref{YWfig}, with the vertical axis corresponding to the spin\footnote{The element of algebra is typically a sum of several spins, in this case, ``spin" means the maximal spin in the sum \cite[sec.3]{MMMP1}.} and the horizontal one, to the grading, and can be generated from three elements: $W_0={1/6}\psi_3$, $e_0$ and $f_0$ having the grading and spin (0,3), (1,0) and (-1,0) correspondingly, and from the central element $\psi_0$. Then, one constructs the elements $e_1=[W_0,e_0]$ and $e_2=[W_0,e_1]$, which generate the whole commutative subalgebra, ray (1,1) as
\be
H_k^{(1,1)}=\hbox{ad}_{e_2}^{k-1}e_1\nn
\ee
Similarly, using $e_n=\hbox{ad}_{W_0}^{n}e_0$, one constructs a generating pair $(e_n,e_{n+1}$, which generates the commutative subalgebra, ray (1,n) as
\be
H_k^{(1,n)}=\hbox{ad}_{e_{n+1}}^{k-1}e_n\nn
\ee
Similarly, one constructs $H_{-k}^{(1,n)}$ associated with rays (-1,r) from $W_0$ and $f_0$.

Commutativity of rays ($\pm 1$,r) is guaranteed by the Serre relations, which are the same in the both algebras \cite{MMMP2} (they were called integer rays in \cite{MMMP1}). Moreover, all these ray can be constructed using ``a rotation operator" ${\cal O}_h$ from the horizontal (evidently commutative) rays $(\pm 1,0)$, see Fig.\ref{YWfig}. This operators maps rays ${\cal O}_h:(-1,r)\to (-1,r+1)$, while ${\cal O}_h^{-1}:(1,r)\to (1,r+1)$.

However, in the case of the $W_{1+\infty}$ algebra, there are more commutative subalgebras: those associated with rays ($\pm$p,r) with an arbitrary pair of coprime integers $p$ and $r$ (they were called rational rays in \cite{MMMP1}, which are constructed as follows: one starts from $e_0^{(p)}=\hbox{ad}_{e_{1}}^{p-1}e_0$ and then repeats the construction above using $e_0^{(p)}$ instead of $e_0=e_0^{(1)}$, and similarly for $f$'s \cite{MMMP1}. These repeated commutators are depicted by points of the integer
$2d$ lattice in Fig.\ref{YWfig}.

A construction of rational rays for the affine Yangian is not known yet.

\begin{figure}[h]
\setlength{\unitlength}{.9pt}
\begin{picture}(350,340)(-285,-170)
{\footnotesize
\put(-30,-150){\line(0,1){300}}
\put(-60,-150){\line(0,1){300}}
\put(30,-150){\line(0,1){300}}
\put(60,-150){\line(0,1){300}}

%\put(-2.5,-2.5){\mbox{$\bullet$}}

\put(-32.5,147.5){\mbox{$\bullet$}}
\put(-32.5,117.5){\mbox{$\bullet$}}
\put(-32.5,87.5){\mbox{$\bullet$}}
\put(-32.5,57.5){\mbox{$\bullet$}}
\put(-32.5,27.5){\mbox{$\bullet$}}
\put(-32.5,-2.5){\mbox{$\bullet$}}
\put(-32.5,-32.5){\mbox{$\bullet$}}
\put(-32.5,-62.5){\mbox{$\bullet$}}
\put(-32.5,-92.5){\mbox{$\bullet$}}
\put(-32.5,-122.5){\mbox{$\bullet$}}
\put(-32.5,-152.5){\mbox{$\bullet$}}
\put(-32.5,27.5){\mbox{$\bullet$}}

\put(-62.5,147.5){\mbox{$\bullet$}}
\put(-62.5,117.5){\mbox{$\bullet$}}
\put(-62.5,87.5){\mbox{$\bullet$}}
\put(-62.5,57.5){\mbox{$\bullet$}}
\put(-62.5,27.5){\mbox{$\bullet$}}
\put(-62.5,-2.5){\mbox{$\bullet$}}
\put(-62.5,-32.5){\mbox{$\bullet$}}
\put(-62.5,-62.5){\mbox{$\bullet$}}
\put(-62.5,-92.5){\mbox{$\bullet$}}
\put(-62.5,-122.5){\mbox{$\bullet$}}
\put(-62.5,-152.5){\mbox{$\bullet$}}
\put(-62.5,27.5){\mbox{$\bullet$}}

\put(-92.5,147.5){\mbox{$\bullet$}}
\put(-92.5,117.5){\mbox{$\bullet$}}
\put(-92.5,87.5){\mbox{$\bullet$}}
\put(-92.5,57.5){\mbox{$\bullet$}}
\put(-92.5,27.5){\mbox{$\bullet$}}
\put(-92.5,-2.5){\mbox{$\bullet$}}
\put(-92.5,-32.5){\mbox{$\bullet$}}
\put(-92.5,-62.5){\mbox{$\bullet$}}
\put(-92.5,-92.5){\mbox{$\bullet$}}
\put(-92.5,-122.5){\mbox{$\bullet$}}
\put(-92.5,-152.5){\mbox{$\bullet$}}
\put(-92.5,27.5){\mbox{$\bullet$}}

\put(-2.5,147.5){\mbox{$\bullet$}}
\put(-2.5,117.5){\mbox{$\bullet$}}
\put(-2.5,87.5){\mbox{$\bullet$}}
\put(-2.5,57.5){\mbox{$\bullet$}}
\put(-2.5,27.5){\mbox{$\bullet$}}
\put(-2.5,-2.5){\mbox{$\bullet$}}
\put(-2.5,-32.5){\mbox{$\bullet$}}
\put(-2.5,-62.5){\mbox{$\bullet$}}
\put(-2.5,-92.5){\mbox{$\bullet$}}
\put(-2.5,-122.5){\mbox{$\bullet$}}
\put(-2.5,-152.5){\mbox{$\bullet$}}
\put(-2.5,27.5){\mbox{$\bullet$}}

\put(27.5,147.5){\mbox{$\bullet$}}
\put(27.5,117.5){\mbox{$\bullet$}}
\put(27.5,87.5){\mbox{$\bullet$}}
\put(27.5,57.5){\mbox{$\bullet$}}
\put(27.5,27.5){\mbox{$\bullet$}}
\put(27.5,-2.5){\mbox{$\bullet$}}
\put(27.5,-32.5){\mbox{$\bullet$}}
\put(27.5,-62.5){\mbox{$\bullet$}}
\put(27.5,-92.5){\mbox{$\bullet$}}
\put(27.5,-122.5){\mbox{$\bullet$}}
\put(27.5,-152.5){\mbox{$\bullet$}}
\put(27.5,27.5){\mbox{$\bullet$}}

\put(57.5,147.5){\mbox{$\bullet$}}
\put(57.5,117.5){\mbox{$\bullet$}}
\put(57.5,87.5){\mbox{$\bullet$}}
\put(57.5,57.5){\mbox{$\bullet$}}
\put(57.5,27.5){\mbox{$\bullet$}}
\put(57.5,-2.5){\mbox{$\bullet$}}
\put(57.5,-32.5){\mbox{$\bullet$}}
\put(57.5,-62.5){\mbox{$\bullet$}}
\put(57.5,-92.5){\mbox{$\bullet$}}
\put(57.5,-122.5){\mbox{$\bullet$}}
\put(57.5,-152.5){\mbox{$\bullet$}}
\put(57.5,27.5){\mbox{$\bullet$}}

\put(87.5,147.5){\mbox{$\bullet$}}
\put(87.5,117.5){\mbox{$\bullet$}}
\put(87.5,87.5){\mbox{$\bullet$}}
\put(87.5,57.5){\mbox{$\bullet$}}
\put(87.5,27.5){\mbox{$\bullet$}}
\put(87.5,-2.5){\mbox{$\bullet$}}
\put(87.5,-32.5){\mbox{$\bullet$}}
\put(87.5,-62.5){\mbox{$\bullet$}}
\put(87.5,-92.5){\mbox{$\bullet$}}
\put(87.5,-122.5){\mbox{$\bullet$}}
\put(87.5,-152.5){\mbox{$\bullet$}}
\put(87.5,27.5){\mbox{$\bullet$}}

\put(150,-75){\line(-2,1){300}}
\put(-150,-75){\line(2,1){300}}
\put(75,-150){\line(-1,2){150}}
\put(-75,-150){\line(1,2){150}}

\put(150,150){\mbox{(1,1) ray}}
\put(150,75){\mbox{(2,1) ray}}
\put(-150,152){\mbox{(-1,1) ray}}
\put(-160,80){\mbox{(-2,1) ray}}
\put(58,155){\mbox{(1,2) ray}}
\put(-95,155){\mbox{(-1,2) ray}}

\put(-40,5){\mbox{$e_{(-1,0)}$}}
\put(-70,5){\mbox{$e_{(-2,0)}$}}
\put(20,5){\mbox{$e_{(1,0)}$}}
\put(50,5){\mbox{$e_{(2,0)}$}}

\put(-35,36){\mbox{$e_{(-1,1)}$}}
\put(32,28){\mbox{$e_{(1,1)}$}}
\put(2,28){\mbox{$e_{(0,1)}$}}
\put(2,58){\mbox{$e_{(0,2)}$}}
\put(32,58){\mbox{$e_{(1,2)}$}}
\put(62,58){\mbox{$e_{(2,2)}$}}

\put(-220,120){\line(0,1){30}}
\put(-220,150){\vector(0,1){3}}
\put(-260,135){\mbox{$[e_{(0,1)},\ast]$}}
\put(-220,120){\line(1,0){30}}
\put(-190,120){\vector(1,0){3}}
\put(-220,110){\mbox{$[e_{(1,0)},\ast]$}}

\linethickness{1.5pt}
\put(150,-150){\line(-1,1){300}}
\put(-150,-150){\line(1,1){300}}
\put(0,-150){\line(0,1){300}}
\put(-150,0){\line(1,0){300}}

\qbezier(10,150)(55,170)(80,110)
\put(80.5,110){\vector(1,-2){3}}
\put(35,158){\mbox{${\cal O}_v$}}

\qbezier(-140,10)(-170,55)(-110,80)
\put(-110,80){\vector(2,1){3}}
\put(-165,35){\mbox{${\cal O}_h$}}

}
\end{picture}
\caption{\footnotesize $2d$ integer lattice of generators of the elliptic Hall/DIM algebra. Each ray (p,r) gives rise to a commutative subalgebra, and each pair of rays (p,r) and (-p,-r) form a Heisenberg subalgebra. Depicted at the left upper corner is the action of commutators with $e_{(1,0)}=\hat E_0$ and $e_{(0,1)}=\hat W_0$ (when the pairs are admissible).
The spin (vertical grading) is now different from Fig.\ref{YWfig} and the second Miki rotation ${\cal O}_v$ emerges.
}
\label{Dimfig}
\end{figure}
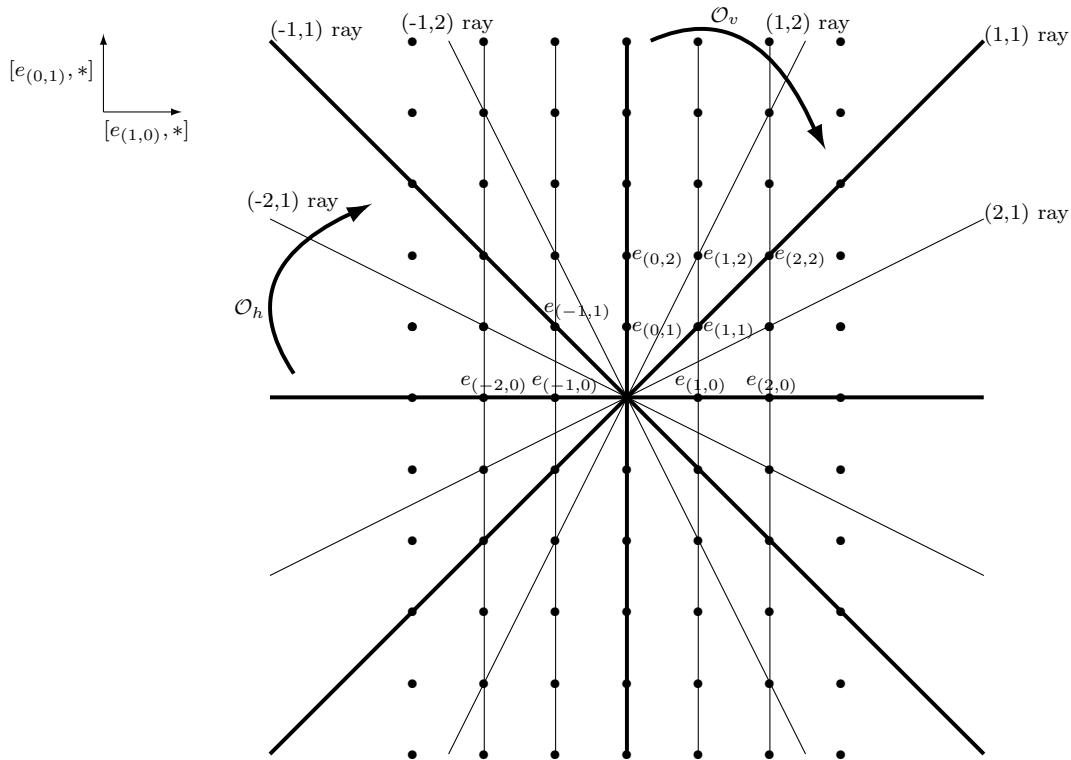

In this paper, we are going to make a next step and consider the DIM algebra \cite{DI,Miki}, or the elliptic Hall algebra \cite{K,BS,S}, which is basically the same \cite{S,Feigin}. In variance with the $W_{1+\infty}$/Yangian algebras, generators of the DIM algebra are associated with points of $2d$ integer lattice (see Fig.\ref{Dimfig}) and is a deformation of the $qW_{1+\infty}$ algebra \cite{KR,Miki} (see also sec.\ref{s.qW}). Though the two figures Fig.\ref{YWfig} and Fig.\ref{Dimfig} are looking very similar, the points of them are much different: in the first figure, the three vertical rows, $f_i$, $\psi_i$ and $e_i$ are associated with generating elements of the algebra, while all points on the rays are just repeated commutators, most of points not having definite spins\footnote{This is because the commutation relations of the algebra do not respect the spin but only the grading.} but only gradings. On the other hand, the points of the second figure do have definite spins as long as gradings (since the commutation relation respects both the spin and the grading) but the points on the rays, excluding those generating the ray are not repeated commutators. One may say that establishing a correspondence between the two pictures is exactly the problem of taking the limit from the DIM algebra to the affine Yangian algebra.

In fact, these distinctions of figures lead to two basic differences in the two cases: first of all, commutators in the upper half of the DIM algebra may not decrease the spin, while those with element $e_0$ of the $W_{1+\infty}$/Yangian algebras can, which is seen at the picture of generator action at the left upper corner of the figures.

The second difference is that, in the DIM algebra case, besides the automorphism ${\cal O}_h$, there is another automorphism, ${\cal O}_v$, which maps the ray (p,1) to ray (p+1,1). These two automorphisms are called Miki automorphisms \cite{Miki1,Miki} and are generators of the $SL(2,\mathbb{Z})$ group. This $SL(2,\mathbb{Z})$ symmetry allows one to obtain immediately the commutativity of all rays, both integer and rational, since different rays are related by Miki automorphisms, and commutativity becomes a kind of ``trivial":
in the elliptic Hall algebra formulation \cite{BS},
it is just one of the basic commutativity relations
\be\label{sb}
\left[ e_{\vec\gamma}, e_{k\vec \gamma}\right] = 0 \ \ \ \ \ \forall \vec\gamma \ {\rm and} \ k\in Z_+
\ee

Still, there is something to discuss:
\begin{itemize}

\item{}
One can wonder what this commutativity means for operators in the $N$-body representation,
where, for the ray $\vec\gamma = (0,1)$, they acquire the form of trigonometric Ruijsenaars Hamiltonians.
It turns out that their ${\cal O}_v$-rotations to rays $(p,1)$ are just substitution $q^{x_i\p_i}\longrightarrow x_i^pq^{x_i\p_i}$,
while ${\cal O}_h$-rotations for the ray $\vec\gamma = (1,0)$ to rays $(1,r)$ are non-trivial transforms, intermixing
coordinates with momenta and providing Hamiltonians with more sophisticated shift operators.

\item Similarly, in the Fock representation, ${\cal O}_v$-rotations are given just by the substitution
$$
  \exp\left(-\sum_k(1-q^k)z^{-k}\frac{\p}{\p p_k}\right)\longrightarrow
           {\frac{q^{\frac{1}{2}}}{z}}\exp\left(-\sum_k(1-q^k)z^{-k}{\frac{\p}{\p p_k}}\right)
$$
while the ${\cal O}_h$-rotation is also more complicated.

\item{} We explain how the DIM algebra can be formulated in such a way that all generators $e_{(n,m)}$ {\bf with coprime} $n$ and $m$ are obtained as repeated commutators. This looks similar to the $W_{1+\infty}$/Yangian case, however, in the latter case, the repeated commutators allowed one to construct commutative families, while, in the DIM case, the commutative elements are constructed with a rather non-linear procedure.

\item{}  This non-linear procedure can be understood from the following oversimplified analogy explaining the difference between generators $e_{(n,m)}$ with coprime and non-coprime $n$ and $m$:
the coprime elements are a kind of counterparts of the generators $E$ and $F$ in the Lie algebra $sl_2$,
while the non-coprime one substitutes the Cartan generator $H$.
Under the quantum group deformation, the basic commutation relation is $[E,F]\sim q^H - q^{-H}$, i.e. the commutator of two coprime generators is exponential of the non-coprime one.
In the loop algebra case, the exponential is expanded into the complete homogeneous symmetric polynomials,
$\exp\left( \sum_n H_n{\frac{z^n}{n}}\right) = \sum_n z^n h_n\{H_k\}$.
Thus one can expect that commutators of the coprime generators have to be related to the complete homogeneous symmetric polynomials of the non-coprime (ray) generators. And this is what actually happens.

\item{} Transition/limit of the whole picture to the Yangian case is quite non-trivial.
This is seen already at the level of the ``one-body" representation of $W_{1+\infty}$,
where one needs to get operators $z^mD^n$ from $z^mq^{nD}$ in the limit $q\longrightarrow 1$.
This requires non-trivial linear transformations, which automatically imply a restriction to $n\geq 0$.
Technically, this means one has to use commutative families uniting a bunch of rays, which were called cones in \cite{MMMP1}.
Thus it is not surprising, that, at the Yangian level, commutativity even of integer rays
was a non-trivial statement.
Moreover, it is still unclear if non-integer rays and cones survive the $\beta$-deformation from the $W_{1+\infty}$ algebra to the affine Yangian one.

\item{} This implies that constructing cones in the DIM algebra is of importance for taking the Yangian limit of the commutative subalgebras. However, the commutative cones are not that simply organized: in variance with the $_{1+\infty}$ algebra, they are non-linear combinations of algebra generators and, technically, the simplest way to construct them is to use a peculiar inner automorphism ${\cal O}(u)$.

\end{itemize}

Our main object of interest in this paper are commutative subalgebras of the elliptic Hall/DIM algebra.
We follow the following scheme in our presentation. In section 2, we discuss a starting model for the quantum toroidal algebra, the $qW_{1+\infty}$ algebra. It possesses many typical features of the quantum toroidal algebra, in particular, the $SL(2,\mathbb{Z})$ group of automorphisms. Then, in section 3, we explain how one can construct generators $e_{(n,m)}$ with coprime $n$ and $m$ of the elliptic Hall/DIM algebra using repeated commutators. After this, in section 4, we already discuss the full-fledged algebra.
In the next four sections, we discuss representations of the elliptic Hall/DIM algebra: one-bode, $N$-body, Fock representations. In particular, we discuss in section 6 a new series of integrable many-body systems given by the commutative families associated with all rays. Section 9 is devoted to the matrix model partition functions that can be associated with commutative families: rays and cones, ni the Fock representation. As a technical tool, we use the description of automorphisms ${\cal O}_h$ and ${\cal O}(u)$ in the basis of Macdonald polynomials. In section 10, we construct the eigenfunctions of the commutative families. At last, in section 11, we briefly touch the problem of limit of the elliptic Hall/DIM algebra to the Yangian algebra.
We end with a summary in the Conclusion (section 12) and Appendices for some side, but closely related stories.

\paragraph{Notation.} In the paper, we denote through $P(x_i)$ symmetric polynomials as functions of $x_i$, and through $P\{p_k\}$, these polynomials as graded polynomials of power sums $p_k=\sum_ix_i^k$. In particular, $M_\lambda$ denotes the Macdonald polynomial labelled by the Young diagram (partition) $\lambda$ with $l_\lambda$ lines ($l_\lambda$ parts): $\lambda_1\ge \lambda_2\ge\ldots\ge \lambda_{l_\lambda}>0$, and the grading of this Macdonald polynomial is $|\lambda|:=\sum_i\lambda_i$.
We use the notation $\overline{M}_\lambda$ for the Macdonald polynomial with parameters $q$ and $t$ permuted, and use the notation $\lambda^\vee$ for the transposed Young diagram.

We denote through $h_n\{p_k\}$'s the complete homogeneous symmetric polynomials defined as
\be\label{h}
\exp\Big(\sum_k{\frac{p_kz^k}{k}}\Big)=\sum_m{h_n\{p_k\}z^n}\nn
\ee
In particular, $h_{k}=0$ at $k<0$.
We also denote through $\mathfrak{e}_n\{p_k\}$' the elementary symmetric polynomials\footnote{They are also known as Schur/Jack/Macdonald functions for totally antisymmetric representations,  when they all are, in fact, the same.} defined as
\be\label{e}
\exp\Big(\sum_k(-1)^{k+1}{\frac{p_kz^k}{k}}\Big)=\sum_m{\mathfrak{e}_n\{p_k\}z^n}\nn
\ee

Rewriting monomials in the form $p_\Delta=\prod_{k=1} p_k^{m_k}$, we define the scalar product for polynomials of $p_k$ induced by
\be
<p_{\Delta},p_{\Delta'}>=z_\Delta\, \delta_{\Delta,\Delta'}
\prod_i{\frac{\{q^{\delta_i}\}}{\{t^{\delta_i}\}}}\nn
\ee
where $z_\Delta:=\prod_k k^{m_k}m_k!$ is the order of automorphism of the Young diagram $\Delta$, $p_\Delta=p_{\Delta_1}p_{\Delta_2}\ldots
p_{\Delta_{l_\Delta}}$,
and we have scaled the original inner product of Macdonald \cite{Mac} by the factor of $(q/t)^{|\Delta|}$.
The orthogonality relations for the Macdonald polynomials with this scalar product are
\be\label{MOR}
\Big<M_\lambda\{p_k\},M_\nu\{p_k\}\Big>=||\lambda||^2\, \delta_{\lambda,\nu}
\ee
where
\be\label{hook}
||\lambda||^2:=\prod_{(i,j)\in\lambda}{1-t^{\lambda_j^\vee-i}q^{\lambda_i-j+1}\over 1-t^{\lambda_j^\vee-i+1}q^{\lambda_i-j}}
\ee
For details related to the notation, see \cite{Mac,Fulton}.

\section{$qW_{1+\infty}$ algebra: an approach to the DIM algebra\label{s.qW}}

One can choose in the completion of the $W_{1+\infty}$ algebra the Kac-Radul basis \cite{KR,Miki} with elements $u_{m,i}$ that commute in the same way as operators $z^mq^{i\hat D}$, where $D:=z{d\over dz}$, i.e. the commutation relations in this basis are
\be\label{KR}
[u_{m,i},u_{n,j}]=(q^{mj}-q^{ni})u_{m+n,i+j}
\ee
This algebra is sometimes called $qW_{1+\infty}$ algebra, and the representation in terms of operators $z^mq^{i\hat D}$ is called one-body representation.

One can also introduce the central extension of this algebra \cite{Miki}:
\be\label{KRc}
[u_{m,i},u_{n,j}]=(q^{mj}-q^{ni})u_{m+n,i+j}+\delta_{m+n}\delta_{i+j}(c_1m+c_2i)q^{mi}
\ee
and it is still a Lie algebra. There is a deformation $U_t(qW_{1+\infty})$ of the universal enveloping $U(qW_{1+\infty})$ of this algebra \cite{Miki}, which is exactly the Ding-Iohara-Miki (DIM) algebra \cite{DI,Miki}. We discuss it in the next sections, and here we consider the Lie algebra $qW_{1+\infty}$.

In the one-body representation of the $qW_{1+\infty}$ algebra, there is no central extension, $c_1=c_2=0$, and the algebra is realized by the operators $u_{m,i}=z^mq^{i\hat D}$ with the commutation relations given by (\ref{KR}). In this case, one can easily enumerate the commutative subalgebras: they are given by arbitrary integer pairs $(p,r)$ and consist of elements $u_{np,nr}$. In other words, they are associated with an arbitrary straight line passing through the origin and through at least one point (and, hence, infinitely many points) of the integer lattice $\mathbb{Z}^2$.

Note that, in the one-body representation, one can generate commutative $(1,r)$ subalgebras by simple rotations from the vertical line $u_{0,n}=q^{n\hat D}$, and $(p,1)$ subalgebras, from the horizontal line $u_{n,0}=z^n$. This is done by the operators
\be\label{O1}
\hat{\cal O}_v:={1\over\sqrt{z}}z^{{1\over 2}\log_q z}\nn\\
\hat{\cal O}_h:=q^{-{1\over 2}\hat D(\hat D-1)}
\label{1bodyrotations}
\ee
so that
\be\label{O2}
\hat{\cal O}_v^{-1}\cdot q^{\hat D}\cdot \hat{\cal O}_v=zq^{\hat D}\nn\\
\hat{\cal O}_v^{-p}\cdot q^{\hat D}\cdot \hat{\cal O}_v^p=z^pq^{\hat D}
\ee
and
\be\label{O3}
\hat{\cal O}_h\cdot z\cdot \hat{\cal O}_h^{-1}=zq^{\hat D}\nn\\
\hat{\cal O}_h^{r}\cdot z\cdot \hat{\cal O}_h^{-r}=zq^{r\hat D}
\ee
i.e.
\be\label{O4}
\hat{\cal O}_v^{-p}\cdot u_{0,n}\cdot \hat{\cal O}_v^p\propto u_{pn,n}\nn\\
\hat{\cal O}_h^{-r}\cdot u_{n,0}\cdot \hat{\cal O}_h^{r}\propto u_{n,-rn}
\ee
The operator $\hat{\cal O}_h$ is nothing but the rotation operator $\hat{\cal O}$ that we discussed in \cite{Ch1,Ch2,MMMP1,MMMP2}.
We will see in the next sections that, in some representations, manifest constructions of the operators $\hat{\cal O}_v$ and $\hat{\cal O}_h$ is available, and is sometimes very simple.

In the case of non-zero central charges, as follows from (\ref{KRc}), the same $(p,r)$ line gives rise to a Heisenberg subalgebra, and only its half describes the commutative subalgebra. As we explain in the next section, this property survives the deformation of the $U(qW_{1+\infty})$ algebra to the DIM algebra.

Note that one can also construct commutative subalgebras of the form
\be\label{cone}
\hat H_1=\sum c_{n_1}u_{1,n_1}\nn\\
\hat H_2=\sum q^{n_2}c_{n_1}c_{n_2}u_{2,n_1+n_2}\nn\\
\hat H_3=\sum q^{n_2+2n_3}c_{n_1}c_{n_2}c_{n_3}u_{3,n_1+n_2+n_3}\nn\\
\ldots
\ee
This kind of commutative subalgebras were called {\bf cones} in \cite{MMMP1}. With a proper choice of the coefficients $c_n$, one can obtain at $q\to 1$ in the leading order the operator $\hat D$ of any degree, reproducing this way all the commutative subalgebras of $W_{1+\infty}$ in \cite{MMMP1}. Unfortunately, this trick seems not to work that immediately in the affine Yangian limit of the DIM algebra below.

\section{Basic construction of DIM algebra}

Now we proceed to a deformation of the $qW_{1+\infty}$ algebra to the DIM algebra  $U_{q,t}(\widehat{\widehat{\mathfrak{gl}}}_1)$.
It has different descriptions,
of which the most interesting for us will be that in terms of elliptic Hall algebra surveyed in sec.\ref{ellHall} below.
Its the main advantage is that the ray commutativity (\ref{sb}) is among the main {\it postulates},
and the only question is an explicit construction of operators $e_{k\vec\gamma}$ associated with point of $2d$ integer points.
Remarkably, for coprime $(\gamma_1,\gamma_2)$ in $\vec\gamma$, there is quite a universal and {\it simple} description
of $e_{\vec\gamma}$, and this is what we begin in this short section.

Introduce
\be\label{ena3}
e_{(1,m)}&=&E_m,\ \ \ \ \ m\ge 0\nn\\
e_{(n,kn+a)}&=&E_k^{(n,a)},\ \ \ \ \  a=1\dots n-1,\ gcd(a,n)=1,\ k\ge 0,\ n>1
\ee
and we are going to express all $E_k^{(n,a)}$ through $E_m$  recursively in $n$. The notation $E_m$ for positive generating elements of algebra follows our notation in \cite{MMMP1}.
For the sake of simplicity, we consider only positive quadrant of the algebra, see Fig.\ref{Dimfig}, i.e. the case of $e_{(n,m)}$ with $n\ge 0$, $m>0$.

At $n=2$ we get just  $E_k^{(2,1)}=[E_{k+1},E_k]$.
Afterwards we construct $E_k^{(3,a)}$, $a=1,2$ from two $E$'s of smaller grading,
then $E_k^{(4,a)}$, $a=1,3$ from two $E$'s of smaller grading, etc:
\be
E^{(2,1)}_k&=&[E_{k+1},E_k]\nn\\
E_k^{(3,1)}&=&[[E_{k+1},E_k],E_k]\nn\\
E_k^{(3,2)}&=&[[E_{k},E_{k+1}],E_{k+1}]\nn\\
 E_k^{(4,1)}&=& [[[E_{k+1},E_k],E_k],E_k]\nn\\
 E_k^{(4,3)}&=&[[[E_{k+1},E_k],E_{k+1}],E_{k+1}]
 \label{Ecoprime}
\ee
This recursion can be easily continued, and we explain in sec.\ref{334} the details.

At the same time, the calculation in the case of $e_{(n,m)}$ with $gcd(n,m)\ne 1$ is more tricky,
and $e_{(kn,km)}$ is {\it not} given just by repeated commutators.
This non-coprime case is the point of our main interest in this paper,
since it describes the commutative rays.

An explanation of why the non-coprime case is more involved is as follows. First of all,  the whole algebra may be given by elements $e_{(\pm 1,m)}$ and by commutative elements $e_{(0,m)}$, how all $e_{(n,m)}$ with coprime $n$ and $m$ are generated from $e_{(1,m)}$ for the positive quadrant we just described. Now, the commutators of $e_{(1,m)}$ and $e_{(-1,m')}$ naturally give rise to the complete homogeneous symmetric polynomials (i.e. the Schur functions corresponding to the symmetric Young diagrams) of elements of the vertical ray just as it happens in Drinfeld's new realization of quantum affine algebras \cite{D,Beck}, or in quantum loop algebras. Basically, it comes from exponentiation of the Cartan elements under quantum deformations. It remains to note that there is a natural $SL(2,\mathbb{Z})$-symmetry acting on the integer lattice of the generators $e_{(n,m)}$. Hence, all $e_{(kn,km)}$ lying on the ray going from the origin can be obtained from the vertical ray by an $SL(2,\mathbb{Z})$-rotation, and they are also related with the repeated commutators like (\ref{Ecoprime}) via the complete homogeneous symmetric polynomials.

Thus, it turns out that the generators with coprime $\gamma_1$ and $\gamma_2$ and their descendants along the rays $k\vec\gamma$
possess a kind of complementary properties: say, a realization like (\ref{Ecoprime}) is rather immediate, but the commutation relations for the elements with coprime $\gamma_1$ and $\gamma_2$ are not always that simple. On contrary, the commutation relations (\ref{sb}) of their descendants along the rays $k\vec\gamma$ are very simple, but a realization of this elements is rather involved (see sec.\ref{ellHall}).

This explains why we need to go deeper in the details of the definitions
and realizations of particular representations.
Still, at some moments the simplicity of (\ref{Ecoprime}) will show up again and again,
pointing to the possibility of further improvements in the study of commutative rays.

\section{Elliptic Hall algebra/DIM algebra: definitions and properties\label{ellHall}}

\subsection{Elliptic Hall algebra\label{eHa}}

Let $q_1$, $q_2$ and $q_3$ be formal parameters satisfying $q_1q_2q_3=1$.

The elliptic Hall algebra {\bf E} is a deformation of the $qW_{1+\infty}$ algebra, which is an associative algebra, multiplicatively generated by the central elements $\mathfrak{q}^{c_1}$, $\mathfrak{q}^{c_2}$ and by the elements\footnote{The parameter $\mathfrak{q}$ is here either of $q_{_{1,2,3}}^{1\over 2}$. We introduce this parameter for the following reason: an important feature of the basis of elements $e_\gamma$ is that it is a deformation of the basis of $u_{n,i}$ in the $qW_{1+\infty}$ algebra, and one can obtain the non-deformed case bringing either of $q_{_{1,2,3}}$ to 1. For instance, if $\mathfrak{q}=q_3\to 1$, then $u_{n,k}=q_1^{nk/2}(q_1^k-1)e_{(n,k)}$. However, the central charges should not be additionally redefined when taking this limit only if one parameterizes them as in (\ref{1}). Besides, this parametrization of the central charges matches earlier notations.} $e_{\vec{\gamma}}$, with $\vec{\gamma} \in \mathbb{Z}^2\setminus \{(0,0)\}$, satisfying a set of commutation relations \cite{Feigin,BS,Zenk}
\begin{itemize}
\item For any $\vec\gamma\in\mathbb{Z}^2/\{(0,0)\}$ such that $gcd(\gamma_1,\gamma_2)=1$
\be\label{1}
\phantom{.}[e_{n\vec\gamma},e_{m\vec\gamma}]={n\over\kappa_n}\delta_{n+m,0}(\mathfrak{q}^{n\vec c\cdot\vec\gamma}-\mathfrak{q}^{-n\vec c\cdot\vec\gamma})
\ee
where
\be
\kappa_n:=(1-q_1^n)(1-q_2^n)(1-q_3^n),\ \ \ \ \ \ \ \ \vec c=(c_1,c_2)
\ee
 This form of $\kappa_n$ is the reason to call the algebra {\it elliptic}.
\item For any $\vec\alpha,\ \vec\beta\in\mathbb{Z}^2/\{(0,0)\}$, $\vec\alpha\ne\vec\beta$ we put $\vec\alpha>\vec\beta$ if $\alpha_1>\beta_1$, or if $\alpha_1=\beta_1$ and $\alpha_2>\beta_2$. We call a pair $(\vec\alpha,\vec\beta)$ admissible if the triangle $(-\vec\beta,0,\vec\alpha)$ does not contain any $\mathbb{Z}^2$ points in the interior and on at least two of its edges\footnote{See Appendix~\ref{sec:admissible-algorithm} for the evaluation procedure.}. Let us remove from the triple the largest and the smallest vectors, and denote the remaining one $\vec m_{(\vec\alpha,\vec\beta)}$. Then, for any admissible pair $\vec\alpha,\ \vec\beta\in\mathbb{Z}^2/\{(0,0)\}$ such that $\alpha_1\beta_2>\alpha_2\beta_1$,
    \be\label{2}
    \phantom{.}[e_{\vec\alpha},e_{\vec\beta}]={1\over\kappa_1}\mathfrak{q}^{\vec c\cdot \vec m_{(\vec\alpha,\vec\beta)}}\mathfrak{h}_{\vec\alpha+\vec\beta}
    \ee
where for any $\vec\gamma\in\mathbb{Z}^2/\{(0,0)\}$ such that $gcd(\gamma_1,\gamma_2)=1$,
\be
\mathfrak{h}_{n\vec\gamma}= h_n\{p_k=-\kappa_ke_{\vec\gamma}\}
\ee
Note that they can be also written in the form involving the Macdonald polynomials naturally associated with this algebra:
\be
h_n\{p_k=-\kappa_ke_{\vec\gamma}\}=(-1)^k\mathfrak{e}_k\{p_k=\kappa_ke_{\vec\gamma}\}=
(-1)^kM_{[1^n]}\{p_k=\kappa_ke_{\vec\gamma}\}\nn
\ee
\end{itemize}
In fact, the very fact of emerging the complete homogeneous symmetric
polynomials of commutative algebra generators is quite expected \cite{D,G} and is similar to the structure of the quantum loop algebra \cite[sec.1.8]{S2}, however, its derivation for the elliptic Hall algebra requires some work (see \cite[sec.5.3]{BS} and previous \cite[sec.6.3]{S2}).

We would like to emphasize that, while the elliptic Hall algebra definition may seem
involved, it is, in fact, reasonably suitable for computerization
already in this form: one can algorithmically
check whether a given pair is admissible (see Appendix~\ref{sec:admissible-algorithm} for the procedure), and calculate the non-linear r.h.s. in non-admissible cases.

A PBW type basis in the algebra {\bf E} can be given as follows: with any element $e_{\vec \gamma}$, one associates an angle $\pi\ge\theta_{\vec\gamma}\ge -\pi$ of $\gamma$ w.r.t. the horizontal axis, and introduces an ordering $\vec\gamma>\vec\gamma'$ if $\theta_\gamma>\theta_{\gamma'}$ or, in the case of $\theta_\gamma=\theta_{\gamma'}$, $\gamma_1>\gamma_1'$. Then, the basis consists of all ordered elements $e_{\vec\gamma_1}e_{\vec\gamma_2}\ldots e_{\vec\gamma_l}$ with arbitrary $l$.

To summarize, the algebra is formed by words made from multiplication of elements $e_{\vec\gamma}$,
and some words are identified by the rules (\ref{1}) and (\ref{2}),
with the second set of relations imposed  {\bf only} on admissible pairs.
No other relations are present in this associative algebra,
and commutator of any non-admissible pair can be obtained through successive commutators of admissible pairs. In particular, when studying the commutative families,
formed by $e_{n\vec \gamma}$ (see Section~\ref{sec:commutative-subalgebras}),
in order to obtain manifest expressions for the Hamiltonians from the
(multiplicative) basis elements $e_{\pm 1, 0}$, $e_{0, \pm 1}$
one has to
deduce a path that it includes only
admissible pairs, see the discussion in the
Section~\ref{sec:commutative-subalgebra-pr}.
These commutation relations are enough to make independent only the ordered words,
making the algebra similar in spirit to the ordinary universal enveloping algebras.
Commutators are expressed by (\ref{2}) non-linearly, with the help of symmetric functions,
and the consistency of the algebra, i.e. independence of the operator from the way/sequence
in which it is expressed/ordered,
is quite sensitive to the choice of these functions.

\subsection{DIM algebra\label{DIM}}

The (quantum toroidal) Ding-Iohara-Miki (DIM) algebra $U_{q_1,q_2,q_3}(\hat{\hat{\mathfrak{gl}_1}})$ \cite{DI,Miki} (without the grading operators \cite{Feigin}) is isomorphic \cite{S} to the elliptic Hall algebra, the isomorphism being given by the identification
\be\nn
X_n^{\pm}&\sim& e_{(\pm 1,n)},\ \ \ \ \ \ \ n\in\mathbb{Z}\nn\\
X_{n}^0&\sim& e_{(0,n)},\ \ \ \ \ \ \  n\in\mathbb{Z}/\{0\}\nn\\
X_0^0&\sim& \mathfrak{q}^{c_1}
 \ee
 so that (\ref{2}) gives rise to the commutation relations
\be
\phantom{.}[X_m^0,X_n^{\pm}]=\pm \mathfrak{q}^{{m\mp |m|\over 2}c_2}\ X_{n+m}^\pm\nn\\
\phantom{.}[X_n^0,X_m^0]={n\over\kappa_n}\delta_{n+m,0}\Big(\mathfrak{q}^{nc_2}-\mathfrak{q}^{-nc_2}\Big)\nn\\
\phantom{.}[X_n^+,X_m^-]={1\over\kappa_1}\bigg(\mathfrak{q}^{c_1-mc_2}h_{n+m}\left(-\kappa_kX^0_k\right)
-\mathfrak{q}^{-c_1+mc_2}h_{-n-m}\left(-\kappa_kX^0_{-k}\right)\bigg)\nn
\ee
and, additionally, there are the Serre relations\footnote{The Serre relations are often imposed in the form
\be
\hbox{Sym}_{ijk}[X_i^{\pm},[X_{j+1}^{\pm},X_{k-1}^{\pm}]]=0\nn
\ee
however, all these relations are reduced to (\ref{Serre}) if one takes into account the commutation relations $[X_n^0,X_m^{\pm}]$ \cite{Ts}.}
\be\label{Serre}
\phantom{.}[X_n^{\pm},[X_{n+1}^{\pm},X_{n-1}^{\pm}]]=0
\ee
and the quadratic relations
\be
\phantom{.}[X^+_ m,X^+_n]-[X^+_ {m+3},X^+_{n-3}]+\xi_+(X^+_{n-1}X^+_{m+1}+X^+_{m+2}X^+_{n-2})
-\xi_-(X^+_{m+1}X^+_{n-1}+X^+_{n-2}X^+_{m+2})=0\nn\\
\phantom{.}[X^-_ m,X^-_n]-[X^-_ {m+3},X^-_{n-3}]+\xi_-(X^-_{n-1}X^-_{m+1}+X^-_{m+2}X^-_{n-2})
-\xi_+(X^-_{m+1}X^-_{n-1}+X^-_{n-2}X^-_{m+2})=0\nn\\
\xi_\pm:=\sum_iq_i^{\pm 1}\nn
\ee
This describes the algebra of elements $e_{(n,\pm 1)}$, $e_{(n,0)}$ and $c_{1,2}$ that are enough to generate the elliptic Hall algebra.
In fact, the elliptic Hall algebra is generated just by 6 elements $e_{(\pm 1,0)}$, $e_{(0,\pm 1)}$ and $c_{1,2}$ \cite{BS}.

Let us now describe how to manifestly construct commutative subalgebras of the whole algebra starting from these elements. It would give us a technical tool to construct the whole series of commuting Hamiltonians starting from a pair of algebra elements in the concrete representation.

\subsection{Commutative subalgebras}\label{sec:commutative-subalgebras}

\subsubsection{The commutative subalgebra $(1,1)$}

As follows from (\ref{1}), any $\vec\gamma$ such that $gcd(\gamma_1,\gamma_2)=1$ gives rise to a Heisenberg subalgebra, and its half is a commutative subalgebra. Hence, we are interested in the quarter of this algebra, $e_{(n,m)}$ with $n,m\ge 0$ and will discuss how to construct the commutative subalgebras $e_{n\vec\gamma}$ manifestly.

Note that commutators in the quarter of the algebra do not involve central charges: commuting $e_{(n_1,m_1)}$ and $e_{(n_2,m_2)}$ with non-negative $n_i$, $m_i$, one constructs the triangle $[(-n_2,-m_2),(0,0),(n_1,m_1)]$ determining if the pair is admissible. By the ordering rule, the three vectors are ordered: $(-n_2,-m_2)<(0,0)<(n_1,m_1)$, i.e. $m_{(n_1,m_1),(n_2,m_2)}=(0,0)$, and $\mathfrak{q}^{\vec c\cdot \vec m_{(n_1,m_1),(n_2,m_2)}}=1$.

Let us first start with the simplest case of $\gamma=(1,1)$ (see Fig.\ref{Dimfig}). Then, one can easily obtain (we use the notation $\hat W_0$ and $\hat E_k$ in order to make a close contact with the notation of \cite{MMMP1}):
\be\label{cr}
\phantom{.}[\underbrace{e_{(0,1)}}_{\hat W_0},\underbrace{e_{(1,0)}}_{\hat E_0}]&=&-{1\over\kappa_1}\mathfrak{h}_{(1,1)}=-{1\over\kappa_1}h_1(-\kappa_ke_{(k,k)})
=\underbrace{e_{(1,1)}}_{\hat E_1}:=\hat H_{1}^{(1,1)}\nn\\
\phantom{.}[e_{(0,1)},e_{(1,1)}]&=&e_{(1,2)}\nn\\
\phantom{.}[e_{(1,0)},e_{(1,2)}]&=&{1\over\kappa_1}\mathfrak{h}_{(2,2)}={1\over\kappa_1}h_2(-\kappa_ke_{(k,k)})
={\kappa_1\over 2}e_{(1,1)}^2-{\kappa_2\over 2\kappa_1}e_{(2,2)}:=\hat H_{2}^{(1,1)}\nn\\
\phantom{.}[e_{(1,2)},e_{(1,1)}]&=&e_{(2,3)}\nn\\
\phantom{.}[e_{(1,0)},e_{(2,3)}]&=&-{\kappa_1^2\over 6}e_{(1,1)}^3+{\kappa_2\over 2}e_{(2,2)}e_{(1,1)}-{\kappa_2\over 3\kappa_1}e_{(3,3)}:=\hat H_{3}^{(1,1)}\nn\\
\phantom{.}[e_{(2,3)},e_{(1,1)}]&=&e_{(3,4)}\nn\\
\phantom{.}[e_{(1,0)},e_{(3,4)}]&=&{\kappa_1^3\over 24}e_{(1,1)}^4-{\kappa_1\kappa_2\over 4}e_{(1,1)}^2e_{(2,2)}+{\kappa_2^2\over 8\kappa_1}e_{(2,2)}^2+{\kappa_3\over 3}e_{(1,1)}e_{(3,3)}-{\kappa_4\over 4\kappa_1}e_{(4,4)}:=\hat H_{4}^{(1,1)}\nn\\
\ldots
\ee
Hence, the elements $\hat H_{1}^{(1,1)}=\hat E_1$ and\footnote{We define {\bf ad} with the right action.}
\be\label{1,1}
\boxed{
\hat H_{k}^{(1,1)}:=[\hat E_0,[\ldots[\hat W_0,\underbrace{\hat E_1],\hat E_1]\ldots,\hat E_1]]}_{k-1\ times}]=
[\hat E_0,\hbox{ad}^{k-1}_{\hat E_1}\hat W_0]={1\over\kappa_1} h_k\{p_k=-\kappa_ke_{(k,k)}\}
\ \ \ \ \ \ \hbox{for}\ k\ge 0
}
\ee
form a commutative family.

One can similarly generate the commutative family $e_{(0,r)}$:
\be\label{1,r-alt}
\phantom{.}[e_{(1,1)},e_{(-1,0)}]&=&{\mathfrak{q}^{c_1}\over\kappa_1}\mathfrak{h}_{(0,1)}=\mathfrak{q}^{c_1}e_{(0,1)}\nn\\
\phantom{.}[e_{(1,1)},e_{(0,1)}]&=&-e_{(1,2)}\nn\\
\phantom{.}[e_{(1,2)},e_{(-1,0)}]&=&{\mathfrak{q}^{c_1}\over\kappa_1}\mathfrak{h}_{(0,2)}\nn\\
\phantom{.}[e_{(1,2)},e_{(0,1)}]&=&-e_{(1,3)}\nn\\
\phantom{.}[e_{(1,3)},e_{(-1,0)}]&=&{\mathfrak{q}^{c_1}\over\kappa_1}\mathfrak{h}_{(0,3)}\nn\\
\ldots
\ee
i.e. \cite{MMgenM}
\be\label{0,r}
\hat H_{k}^{(0,1)}:=[e_{(-1,0)},\hbox{ad}_{e_{(0,1)}}^{n-1}e_{(1,1)}]={\mathfrak{q}^{c_1}\over \kappa_1}\mathfrak{h}_{(0,n)}\{-\kappa_ke_{(0,k)}\}
\ \ \ \ \ \ \hbox{for}\ n>1
\ee

Note that this construction gives rise to a different combination of the Hamiltonians constructed from elements $e_{(k,k)}$ as compared with those discussed in \cite{Ch3}: there we used the relations
\be
\phantom{.}[e_{(2,2)},e_{(0,1)}]=e_{(2,3)}\nn\\
\phantom{.}[e_{(3,3)},e_{(0,1)}]=e_{(3,4)}\nn\\
\ldots\nn
\ee
in order to construct
\be
\boxed{
[{\hat {\bar H}}_{k}^{(1,1)},\hat W_0]:=[\ldots[\hat W_0,\underbrace{\hat E_1],\hat E_1]\ldots,\hat E_1]]}_{k-1\ times}\ \ \ \ \ \ \hbox{for}\ k>0
}\nn
\ee
so that ${\hat {\bar H}}_{k}^{(1,1)}=e_{(k,k)}$ up to an entry commutative with $\hat W_0$, i.e. up to a polynomial of many variables $e_{(0,j)}$. However, requiring commutativity of ${\hat {\bar H}}_{k}^{(1,1)}$, one can exclude this polynomial.

Technically, the first way of doing, (\ref{1,1}) is more direct,
and we use it throughout the paper.

The reason for such a construction in variance with \cite{MMMP1} is that the DIM algebra in the non-deformed limit, $qW_{1+\infty}$, as we discussed in the previous section, is generated by the
one-body operators $u_{\vec\alpha}=z^{\alpha_1}q^{\alpha_2\hat D}$ with $\alpha_{1,2}\in\mathbb{Z}$, while the $W_{1+\infty}$ algebra considered in \cite{MMMP1}, by $w_{\vec\alpha}=z^{\alpha_1}\hat D^{\alpha_2}$ with $\alpha_1\in\mathbb{Z}$, $\alpha_2\in\mathbb{Z}_{>0}$ (i.e. $qW_{1+\infty}$ corresponds to the doubled $W_{1+\infty}$ algebra) so that the commutation relations are (see (\ref{KR}))
\be
[u_{\vec\alpha},u_{\vec\beta}]=(q^{\alpha_1\beta_2}-q^{\beta_1\alpha_2})u_{\vec\alpha+\vec\beta}\nn\\
\phantom{.}[w_{\vec\alpha},w_{\vec\beta}]=\sum_{k=2}^{\alpha_1+\beta_1}
C_k(\vec\alpha,\vec\beta)w_{(\alpha_1+\beta_1-k,\alpha_2+\beta_2)}\nn
\ee
where $C_k(\vec\alpha,\vec\beta)$ are some (binomial) coefficients.
Thus, in the $W_{1+\infty}$ case, the spin of the commutator is less than the sum of spins of the operators in the commutator by two or more, while, in the $qW_{1+\infty}$ case, the spin of the commutator is just equal to the sum of spins of the operators in the commutator.

\subsubsection{The commutative subalgebra $(1,2)$}

Similarly, one can construct the commutative subalgebra given by the ray $(1,2)$:
\be
\phantom{.}[e_{(1,1)},e_{(0,1)}]=[\hat E_1,\hat W_0]=-\underbrace{e_{(1,2)}}_{\hat E_2}=H_1^{(1,2)}\nn\\
\phantom{.}[e_{(0,1)},e_{(1,2)}]=e_{(1,3)}=[\hat W_0,\hat E_2]\nn\\
\phantom{.}[e_{(1,1)},e_{(1,3)}]={1\over\kappa_1}\mathfrak{h}_{(2,4)}=[\hat E_1,[\hat W_0,\hat E_2]]=H_2^{(1,2)}\nn\\
\phantom{.}[e_{(1,3)},e_{(1,2)}]=-e_{(2,5)}\nn\\
\phantom{.}-[e_{(1,1)},e_{(2,5)}]={1\over\kappa_1}\mathfrak{h}_{(3,6)}=[\hat E_1,[[\hat W_0,\hat E_2],\hat E_2]]=H_3^{(1,2)}\nn
\ee
and generally
\be\label{1,2}
\boxed{
\hat H_{k}^{(1,2)}:=[\hat E_1,\hbox{ad}^{k-1}_{\hat E_2}\hat W_0]={1\over\kappa_1} h_k\{p_k=-\kappa_ke_{(k,2k)}\}
\ \ \ \ \ \ \hbox{for}\ k\ge 0
}
\ee
Again, one can construct the Hamiltonians ${\hat {\bar H}}_{k}^{(1,2)}=e_{(k,2k)}$ using the formula
\be
\boxed{
[e_{(k,2k)},\hat W_0]=\hbox{ad}^{k-1}_{\hat E_2}\hat W_0=[\hat{\bar H}_k^{(1,2)},\hat W_0]
\ \ \ \ \ \ \hbox{for}\ k>0
}\nn
\ee

\subsubsection{The commutative subalgebra $(p,r)$}\label{sec:commutative-subalgebra-pr}

These Hamiltonians are immediately extended to the general ray $(1,r)$. First of all, one constructs the generating elements
\be
\hat E_r=e_{(1,r)}=\underbrace{[\hat W_0,\ldots[\hat W_0}_{r-1\ times},\hat E_1]\ldots]\ \ \ \ \ \ \hbox{for}\ r\ge 1\nn
\ee
Then, the Hamiltonians are given by the formula
\be\label{1,r}
\boxed{
\hat H_{k}^{(1,r)}:=[\hat E_{(r-1)},\hbox{ad}_{\hat E_r}^{k-1}\hat W_0]={1\over\kappa_1} h_k\{p_k=-\kappa_ke_{(k,rk)}\}
\ \ \ \ \ \ \hbox{for}\ k\ge 1
}
\ee
In fact, the general structure of the Hamiltonians is always
\be\label{ABC}
\boxed{
\hat H_{k}^{(p,r)}:=[\hat A,\hbox{ad}_{\hat B}^{k-1}\hat C]={1\over\kappa_1} h_k\{p_k=-\kappa_ke_{(pk,rk)}\}
\ \ \ \ \ \ \hbox{for}\ k\ge 1
}
\ee
where
\be
\hat B=[\hat A,\hat C]=e_{(p,r)}\nn
\ee
and $\hat A$ and $\hat C$ can be chosen in different ways\footnote{This may result in a sign of the expression, see also \cite[sec.5]{S}.}. For instance, for the ray (2,3), one can choose $\hat A=e_{(1,1)}$, $\hat C=e_{(1,2)}$ or $\hat A=e_{(1,2)}$, $\hat C=e_{(1,1)}$.

One foolproof way\footnote{That we, without loss of generality, describe for $p>0$, $r>0$}
to choose a splitting is to utilize the Euclid algorithm.
The elliptic Hall algebra has two automorphisms (see sec.\ref{Miki} below): $\mathcal{O}_h$ and $\mathcal{O}_v$ that act on generators
as follows
\begin{align}\label{Oh1}
  \mathcal{O}_h : e_{(p,r)} \rightarrow e_{(p, p-r)} \\
  \mathcal{O}_v : e_{(p,r)} \rightarrow e_{(p+r, r)}
\label{Ov1}
\end{align}
We will use the fact that, since these maps are automorphisms, they do not change the admissibility of pairs.

Hence, for any given element $\hat{B} = e_{(p,r)}$ with coprime $p$ and $r$, we
write down a sequence of $\mathcal{O}_h^{-1}$ and $\mathcal{O}_v$ that constructs it
from $e_{(1,1)}$ (this is always possible thanks to them being coprime). In order to produce $e_{1,1}$, we use the admissible pair
\begin{align}\label{ea2}
e_{(1,1)} =[e_{(0,1)}, e_{(1,0)}]
\end{align}
Now we apply the sequence of $\mathcal{O}_h^{-1}$ and $\mathcal{O}_v$ moves in reverse in order to return back $e_{(1,1)}$ to $\hat B$, also obtaining from $e_{(0,1)}$ and $e_{(1,0)}$ the admissible pair of $\hat A$ and $\hat C$.

For instance, for $e_{(5,7)}$ we write
\begin{align}\label{ea1}
  e_{(5,7)} \mathop{\longleftarrow}_{\mathcal{O}_h} e_{(5,2)}
  \mathop{\longleftarrow}_{\mathcal{O}_v^2} e_{(1,2)}
  \mathop{\longleftarrow}_{\mathcal{O}_h} e_{(1,1)}
\end{align}
Now moving back to $e_{(5,7)}$, we produce
\begin{align}\label{ea3}
  (e_{(1,0)}, e_{(0,1)})
  \mathop{\longrightarrow}_{\mathcal{O}_h}
  (e_{(1,1)}, e_{(0,1)})
  \mathop{\longrightarrow}_{\mathcal{O}_v^2}
  (e_{(3,1)}, e_{(2,1)})
  \mathop{\longrightarrow}_{\mathcal{O}_h}
  (e_{(3,4)}, e_{(2,3)})
\end{align}
the desired decomposition of $e_{(5,7)}$. Analogously for, say, $e_{5,12}$ the decomposition
would be $(e_{(3,7)}, e_{(2,5)})$.

One can make similar expansion for other quadrants.\\ Say, $(-5,11)  \stackrel{{\cal O}_h^{-1}}{\longrightarrow}  (-5,6)
\stackrel{{\cal O}_h^{-1}}{\longrightarrow}   (-5,1) \stackrel{{\cal O}_v^4}{\longrightarrow}  (-1,1)$
implies that $e_{(-5,11)} = {\cal O}_h^{2} {\cal O}_v^{-4}  \, e_{(-1,1)}$,\\ while
$(-7,4) \stackrel{ {\cal O}_v}{\longrightarrow} (-3,4) \stackrel{{\cal O}_h^{-1}}{\longrightarrow}(-3,1)
\stackrel{{\cal O}_v^2}{\longrightarrow} (-1,1)$, that
$e_{(-4,7)} =  {\cal O}_v^{-1} {\cal O}_h {\cal O}_v^{-2} \, e_{(-1,1)}$.

\bigskip

\centerline{\fbox{We call this operation of making $\hat B$ from $\hat A$ and $\hat C$ {\bf $ABC$-procedure}.}}

\bigskip

\subsection{Constructing generators from grading one elements\label{334}}

Now we explain how to construct the whole quadrant of the elliptic Hall algebra, say, $e_{(n,m)}$ with $n,m>0$ from the grading one elements $e_{(1,m)}$. In other words, we explain how to use the input of the DIM algebra in order to construct the elliptic Hall algebra.

\paragraph{Constructing $e_{(n,m)}$ such that $gcd(n,m)=1$.} Let us first denote
\be
e_{(1,m)}&=&E_m,\ \ \ \ \ m\ge 0\nn\\
e_{(n,kn+a)}&=&E_k^{(n,a)},\ \ \ \ \  a=1\dots n-1,\ gcd(a,n)=1,\ k\ge 0,\ n>1\nn
\ee
and we are going to express all $E_k^{(n,a)}$ through $E_m$, and will do this for recursively in $n$.

The simplest case is $n=2$. Then, one easily gets $E_k^{(2,1)}=[E_{k+1},E_k]$.
To simplify notations, we denote the commutator of two elements $E$ by a product sign $\circ$: $E_k^{(2,1)}=E_{k+1}\circ E_k$.
One can further continue procedure applying the $ABC$-procedure, i.e. constructing $E_k^{(3,a)}$, $a=1,2$ from two $E$'s of smaller grading, then $E_k^{(4,a)}$, $a=1,3$ from two $E$'s of smaller grading, etc:
\be
E^{(2,1)}_k&=&E_{k+1}\circ E_k\nn\\
E_k^{(3,1)}&=&E^{(2,1)}_k\circ E_k=E_{k+1}\circ E_k\circ E_k\nn\\
E_k^{(3,2)}&=&-E^{(2,1)}_k\circ E_{k+1}=- E_{k+1}\circ E_k\circ E_{k+1}=
E_{k}\circ E_{k+1}\circ E_{k+1}\nn\\
 E_k^{(4,1)}&=& E^{(3,1)}_k\circ E_k= E_{k+1}\circ E_k\circ E_k\circ E_k\nn\\
 E_k^{(4,3)}&=&- E^{(3,2)}_k\circ E_{k+1}= E_{k+1}\circ E_k\circ E_{k+1}\circ E_{k+1}
\label{subsequentcom}
\ee
and commutators are read from the left to the right: $A\circ B\circ C=[[A,B],C]$.

The next level is already more complicated:
\be
 E_k^{(5,1)}&=& E^{(4,1)}_k\circ E_k\nn\\
 E_k^{(5,2)}&=&- E^{(3,1)}_k\circ E_k^{(2)}\nn\\
 E_k^{(5,3)}&=& E^{(3,2)}_k\circ E_k^{(2)}\nn\\
 E_k^{(5,4)}&=& E^{(4,3)}_k\circ E_{k+1}\nn
\ee
Further examples are:
\be
 E_k^{(6,1)}&=& E^{(5,1)}_k\circ E_k\nn\\
 E_k^{(6,5)}&=&- E^{(5,4)}_k\circ E_{k+1}\nn\\
 E_k^{(7,1)}&=& E^{(6,1)}_k\circ E_k\nn\\
 E_k^{(7,2)}&=&- E^{(4,1)}_k\circ E_k^{(3,1)}\nn\\
 E_k^{(7,3)}&=&- E^{(5,2)}_k\circ E_k^{(2)}\nn\\
 E_k^{(7,4)}&=& E^{(5,3)}_k\circ E_k^{(2)}\nn\\
 E_k^{(7,5)}&=& E^{(4,3)}_k\circ E_k^{(3,2)}\nn\\
 E_k^{(7,6)}&=& E^{(6,5)}_k\circ E_{k+1}\nn
\ee
All these formulas being written in terms of $E_k$ only, can be also presented in a very compact form: denote $E_{k+1}\circ E_k$ as [10], $E^{(3,1)}_k\circ E_k^{(2)}=E_k\circ E_{k+1}\circ E_k\circ (E_{k+1}\circ E_k)$ as [010(10)], etc. and denote $E_k^{(n,a)}$ just as $\{n,a\}$. Then, we have
\be\label{01}
\{2,1\}&=&[10]\nn\\
\{3,1\}&=&[100],\ \ \ \ \ \{3,2\}=[011]\nn\\
\{4,1\}&=&[1000],\ \ \ \ \{4,3\}=[1011]\nn\\
\{5,1\}&=&[10000],\ \ \{5,2\}=[010(10)],\ \ \ \  \{5,3\}=[011(10)],\ \ \{5,4\}=[10111]\nn\\
\{6,1\}&=&[100000], \ \{6,5\}=[011111]\nn\\
\{7,1\}&=&[1000000],\{7,2\}=[0100(100)],\{7,3\}=[10010(10)],\{7,4\}=[01110(10)], \{7,5\}=[1011(011)],\{7,6\}=[0111111]\nn\\
&\ldots &
\ee
What is important, we have the same structure of answer for any $k$. This reminds a similar property in the $W_{1+\infty}$ algebra \cite{MMMP1}, which allows one  to construct the commutative rays only for the horizontal ray. The reason is that the Serre relations, which are in charge of commutativity are invariant w.r.t. a simultaneous shift of all indices by the same integer \cite{MMMP1}. In the DIM algebra case, the Serre relations (\ref{Serre}) also celebrate this property. To put it differently, there is the automorphism of algebra that shift this index $k$. In the case under consideration, this automorphism is the automorphism $\hat {\cal O}_h^{-1}$ (\ref{Oh1}), which shifts $E_k^{(n,a)}\to E_{k+1}^{(n,a)}$.

\paragraph{Constructing commutative rays.} Now having a generic element $e_{(n,m)}$ with $gcd(n,m)=1$ at hands, one can construct the ray that is generated by this element following the procedure described in sec.\ref{sec:commutative-subalgebra-pr}, i.e. one first constructs the Hamiltonians $\mathfrak{h}_{(kn,km)}$ and then the elements $e_{(kn,km)}$ using formula (\ref{ABC}). Technically, we represented here every element $E_k^{(n,a)}$ as a commutator of two polynomials at smaller $n$'s using the $ABC$-procedure: $\underbrace{E_k^{(\overbrace{n_1+n_2}^n,\overbrace{a_1+a_2}^a)}}_B=\underbrace{E_k^{(n_1,a_1)}}_A\circ \underbrace{E_k^{(n_2,a_2)}}_C$ so that the commutative ray consists of Hamiltonians associated with polynomials
\be
\mathfrak{h}_{(rn,rkn+ra)}=\kappa_1\hat C\circ\underbrace{\hat B\circ\ldots\circ\hat B}_{r-1\ times}\circ\hat A\nn
\ee
which can be rewritten in terms of elements $e_{(kn,km)}$ using
\be
\sum_r\kappa_re_{(rn,rkn+ra)}{z^r\over r}=-\log\left(1-\sum_rz^r\mathfrak{h}_{(rn,rkn+ra)}\right)\nn
\ee
which can be equivalently written as
\be
e_{(rn,rkn+ra)}={1\over\kappa_r}\sum_{s=1}^r\sum_{\sum r_i=r}{(s-1)!(r-s)!\over r!}\prod_{i=1}^s\mathfrak{h}_{(r_in,r_ikn+ra)}\nn
\ee
On one hand, this is the standard relation in quantum deformed algebra connecting algebra commutators with the Cartan elements.
On the other hand, this is a standard QFT relation between \textit{connected} quantities
(coefficients of the free energy $\mathcal{F}$) and \textit{disconnected}
quantities (coefficients of the partition function $\mathcal{Z} = e^{\mathcal{F}}$).
They are manifested here at the abstract level of operators, from which concrete
exactly solvable QFTs have to be constructed yet.

\subsection{More commutative subalgebras: cones\label{cones}}

Now one may ask if it is possible to deform the commutative subalgebras of $qW_{1+\infty}$ given by cones as in (\ref{cone}). In order to understand it, consider a simple example of $\hat H_1=e_{(1,1)}+\alpha e_{(1,2)}$. Then, one notes that the combination
\be \label{eq:first-cone-naive-attempt}
\hat H_2=e_{(2,2)}+\alpha e_{(2,3)}+\alpha^2e_{(2,4)}
\ee
commutes with $\hat H_1$:
\be
\phantom{.}[\hat H_1,\hat H_2]=\underbrace{[e_{(1,1)},e_{(2,2)}]}_{\stackrel{(\ref{1})}{=}0}+\alpha
\underbrace{[e_{(1,1)}, e_{(2,3)}]}_{\stackrel{(\ref{2})}{=}-e_{(3,4)}}+\alpha^2
\underbrace{[e_{(1,1)}, e_{(2,4)}]}_{\stackrel{(\ref{2})}{=}-e_{(3,5)}}+
\alpha \underbrace{[e_{(1,2)},e_{(2,2)}]}_{\stackrel{(\ref{2})}{=}e_{(3,4)}}+
\alpha^2 \underbrace{[e_{(1,2)},e_{(2,3)}]}_{\stackrel{(\ref{2})}{=}e_{(3,5)}}+
\alpha^3 \underbrace{[e_{(1,2)},e_{(2,4)}]}_{\stackrel{(\ref{1})}{=}0}=0\nn\\
\nn
\ee
and all these pairs are admissible. The next Hamiltonian $\hat H_3=e_{(3,3)}+\alpha e_{(3,4)}+\alpha^2e_{(3,5)}+\alpha^3e_{(3,6)}$ also commutes with the previous ones, and all pairs of $e_{(n,m)}$ involved are admissible. However, checking commutativity of already the fourth Hamiltonian,
$\hat H_4=e_{(4,4)}+\alpha e_{(4,5)}+\alpha^2e_{(4,6)}+\alpha^3e_{(4,7)}+\alpha^4e_{(4,8)}$ with $\hat H_1$ involves a non-admissible pair $(e_{(1,2)},e_{(4,5)})$, their commutator is not equal to $e_{(5,7)}$ ($\{\ldots\}$ denotes the anticommutator):
\be
\phantom{.}[e_{(1,2)},e_{(4,5)}]=[e_{(1,2)},\underbrace{[e_{(1,1)},e_{(3,4)}]}_{\stackrel{(\ref{2})}{=}-e_{(4,5)}}]
\stackrel{Jacobi}{=}-[e_{(3,4)},\underbrace{[e_{(1,2)},e_{(1,1)}]}_{\stackrel{(\ref{2})}{=}e_{(2,3)}}]
-[e_{(1,1)},\underbrace{[e_{(3,4)},e_{(1,2)}]}_{\stackrel{(\ref{2})}{=}{1\over\kappa_1}\mathfrak{h}_{(4,6)}}]=\nn\\
=-\underbrace{[e_{(3,4)},e_{(2,3)}]}_{\stackrel{(\ref{2})}{=}-e_{(5,7)}}-
{1\over\kappa_1}[e_{(1,1)},\underbrace{\mathfrak{h}_{(4,6)}}_{\stackrel{(\ref{2})}{=}{\kappa_1^2\over 2}e_{(2,3)}^2-
{\kappa_2\over 2}e_{(4,6)}}]=e_{(5,7)}+{\kappa_2\over 2\kappa_1}\underbrace{[e_{(1,1)},e_{(4,6)}]}_{\stackrel{(\ref{2})}{=}-e_{(5,7)}}-{\kappa_1\over 2}[e_{(1,1)},e_{(2,3)}^2]=\nn\\
=\Big(1-{\kappa_2\over 2\kappa_1}\Big)e_{(5,7)}-
{\kappa_1\over 2}\Big\{\underbrace{[e_{(1,1)},e_{(2,3)}]}_{\stackrel{(\ref{2})}{=}-e_{(3,4)}},e_{(2,3)}\Big\}
=\Big(1-{\kappa_2\over 2\kappa_1}\Big)e_{(5,7)}+{\kappa_1\over 2}\{e_{(2,3)},e_{(3,4)}\}
\label{H4cone}
\ee
Hence, $\hat H_4$ does not commute with $\hat H_1$, and we have only three commutative Hamiltonians, and are not able to reproduce the whole commutative cone family as in \cite{MMMP1}.

Similarly, commuting are the Hamiltonians
\be
\hat H_1'=e_{(2,1)}+\alpha e_{(2,2)}\nn\\
\hat H_2'=e_{(4,2)}+\alpha e_{(4,3)}+\alpha^2 e_{(4,4)}\nn
\ee
since
\be
\phantom{.}[\hat H_1',\hat H_2']=\underbrace{[e_{(2,1)},e_{(4,2)}]}_{\stackrel{(\ref{1})}{=}0}+\alpha
\underbrace{[e_{(2,1)}, e_{(4,3)}]}_{\stackrel{(\ref{2})}{=}\mathfrak{h}_{(6,4)}}+\alpha^2
\underbrace{[e_{(2,1)}, e_{(4,4)}]}_{\stackrel{(\ref{2})}{=}-e_{(6,5)}}+
\alpha \underbrace{[e_{(2,2)},e_{(4,2)}]}_{\stackrel{(\ref{2})}{=}-\mathfrak{h}_{(6,4)}}+
\alpha^2 \underbrace{[e_{(2,2)},e_{(4,3)}]}_{\stackrel{(\ref{2})}{=}e_{(6,5)}}+
\alpha^3 \underbrace{[e_{(2,2)},e_{(4,4)}]}_{\stackrel{(\ref{1})}{=}0}=0\nn\\
\nn
\ee
However, already checking commutativity of the third Hamiltonian, $\hat H_3'=e_{(6,3)}+\alpha e_{(6,4)}+\alpha^2 e_{(6,5)}+\alpha^3 e_{(6,6)}$ with $\hat H_1'$ involves a non-admissible pair $\{e_{(2,1)},e_{(6,5)}\}$, and, hence, they do not commute.

Generally, the cone commutative subalgebra generated by an element
\be\label{n1}
\hat H_1^{(2)}=e_{(n,m)}+\alpha e_{(n,m+1)}
\ee
is given by the commuting Hamiltonians
\be\label{n2}
\hat H_k^{(2)}=\sum_{i=0}^k \alpha^i e_{(kn,km+i)}
\ee
and $\hat H_k$ and $\hat H_n$ commute {\bf if the pairwise commutators of their elements are admissible pairs}, which is generally not the case.

More generally, the cone commutative subalgebra of width $n$ generated by an element
\be
\hat H_1^{(n)}=e_{(n,m)}+\sum_{j=1}^n\alpha_je_{(n,m+j)}\nn
\ee
is given by the elements
\be
\hat H_k^{(n)}=\sum_{i=0}^k\sum_{j=1} \alpha_j^i e_{(kn,km+ji)}\nn
\ee
and again $\hat H_k^{(n)}$ and $\hat H_m^{(n)}$ commute {\bf if the pairwise commutators of their elements are admissible pairs}.
Again, it does not work even in the simplest examples: the sums $e_{(1,1)}+\alpha e_{(1,3)}$ and $e_{(2,2)}+\alpha e_{(2,4)}+\alpha^2e_{(2,6)}$ do not commute since pair $\{e_{(1,1)},e_{(2,6)}\}$ is not admissible.

However, there is a trick that allows one to construct more involved commutative series of operators preserving the grading. This trick generalizes the cone construction. We illustrate it in a simple example: consider an inner automorphism of the algebra {\bf E} given by the operator
\be\label{O}
\hat{\cal O}(u):=\exp\left(-\sum_n e_{(0,n)}{u^n\over n}\right)
\ee
It acts on the grading one elements of the algebra as
\be \label{eq:calO-first-cone-ham}
\hat{\cal O}(u)\cdot e_{(1,n)}\cdot\hat{\cal{O}}(u)^{-1}=e_{(1,n)}-ue_{(1,n+1)}
\ee
and, similarly,
\be
\hat{\cal O}(u)^{-1}\cdot e_{(-1,n)}\cdot\hat{\cal{O}}(u)=e_{(-1,n)}-ue_{(-1,n+1)}\nn
\ee
This looks like a cone case, however, the adjoint action of this operator on other elements of ray $(1,n)$ produces more involved expressions non-linear in $e_{(n,m)}$. Still, it maps any commutative ray $(\pm 1,n)$ onto a commutative family of elements. In \cite{Ch3}, we used this operator in order to generate a commutative family of elements $\mathfrak{H}_{(k,km)}$:
\be\label{OH}
\hat{\mathfrak{H}}_{(\pm k,kn)}=\hat{\cal O}(u)^{\pm n}\cdot e_{(\pm k,0)}\cdot\hat{\cal{O}}(u)^{\mp n}
\ee

\bigskip

\noindent Applying this operator to the ray $(1,n)$ is actually rather instructive.
Continuing \eqref{eq:calO-first-cone-ham}
\begin{align}
  \hat{\cal O}(u)\cdot e_{(2,2 n)}\cdot\hat{\cal{O}}(u)^{-1}
  = & \ e_{(2,2 n)} - u e_{(2,2 n+1)} + u^2 e_{(2,2 n+2)}\nn
  \\ \notag
  \hat{\cal O}(u)\cdot e_{(3,3 n)}\cdot\hat{\cal{O}}(u)^{-1}= & \
  e_{(3,3 n)}- u e_{(3,3 n+1)}
  + u^2 e_{(3,3 n+2)} - u^3 e_{(3,3 n+3)}
\end{align}
we observe that first three Hamiltonians, indeed, coincide with the naive attempt
\eqref{eq:first-cone-naive-attempt}. However, the fourth Hamiltonian
(after some tedious algebra) \textit{is} different from the naive one
\begin{align}
  \hat{\cal O}(u)\cdot e_{(4,4 n)}\cdot\hat{\cal{O}}(u)^{-1}= & \
  e_{(4,4 n)}-ue_{(4,4 n+1)}
  + u^2 \underbrace{\left (\left(\frac{\kappa_2}{2 \kappa_1} - 1\right) e_{(4,4 n+2)}
    + \left(-\frac{\kappa_1}{2}\right) e_{(2,2 n+1)}^2 \right )}
  _{-\mathfrak{h}_{4, 4 n + 2} - e_{4, 4 n + 2}}
  - u^3 e_{(4,4 n+3)}
  + u^4 e_{(4,4 n+4)}\nn
\end{align}
This $u^2$-term is exactly the one cancelling (\ref{H4cone}).

We return to action of the operator $\hat{\cal O}(u)$ on the generators of algebra {\bf E} and to the Hamiltonians $\hat{\mathfrak{H}}_{(\pm k,kn)}$ in sec.\ref{pent}.

\subsection{Miki automorphism: why one commutative ray implies the others\label{Miki}}

We explained in sec.2 that there is an algebra of automorphisms of the $qW_{1+\infty}$ algebra generated by ${\cal O}_v$ and ${\cal O}_h$, which maps commutative subalgebras to each other. This automorphism is described by the $SL(2,\mathbb{Z})$ algebra naturally associated with the two-dimensional integer lattice. In these terms, ${\cal O}_v$ and ${\cal O}_h$ are described as
\be
{\cal O}_v=\left(\begin{array}{cc}
1&1\cr
0&1
\end{array}
\right)\nn
\ee
and
\be
{\cal O}_h=\left(\begin{array}{cc}
1&0\cr
-1&1
\end{array}
\right)\nn
\ee

An extension of this automorphism to the elliptic Hall algebra is called Miki automorphism \cite{Miki,BS}, and, in the case of zero central charges, is given by
\be
g(e_{\vec\alpha})=e_{g(\vec\alpha)},\ \ \ \ \ \ \ \ \ g=
\left(\begin{array}{cc}
a&b\cr
c&d
\end{array}
\right)\in SL(2,\mathbb{Z})\nn
\ee
where for $\vec\alpha=(n,m)$, $g(\vec\alpha)=(an+bm,cn+dm)$,

When the central charges are non-zero, the algebra of automorphisms changes to the universal covering $\widetilde{SL(2,\mathbb{Z})}$ of $SL(2,\mathbb{Z})$ \cite{BS}, and acts projectively on the generators:
\be
g(e_{\vec\alpha})=\mathfrak{q}^{p_{g,g(\vec\alpha)}\vec c\cdot g(\vec\alpha)} e_{g(\vec\alpha)}\nn
\ee
where $p_{g,g(\vec\alpha)}$ is an integer, see \cite[Eq.(6.16)]{BS}.

In particular, one can choose
\be\label{MikiST}
{\cal O}_v^{-1}\Big(e_{(n,m)}\Big)&=&\left\{\begin{array}{cl}
e_{(n-m,m)}\ \ \ \ \ \ \ &n\cdot m\ge 0,\ \  m\ne 0\cr
&\cr
\mathfrak{q}^{(n-m)c_1+mc_2}e_{(n-m,m)}\ \ \ \ \ \ \ &n\cdot m<0
\end{array}
\right.\nn\\
{\cal O}_h\Big(e_{(n,m)}\Big)&=&e_{(n,m-n)}
\ee

We demonstrate below that the first automorphism ${\cal O}_v$ is simply realized in concrete representations, while the second one ${\cal O}_h$ is not that simple to realize (see sec.\ref{Oh}).

Note that when the automorphisms are realized as operators with the adjoint action on elements of the algebra, we denote them with an additional hat:
\be
{\cal O}(e_{(n,m)})=\hat{\cal O}^{-1}\cdot e_{(n,m)}\cdot\hat{\cal O}\nn
\ee

\section{One-body representation\label{s1body}}

The simplest representation of the elliptic Hall algebra is a one-body representation, with the generators of the form
\be
e_{(n,m)}\sim x^nq^{m\hat D},\ \ \ \ \ \ \ \hat D=x{d\over d x}\nn
\ee
and with the both zero central charges. Hence, the one-body representation of the algebra {\bf E} looks like that of the $qW_{1+\infty}$ algebra of sec.2. However, in order to guarantee the commutation relations (\ref{2}) of the elliptic Hall algebra, one has to choose some peculiar normalization of generators of the algebra. That is, for the element $e_{(kp,kr)}$ with $gcd(p,r)=1$,
\be\label{1body}
e_{(kp,kr)}={q^{(3pr+1)k/2-pr}\over q^k-1}x^{kp}q^{kr\hat D}
\ee
With this identification, one immediately comes to the commutation relations (\ref{1}) and (\ref{2}).

However, there is a subtlety: one still should not use formula (\ref{2}) for non-admissible pairs. The reason is that, in variance with the $qW_{1+\infty}$ algebra, which is a Lie algebra, the elliptic Hall algebra even in the one-body representation is not a Lie algebra, but an associative algebra. This means that the element of the form $x^{2}q^{3\hat D}$ may be $e_{(2,3)}$, or maybe, for instance, $e_{(1,2)}e_{(1,1)}$. Consider, for instance, the commutator of a non-admissible pair
\be
\phantom{.}[e_{(1,1)},e_{(4,1)}]={q^{9/2}(q^3-1)\over (q-1)^2}x^5q^{2\hat D}\nn
\ee
The r.h.s. is not an element $e_{(5,2)}$ (\ref{1body}). Similarly, if one considers the commutator
\be
\phantom{.}[e_{(5,1)},e_{(1,1)}]=-{q^5(q^4-1)\over (q-1)^2}x^6q^{2\hat D}\nn
\ee
it cannot be presented as
\be
{1\over\kappa_1}\mathfrak{h}_{(6,2)}={1\over\kappa_1}\left(-{\kappa_2\over 2}e_{(6,2)}+{\kappa_1^2\over 2}e_{(3,1)}^2\right)\nn
\ee
with elements of the form (\ref{1body}), their normalization would become depending on $t$, not only on $q$.

Technically, this works in the following way. Consider the commutator of $e_{\vec\alpha}$ and $e_{\vec\beta}$ with $gcd(\alpha_1,\alpha_2)=1$, $gcd(\beta_1,\beta_2)=1$:
\be
\phantom{.}[e_{\vec\alpha},e_{\vec\beta}]\sim{\left(q^{\alpha_1\beta_2-\beta_1\alpha_2}-1\right)\over (q-1)^2}
x^{\alpha_1+\beta_1}q^{(\alpha_2+\beta_2)\hat D}\nn
\ee
In order to have an expression of the form (\ref{1body}) in the case of $gcd(\alpha_1+\beta_1,\alpha_2+\beta_2)=1$, the difference $q^{\alpha_1\beta_2-\beta_1\alpha_2}-1$ has to be proportional to $q-1$, i.e. $\alpha_1\beta_2-\beta_1\alpha_2=\pm 1$. This is one of examples of admissible pairs, e.g., these are $e_{(1,3)}$ and $e_{(1,2)}$. A more involved example is when $gcd(\alpha_1+\beta_1,\alpha_2+\beta_2)\ne 1$, in this case, the analysis has to be subtler. A typical example is the admissible pair $e_{(1,3)}$ and $e_{(1,1)}$.

\section{$N$-body representation}

\subsection{Elements of the representation}

A natural generalization of the one-body representation is an $N$-body representation, which was associated in \cite{MMMP1} with the eigenvalue representation, and which is related to the trigonometric Ruijsenaars system. In fact, this representation can be also connected with the Fock representation in the limit of $N\to \infty$ (see \cite{MMMP1} and references therein for a similar procedure in the $W_{1+\infty}$ algebra case).

Denote $(q_1,q_2,q_3)=(q, t^{-1},q^{-1}t)$. The $N$-particle representation of the algebra {\bf E} has zero central charges, and the space of representation is the space of all symmetric polynomials of $N$ variables $x_i$. The representation is generated by the four operators
\be\label{eb}
e_{(0,1)}={q^{1/2}\over q-1}\sum_{i=1}^N\prod_{j\ne i}{tx_i-x_j\over x_i-x_j}q^{\hat D_i}\nn\\
e_{(0,-1)}={q^{1/2}\over q-1}\sum_{i=1}^N\prod_{j\ne i}{t^{-1}x_i-x_j\over x_i-x_j}q^{-\hat D_i}\nn\\
e_{(\pm 1,0)}={q^{1/2}\over q-1}\sum_{i=1}^Nx_i^{\pm 1}
\ee
The operators $e_{(\pm 1,0)}$ are related with each other by the transform $x_i\to x_i^{-1}$, and similarly do the operators $e_{(0,\pm 1)}$. The latter are equivalently related by the transform $(q,t)\to (q^{-1},t^{-1})$. The operator $e_{(0,1)}$ is nothing but the Macdonald operator \cite{Mac}, i.e. first Hamiltonian of the trigonometric Ruijsenaars system\footnote{There is a simpler Hamiltonian
\be
\hat H_0=q^{\sum_{i=1}^N\hat D_i}\nn
\ee}.

Now one can construct other elements of the algebra. For instance,
\be\label{ee}
e_{(\pm n,0)}&=&{q^{n/2}\over q^n-1}\sum_{i=1}^Nx_i^{\pm n}\ \ \ \ \ \ \ n\in\mathbb{Z}_{> 0}\nn\\
e_{(\pm n,1)}&=&{q^{1\pm n\over 2}\over q-1}\sum_{i=1}^N\prod_{j\ne i}{tx_i-x_j\over x_i-x_j}x_i^{\pm n}q^{\hat D_i}
\ \ \ \ \ \ \ n\in\mathbb{Z}_{\ge 0}\nn\\
e_{(\pm n,-1)}&=&{q^{1\mp n\over 2}\over q-1}\sum_{i=1}^N\prod_{j\ne i}{t^{-1}x_i-x_j\over x_i-x_j}x_i^{\pm n}
q^{-\hat D_i}\ \ \ \ \ \ \ n\in\mathbb{Z}_{\ge 0}\nn\\
\ldots
\ee

\subsection{Commutative subalgebras}

\subsubsection{Ray (0,1)}

Because of the symmetries relating all four quadrants, it is enough to deal with only one quadrant. We consider here the elements $e_{(n,m)}$ with non-negative $n$ and $m$. Then, the two distinguished rays are the horizontal (rather trivial) ray (1,0) consisting of elements $e_{(n,0)}$ (see (\ref{ee}) and the vertical ray (0,1) consisting of elements $e_{(0,n)}$. This latter ray we are going to discuss in detail, since it describes Hamiltonians of the trigonometric Ruijsenaars system \cite{Mac}:
\be\label{RH}
\hat {\mathfrak D}_k=\sum_{i_1<i_2<\ldots<i_k}{\left[\prod_{m=1}^kt^{\hat D_{i_m}}\Delta(x)\right]\over\Delta(x)}
\prod_{m=1}^k q^{\hat D_{i_m}}
\ee
where $\Delta(x)=\prod_{i<j}(x_i-x_j)$ is the Vandermonde determinant. Note that $\hat {\mathfrak D}_k=0$ at $k>N$. Using the repeated commutators (\ref{0,r}),
one obtains that
\be
e_{(0,1)}&=&{q^{1/2}\over q-1}\hat {\mathfrak D}_1\nn\\
e_{(0,2)}&=&{q\over q^2-1}\left(\hat {\mathfrak D}_1^2-2\hat {\mathfrak D}_2\right)\nn\\
e_{(0,3)}&=&{q^{3/2}\over q^3-1}\left(\hat {\mathfrak D}_1^3-3\hat {\mathfrak D}_1\hat{\mathfrak D}_2
+3\hat{\mathfrak D}_3\right)\nn\\
\nn\\
&\ldots&\nn
\ee
\be
\boxed{
e_{(0,n)}={(-1)^{n+1}nq^{n/2}\over q^n-1}\left[\log \Big(1+\sum_k\hat {\mathfrak D}_kz^k\Big)\right]_{z^k}
={(-1)^{k+1}k\over (q^{k/2}-q^{-k/2})}\oint_0{dz\over z^{k+1}}\log \hat {\mathfrak D}(z)
}
\ee
where $[\ldots]_{z^k}$ denotes the coefficient in front of the term $z^k$ of the Taylor expansion, and
\be\label{Mgf}
\hat {\mathfrak D}(z):=1+\sum_k\hat {\mathfrak D}_kz^k
\ee
is the generating function first introduced by I.G. Macdonald \cite[sec.3]{Mac}.

Thus,
\be\label{De}
\hat {\mathfrak D}(z)=\exp\left(-\sum_k(-1)^k(q^{k/2}-q^{-k/2})e_{(0,k)}{z^k\over k}\right)
\ee
and one immediately obtains, using the Cauchy identity, that
\be
\hat {\mathfrak D}_k=\mathfrak{e}_k\left\{(q^{k/2}-q^{-k/2})e_{(0,k)}\right\}\nn
\ee

\subsubsection{Ray (p,1)}

This kind of rays is constructed from the Ruijsenaars Hamiltonians (\ref{RH}) with a simple replacement $q^{\hat D_i}\to x_i^pq^{\hat D_i}$. Indeed, consider
\be\label{HpR}
\hat {\mathfrak D}^{(p)}_k=\sum_{i_1<i_2<\ldots<i_k}{\left[\prod_{m=1}^kt^{\hat D_{i_m}}\Delta(x)\right]\over\Delta(x)}
\prod_{m=1}^k x_i^pq^{\hat D_{i_m}}
\ee
Then, one immediately obtains that
\be
e_{(p,1)}&=&{q^{p+1\over 2}\over q-1}\hat {\mathfrak D}^{(p)}_1\nn\\
e_{(2p,2)}&=&{q^{2(p+1)\over 2}\over q^2-1}\left(\Big(\hat {\mathfrak D}^{(p)}_1\Big)^2-2\hat {\mathfrak D}^{(p)}_2\right)\nn\\
e_{(3p,3)}&=&{q^{3(p+1)\over 2}\over q^3-1}\left(\Big(\hat {\mathfrak D}^{(p)}_1\Big)^3-3\hat {\mathfrak D}^{(p)}_1\hat{\mathfrak D}^{(p)}_2
+3\hat{\mathfrak D}^{(p)}_3\right)\nn\\
\nn\\
&\ldots&\nn
\ee
\be
\boxed{
e_{(pn,n)}={(-1)^{n+1}nq^{n(p+1)\over 2}\over q^n-1}\left[\log \Big(1+\sum_k\hat {\mathfrak D}^{(p)}_kz^k\Big)\right]_{z^k}}\nn
\ee

Note that the replacement is given by a the simple rotation:
\be \label{eq:ov-action}
\hat {\mathfrak D}^{(p)}_k=\hat{\cal O}_v^{-p}\cdot\hat {\mathfrak D}_k\cdot \hat{\cal O}_v^p
\ee
with $\hat{\cal O}_v$ as in (\ref{O1})-(\ref{O2}):
\be\label{NbOv}
\hat{\cal O}_v=\prod_{i=1}^N\hat{\cal O}_v(x_i)\nn\\
\hat{\cal O}_v^{-1}(x_i)\cdot q^{\hat D_i}\cdot \hat{\cal O}_v(z_i)=x_iq^{\hat D_i}\nn\\
\ee
and $\hat{\cal O}_v(x_i)$ is given by formula (\ref{O1}).

In other words, one of the two rotations relating distinct commutative subalgebras survives the deformation from the $qW_{1+\infty}$ to the algebra {\bf E}. This rotation is just multiplying with a function of the coordinates only\footnote{At the classical level, in terms of the Ruijsenaars coordinates $\xi_i=\log x_i$, this corresponds to the canonical transformation involving the momenta $\pi_i$ only: $\pi_i\to\pi_i+\xi_i$.}. In particular, this means that these (p,1) rays give rise to the same trigonometric Ruijsenaars system.

\subsubsection{Cherednik operator: ray (-1,r)\label{Cho}}

In order to construct commutative subalgebras associated with other rays, we need a more involved construction based on the Cherednik operator \cite{book:Ch-daha}. We choose here the negative gradation ray in order to achieve closer similarity with Calogero model formulas \cite{MMMP1}. The Cherednik operators for $N$ particles naturally emerge within the framework of the $GL_N$ affine Hecke algebra (see Appendix B), and are defined as follows
  \begin{align}
    C_k = t^{1 - k} \left(\prod_{i = k + 1}^N R_{k,i}\right)
    \ttop{k}
    \left(\prod_{i = 1}^{k - 1} R^{-1}_{i,k}\right)\nn
  \end{align}
  where the $R$-operators are equal to
  \begin{align}
    R_{i,j} = \frac{(x_i - t^{-1} x_j)}{(x_i - x_j)}
    + \left(t^{-1} - 1\right)\frac{x_j}{(x_i - x_j)} s_{i,j}\nn
    \\ \notag
    R^{-1}_{i,j} =
    t \frac{(t^{-1} x_i - x_j)}{(x_i - x_j)}
    + \left(t - 1 \right)\frac{x_j}{(x_i - x_j)} s_{i,j}
  \end{align}
and one has
  \begin{align}
    R^{-1}_{i,j} R_{i,j} = & \ \text{Id}\nn
  \end{align}

  \bigskip

  Here $s_{i,j}$, as is customary, is the operator of interchanging $i$-th
  and $j$-th particle. For instance,
  \begin{align}
    x_i s_{i,j} = s_{i,j} x_j,\ \ \ \ \ x_j s_{i,j} = s_{i,j} x_i\nn
    \\ \notag
    \ttop{i} s_{i,j} = s_{i,j} \ttop{j},\ \ \ \ \ \ttop{j} s_{i,j} = s_{i,j} \ttop{i}
  \end{align}
  One may think of $s_{i,j}$ as transpositions
  $\sigma_{(j,i)}$, $i \leftrightarrow j$
  in the permutation group $S_N$. Therefore, their product equal more complicated
  permutations. For instance
  \begin{align}
    s_{1,2} s_{2,3} = \sigma_{(2,3,1)}\nn
  \end{align}
  with $\sigma_{(2,3,1)}$ being the cyclic permutation of indices
  $(1,2,3) \rightarrow (2,3,1)$

  Thus defined Cherednik operators $C_k$ pairwise commute \cite{paper:NS-cherednik}.
  Moreover, any symmetric combination of $C_k$ maps the space $\Lambda_N$
  of polynomials in $N$ variables $x_1, \dots x_N$ onto itself.

  \bigskip

The generating function for Macdonald operators (\ref{Mgf}) is expressible
  through the Cherednik operators \cite[Subsection~1.3.5]{book:Ch-daha}
  \begin{align}
    \mathfrak{D}(u) = \prod_{i=1}^N \left(1 + u t^{N-1} C_i\right)\nn
  \end{align}

  \bigskip

  For our consideration, crucial is the transformation property of $C_k$
  under the canonical change of variables $\ttop{i} \rightarrow \frac{1}{x_i} \ttop{i}$, which is nothing but the action of automorphism ${\cal O}_v^{-1}$
  \begin{align} \label{eq:cher-canon-trans}
    \widetilde{C}_k :=
    C_k \Bigg{|}_{\ttop{i} \rightarrow \frac{1}{x_i} \ttop{i}} = & \
    \ \frac{1}{x_k} C_k + \sum_{l > k}
    \left(\frac{1}{x_k} C_l
    - C_l \frac{1}{x_k}
    \right)
    \\ \notag
    = & \
    \ q \cdot C_k \frac{1}{x_k} + q \cdot \sum_{l < k}
    \left( C_l \frac{1}{x_k} - \frac{1}{x_k} C_l \right)
  \end{align}

 The Cherednik operator provides a discrete counterpart of the Dunkl operator \cite{Dunkl}, hence, our construction of the commuting Hamiltonians is much similar to that in the case of the $W_{1+\infty}$ algebra and of the affine Yangian algebra \cite{MMCal,MMMP1}.
Denote for a moment the Hamiltonians in the $W_{1+\infty}$/Yangian case with letter $H$, whereas the Hamiltonians in the elliptic Hall/DIM case with letter $\mathcal{H}$.

Then, there are quite parallel and explicit formulas:
    \begin{align} \label{eq:ray-1r-hamilts}
    \begin{array}{|c||c|}
    \hline
    {\rm Yangian}  &  {\rm DIM} \\ \hline  &  \\
      H^{(-1,r)}_k = \sum_{i=1}^N \left( \frac{1}{x_i}
      \mathcal{D}_i^r \right)^k,
      \ \ \mathcal{D}_i = x_i \tilde{\mathcal{D}}_i
      \ \ & \ \
e_{(-k, r k)} = \alpha_{r,k}
      \sum_{i=1}^N \left( \frac{1}{x_i}
      \mathcal{C}_i^r \right)^k,
      \ \ \mathcal{C}_i = x_i \widetilde{C}_i\\
      \hline
      \end{array}
    \end{align}
    where $\tilde{\mathcal{D}}_i$
    are the Dunkl differential operators
    \begin{align}
      \tilde{\mathcal{D}}_i = \frac{\partial}{\partial x_i}
      + \beta \sum_{j \neq i} \frac{1}{(x_i - x_j)}
      \left(
      1 - s_{i,j}
      \right)\nn
    \end{align}
    where the normalization coefficient $\alpha_{r,k}$
    is equal to
    \begin{align}
      \alpha_{r,k} = \left(\frac{t^{N-1}}{q^{1/2}}\right)^{r k}
      \frac{1}{q^{k/2} - q^{-k/2}}\nn
    \end{align}
Here the Hamiltonians $\sum_{i=1}^N \left( \frac{1}{x_i}\mathcal{C}_i^r \right)^k$ are commutative
since $\frac{1}{x_i} \mathcal{C}_i^r$ are commutative (similarly to the Cherednik operators themselves):
    \begin{align}
      \left[\frac{1}{x_i} \mathcal{C}_i^r, \frac{1}{x_j} \mathcal{C}_j^r\right] = 0,
      \ \ \ \forall i,j\nn
    \end{align}

    In particular, for the first two simplest rays, the expressions are
   \be \label{eq:first-two-hamilt-fams}
    \begin{array}{|c||c|}
    \hline
    {\rm Yangian}  &  {\rm DIM} \\ \hline  &  \\
      H^{(-1,1)}_k = \sum_{i=1}^N
      \tilde{\mathcal{D}}_i^k  &
e_{(-k,k)} =
      \frac{t^{N-1}}{q^k-1}
      \cdot
      \sum_{i=1}^N
      \widetilde{C}_i^k
      \\ &
      \\
      H^{(-1,2)}_k = \sum_{i=1}^N \left(
      \tilde{\mathcal{D}}_i x_i \tilde{\mathcal{D}}_i
      \right)^k  &
e_{(-k,2k)}=
      \frac{t^{2 k (N-1)}}{q^{k/2} (q^k - 1)}
      \cdot
      \sum_{i=1}^N \left(
      \widetilde{C}_i x_i \widetilde{C}_i
      \right)^k\\
      \hline
   \end{array}
   \ee

    \bigskip

    This involved picture becomes rather trivial in the \textbf{one-body} case of $N = 1$,
    but useful for dimension counting arguments.
    Indeed, in this case, there is a single Cherednik operator $C_1 = \ttop{1}$.
    The tilde-operator is equal to $\widetilde{C}_1 = \frac{1}{x} \ttop{1}$, and the ray (-1,$r$) Hamiltonians are equal to
    \begin{align} \label{eq:hams-1r-one-body}
e_{(-k, r k)} \Bigg{|}_{N=1} =
      \frac{q^{\frac{(1-r)}{2} k}}{(q^k-1)} \cdot \left (
      \frac{1}{x} \left(\ttop{1}\right)^r
      \right)^k = \frac{q^{\frac{(1-r)}{2} k}}{(q^k-1)}
      q^{- \frac{1}{2} r k(k-1)} \cdot \frac{1}{x^k} \left(\ttop{1}\right)^{r k}
    \end{align}

    \bigskip

    Quite \textbf{a subtle point} here is that equations
    \eqref{eq:ray-1r-hamilts}, \eqref{eq:first-two-hamilt-fams}
    hold only when acting on the space
    of \textit{symmetric} functions in $x_1, \dots x_N$.
    However, the left hand sides are commutative as differential/difference operators
    on the space of all (not necessarily symmetric) functions of $x_1, \dots x_N$.
    And, similarly, the right hand sides (defined as explicit sums of products of the
    Dunkl/Cherednik operators) mutually commute on the space of all functions of
    $x_1,\dots x_N$. That is, we have, for every ray,
    two \textit{different} sets of mutually commuting operators
    (each, in principle, defining a separate integrable system),
    which, however, coincide when projected (reduced) onto the space of
    symmetric functions in $x_1,\dots,x_N$.

    \subsubsection{Generic rays $(p,r)$ through Cherednik operators\label{Chpq}}

In the $W_{1+\infty}$ algebra, i.e. at $\beta = 1$, it was possible to generalize
    \eqref{eq:ray-1r-hamilts} to the case of Hamiltonians on an arbitrary ray ($-p,r$)
    \cite{MMMP1}
    \begin{align}
      H^{(-p,r)}_{k} = r^{p k} \sum_{i=1}^N
      \left(
      x_i^{\frac{1-r}{2}}
      \cdot x_i^{-1} \left(x_i \mathfrak{d}_i \right)^p \cdot
      x_i^{\frac{1-r}{2}}
      \right)^k\nn
    \end{align}
    but utilizing a somewhat simpler variant of the Dunkl operator $\mathfrak{d}_i$,
    which \textit{does not}
    feature the permutation operator $s_{i,j}$
    \begin{align}
      \mathfrak{d}_i = \frac{\partial}{\partial x_i} +
      \sum_{j \neq i} \frac{1}{(x_i - x_j)}\nn
    \end{align}

    \bigskip

    Surprisingly and elegantly, on the algebra {\bf E} side, due to presence of $\mathcal{O}_h$ and $\mathcal{O}_v$ symmetries, it is also possible to obtain expressions for Hamiltonians on an
    arbitrary ray $(-p,r)$. That is, the generalization
    has the following form
    \begin{align} \label{eq:ham-through-cher}
      \boxed{
        e_{(-kp,kr)}= \sum_{i=1}^N \left(
        C^{(\alpha_1)}_i x_i C^{(\alpha_2)}_i x_i \dots x_i C^{(\alpha_r)}_i
        \right)^k
      },
    \end{align}
    where $W_{\vec\alpha} := \alpha_1\alpha_2\dots\alpha_r$
    is a certain \textit{palindromic} word
    of length $r$ of numbers, such that $\sum \alpha_i = p + r - 1$.
    Clearly, to the Hamiltonians on integer rays $(-1,r)$ correspond words
    $\underbrace{1\dots1}_{r} \equiv 1^r$. $C^{(\alpha)}_i$ are Cherednik
    operators, canonically transformed $\alpha$-times
    \begin{align}
      C^{(\alpha)}_i=\hat{\cal O}_v^{\alpha}C_i \hat{\cal O}_v^{-\alpha}= C_i \Bigg{|}_{\ttop{i} \rightarrow x^{-\alpha} \ttop{i}},
      \ \ \  C^{(1)}_i = \widetilde{C}_i, \ \ C^{(2)}_i =
      \widetilde{\widetilde{C}}_i\nn
    \end{align}

    \bigskip

    In general, words transform
    under action of $\mathcal{O}_h$ and $\mathcal{O}_v$ operators as follows.
    Clearly, under the action of $\mathcal{O}_v$ (c.f. \eqref{eq:ov-action})
    every number in a word is increased by $1$
    \begin{align} \label{eq:ov-oper-words}
      \mathcal{O}_v(\alpha_1 \dots \alpha_r) = (\alpha_1+1)\dots(\alpha_r + 1)
    \end{align}

    The action of $\mathcal{O}_h$ operator is more tricky.
    Write a given word $\alpha_1 \dots \alpha_r$, explicitly writing out $1$'s
    \begin{align}
      W_{\vec\alpha} = \alpha_1\dots \alpha_r = 1^{m_0} A_1 1^{m_1} A_2 1^{m_2} \dots 1^{m_{r-1}} A_s 1^{m_r}, \ \ \ A_i > 1 \ \ \forall i=1..s,\nn
    \end{align}
    where some of the multiplicities $m_i$ may be zero.
    Then $\mathcal{O}_h$ increases the length
    of every group of $1$'s by $1$, and, furthermore, changes every $A_i$ to $A_i - 1$ $2$'s
    \begin{align} \label{eq:oh-oper-words}
      \mathcal{O}_h(W_{\vec\alpha}) = 1^{m_0 + 1} 2^{A_1 - 1}
      1^{m_1 + 1} \dots 1^{m_{s-1} + 1} 2^{A_s - 1} 1^{m_s + 1}
    \end{align}

    \bigskip

    Therefore, in order to figure out the Cherednik formula for particular
    ray of Hamiltonians, one needs, with the help of the Euclid algorithm, figure out
    sequence of $\mathcal{O}_h$ and $\mathcal{O}_v$ transformations, that
    make it from the diagonal ray $(1,1)$.
    For instance for the ray $(-7,11)$ we have
    \begin{align}
      (-7,11) \mathop{\leftarrow}^{\mathcal{O}_h^{-1}}
      (-7,4) \mathop{\leftarrow}^{\mathcal{O}_v^{-1}}
      (-3,4) \mathop{\leftarrow}^{\mathcal{O}_h^{-1}}
      (-3,1) \mathop{\leftarrow}^{\mathcal{O}_v^{-2}}
      (-1,1)\nn
    \end{align}
    Therefore, the corresponding word is
    \begin{align}
      12122122121 \mathop{\leftarrow}^{\mathcal{O}_h}
      2332 \mathop{\leftarrow}^{\mathcal{O}_v}
      1221 \mathop{\leftarrow}^{\mathcal{O}_h}
      3 \mathop{\leftarrow}^{\mathcal{O}_v^2}
      1\nn
    \end{align}
    and hence the Hamiltonians on this ray are manifestly equal to
    \begin{align}
      \mathcal{H}_k^{(-7,11)} = \sum_{i=1}^N
      \left(
      C_i^{(1)} x_i
      C_i^{(2)} x_i
      C_i^{(1)} x_i
      C_i^{(2)} x_i
      C_i^{(2)} x_i
      C_i^{(1)} x_i
      C_i^{(2)} x_i
      C_i^{(2)} x_i
      C_i^{(1)} x_i
      C_i^{(2)} x_i
      C_i^{(1)}
      \right)^k\nn
    \end{align}

Similarly,
\be
e_{(-5,11)} \sim {\cal O}_h^{-2} {\cal O}_v^{-4} \cdot 1 =  {\cal O}_h^{-2}  \cdot 5 = {\cal O}_h^{-1}\cdot 12^41
= 11212121211 \nn \\
e_{(-7,4)} \sim {\cal O}_v^{-1} {\cal O}_h^{-1} {\cal O}_v^{-2} \cdot 1 =
 {\cal O}_v^{-1} {\cal O}_h^{-1}\cdot 3 =   {\cal O}_v^{-1}\cdot 12^21 =   23^22
\label{Nbodyexas}
\ee
in agreement with (\ref{eq:table-of-ham-words}) below.

    \noindent Further, we give an extensive
    table of examples \eqref{eq:table-of-ham-words} to illustrate further how the construction works.

\subsubsection{Examples\label{Nex}}

We will use an \textit{empty} word $\emptyset$ to consistently denote $\frac{1}{x_i}$,
that is, something which, after adding one $x_i \widetilde{C}_i$,
according to the above prescription becomes $\widetilde{C}_i$.

The exponent of the \textit{square brackets} denotes repeating sequences inside a word, so that
  \begin{align}
    [1 2]^2 1 \equiv 1 2 1 2 1\nn
  \end{align}
  and the exponent of round braces is reserved to the usual exponent of an expression
  \begin{align}
    (1)^2 = \widetilde{C}_i^2 \ne \widetilde{C}_i x_i \widetilde{C}_i
    = 1^2 \equiv [1]^2\nn
  \end{align}

  After a word is transformed into a monomial in (suitably transformed)
  Cherednik operators and $x_i$'s, all that is left is to take the sum
  $\sum_{i=1}^N$ to obtain the Hamiltonian.

With this notation, one can straightforwardly write down the first elements of the algebra:

\be \label{eq:table-of-ham-words}
  e_{(-n,m)} =
  \ee
{\footnotesize
\be
   %\!\!\!\!\!\!
   \!\!\!\!\!\!\!\!\!\!\!\!\!\!\!\!\!\!
   \!\!\!\!\!\!\!\!\!\!\!\!\!\!\!\!\!\!\!\!\!\!
  \begin{array}{
%c|
  c|c|c|c|c|c|c|c|c|c|c|c|c||c|}
\ldots
%&
&&&&&&&&&&&&\ldots \\
\hline
%&{\ba (1)^{12}}
&{\bb 12^{10}\,1}&{\ba (12^4\,1)^2}&{\ba (12^3\,1)^3}&{\ba (121)^4} &{\bb 121212^212121}
&{\ba (1^2)^6}&{\bb 1^2 2121^2 212 1^2}&{\ba (1^3)^4}&{\ba (1^4)^3}&{\ba (1^6)^2}&{\bb 1^{12}}&{\ba (0)^{12}}&12 \\
\hline
%&{\bb 2^{11}}
&{\ba (1)^11}&{\bb 12^91}&{\bb 12^412^41}&{\bb 12^212^312^21}&{\bb 1212^212^2121}
&{\bb 12121212121}&{\bb 1^221212121^2}&{\bb 1^221^221^221^2}&{\bb 1^321^321^3}&{\bb 1^521^5}&{\bb 1^{11}}&{\ba (0)^{11}}&11 \\
\hline
%&{\ba (2^5)^2}
&{\bb 2^{10}}&{\ba (1)^{10}}&{\bb 12^81}&{\ba (12^31)^2}&{\bb 12^212^212^21}
&{\ba (12121)^2}&{\ba (1^2)^5}&{\ba (1^221^2)^2}&{\bb 1^321^221^3}&{\ba (1^5)^2}&{\bb 1^{10}}&{\ba (0)^{10}}&10 \\
\hline
%&{\ba (2^3)^3}
&{\bb 2^432^4}&{\bb 2^9}&{\ba (1)^9}&{\bb 12^71}&{\bb 12^312^31}
&{\ba 121)^3}&{\bb 121212121} &{\bb 1^2212121^2}&{\ba (1^3)^3}&{\ba 1^421^4}&{\bb 1^9}&{\ba (0)^9}&9 \\
\hline
%&{\ba (2^2)^4}
&{\bb 2^232^232^2}&{\ba (2^4)^2}&{\bb 2^8}&{\ba (1)^8}&{\bb 12^61}
&{\ba (12^21)^2}&{\bb 1212^2 121}&{\ba (1^2)^4}&{\bb 1^221^221^2}&{\bf (1^4)^2}&{\bb 1^8}&{\ba (0)^8}&8 \\
\hline
%&{\bb 23^223^23}
&{\bb 2323232}&{\bb 2^23232^2}&{\bb 2^332^3}&{\bb 2^7}&{\ba (1)^7}
&{\bb 12^51}&{\bb 12^212^21}&{\bb 1212121}&{\bb 1^22121^2}&{\bb 1^321^3}&{\bb 1^7}&{\ba (0)^7}&7 \\
\hline
%&{\bb (2)^6}
&{\bb 23^42}&{\ba (232)^2} &{\ba (2^2)^3}&{\ba (2^3)^2}&{\bb 2^6}&{\ba (1)^6}
&{\bb 12^4\,1}&{\ba (121)^2}&{\ba (1^2)^3}&{\ba (1^3)^2}&{\bb 1^6}&{\ba (0)^6}&6 \\
\hline
%&{\bb 3^243^2}
&{\bb 3^5}&{\ba (2)^5} &{\bb 23^2 2}&{\bb 23232}&{\bb 2^232^2}&{\bb 2^5}&
{\ba (1)^5} & {\bb 12^3\,1}&{\bb 12121}&{\bb 1^221^2}&{\bb 1^5}&{\ba (0)^5}&5 \\
\hline
%&{\ba (3)^4}
&{\bb 34^23}&{\ba (3^2)^2}&{\bb 3^4}&{\ba (2)^4}&{\bb 23^22}
&{\ba (2^2)^2}&{\bb 2^4}&{\ba (1)^4}&{\bb 12^21}&{\ba (1^2)^2}&{\bb 1^4}&{\ba (0)^4}&4 \\
\hline
%&{\ba (4)^3}
&{\bb 454}&{\bb 4^3}&{\ba (3)^3}&{\bb 343}&{\bb 3^3}&{\ba (2)^3}
&{\bb 232}&{\bb 2^3}&{\ba (1)^3}&{\bb 121}&{\bb 1^3}&{\ba (0)^3}&3 \\
\hline
%&{\ba (6)^2}
&{\bb 6^2} &{\ba (5)^2}&{\bb 5^2}&{\ba (4)^2}&{\bb 4^2}
&{\ba (3)^2}&{\bb 3^2}&{\ba (2)^2}&{\bb 2^2}&{\ba (1)^2}&{\bb 1^2}&{\ba (0)^2}&2 \\
\hline
%&12
&11&10&9&8&7&6&5&4&3&2&{\bb 1 }&{\bb 0 }&1 \\
\hline
%&
&{\ba (\emptyset)^{11}}&{\ba (\emptyset)^{10}}&{\ba (\emptyset)^9}&{\ba (\emptyset)^8}&{\ba (\emptyset)^7}
&{\ba (\emptyset)^6}&{\ba (\emptyset)^5}&{\ba (\emptyset)^4}&{\ba (\emptyset)^3}&{\ba (\emptyset)^2}&\emptyset& 1 &0 \\
\hline\hline
        \ldots
%& -12
        & -11 & -10 & -9 & -8 & -7 & -6 & -5 & -4 & -3 & -2 & -1 & 0 &  -n \backslash m \\
 \end{array}
\nn
\ee
}
i.e.
{\footnotesize
\be
  % \!\!\!\!\!\!\!\!\!\!\!\!\!\!\!\!\!\!\!\!\!\!\!\!\!\!\!\!\!\!\!
   %\!\!\!\!\!\!\!\!\!\!\!\!\!\!\!\!\!\!\!\!\!\!\!\!\!\!\!\!\!\!\!\!\!\!\!\!\!\!\!\!\!\!\!\!\!\!\!\!\!\!\!\!\!\!\!\!\!\!\!\!
   \!\!\!\!\!\!\!\!\!\!\!\!\!\!\!\!\!\!\!\!\!\!
  \begin{array}{
%c|c|c|
  c|c|c|c|c|c|c|c|c||c|}
%&&&
&&&&&&&&&\ldots \\
\hline
%&{\bb (2)^6}&{\bb 23^42}&{\ba (232)^2}
%&{\ba (\tilde{\tilde C} x\tilde{\tilde C} )^3}
%&{\ba (\tilde{\tilde C} x\tilde{\tilde C} x\tilde{\tilde C})^2}
&{\bb \tilde{\tilde C} x\tilde{\tilde C} x\tilde{\tilde C} x\tilde{\tilde C} x\tilde{\tilde C} x\tilde{\tilde C}}
&{\ba \tilde C^6} &{\bb \tilde C x \tilde{\tilde C} x\tilde{\tilde C} x\tilde{\tilde C} x\tilde{\tilde C} x \tilde C}
&{\ba (\tilde C x \tilde{\tilde C}x \tilde C)^2} &{\ba (\tilde C x \tilde C)^3}&{\ba (\tilde C x \tilde C x \tilde C)^2}
&{\bb \tilde C x \tilde C x \tilde C x \tilde C x \tilde C x \tilde C}&{\ba C^6}&6 \\
\hline
%&{\bb 3^243^2}&{\bb 3^5}&{\ba (2)^5}
%&{\bb \tilde{\tilde C} x\tilde{\tilde{\tilde C}} x\tilde{\tilde{\tilde C}} x\tilde{\tilde C}}
%&{\bb \tilde{\tilde C} x\tilde{\tilde{\tilde C}} x\tilde{\tilde C} x\tilde{\tilde{\tilde C}} x\tilde{\tilde C}}
&{\bb \tilde{\tilde C} x\tilde{\tilde C} x\tilde{\tilde{\tilde C}} x\tilde{\tilde C} x\tilde{\tilde C}}
&{\bb \tilde{\tilde C} x\tilde{\tilde C} x\tilde{\tilde C} x\tilde{\tilde C} x\tilde{\tilde C} }&
{\ba \tilde C^5} & {\bb \tilde C x \tilde{\tilde C} x \tilde{\tilde C} x\tilde{\tilde C} x \tilde C}
&{\bb \tilde C x \tilde{\tilde C} x \tilde C x \tilde{\tilde C} x \tilde C}&{\bb \tilde C x \tilde C x \tilde{\tilde C} x \tilde C x \tilde C}
&{\bb  \tilde C x \tilde C x \tilde C x \tilde C x \tilde C}&{\ba C^5}&5 \\
\hline
%&{\ba (3)^4}&{\bb 34^23}&{\ba (3^2)^2}
%&{\bb \tilde{\tilde{\tilde C}} x \tilde{\tilde{\tilde C}} x\tilde{\tilde{\tilde C}} x\tilde{\tilde{\tilde C}} }
%&{\ba (\tilde{\tilde C})^4}
&{\bb \tilde{\tilde C} x \tilde{\tilde{\tilde C}} x\tilde{\tilde{\tilde C}} x\tilde{\tilde C} }
&{\ba (\tilde{\tilde C}x\tilde{\tilde C})^2}
&{\bb \tilde{\tilde C} x \tilde{\tilde C} x \tilde{\tilde C} x \tilde{\tilde C}}&{\ba \tilde C^4}
&{\bb \tilde C}x{\tilde{\tilde C} x \tilde{\tilde C} x \tilde C}&{\ba (\tilde C x \tilde C)^2}
&{\bb  \tilde C x \tilde C x \tilde C x \tilde C  }&{\ba C^4}&4 \\
\hline
%&{\ba (4)^3}&{\bb 454}&{\bb 4^3}
%&{\ba (\tilde{\tilde{\tilde C}})^3}
%&{\bb \tilde{\tilde{\tilde C}} x\tilde{\tilde{\tilde{\tilde C}}} x\tilde{\tilde{\tilde C}} }
&{\bb \tilde{\tilde{\tilde C}} x\tilde{\tilde{\tilde C}} x\tilde{\tilde{\tilde C}} }
&{\ba \tilde{\tilde C}^3}
&{\bb \tilde {\tilde C} x \tilde{\tilde{\tilde C}} x \tilde{\tilde C}}
&{\bb \tilde {\tilde C} x \tilde{\tilde C} x \tilde{\tilde C}}&{\ba \tilde C^3}
&{\bb \tilde C x \tilde{\tilde C} x \tilde C}&{\bb \tilde C x \tilde C x \tilde C}&{\ba C^3}&3 \\
\hline
%&{\ba (6)^2}&{\bb 6^2} &{\ba (5)^2}
%&{\bb \tilde{\tilde{\tilde{\tilde{\tilde C}}}} x \tilde{\tilde{\tilde{\tilde{\tilde C}}}}}
%&{\ba \tilde{\tilde{\tilde{\tilde C}}}^2}
&{\bb \tilde{\tilde{\tilde{\tilde C}}} x \tilde{\tilde{\tilde{\tilde C}}}}
&{\ba \tilde{\tilde{\tilde C}}^2}&{\bb \tilde{\tilde{\tilde C}}x\tilde{\tilde{\tilde C}}}&{\ba (\tilde{\tilde C})^2}&{\bb \tilde{\tilde C}x\tilde{\tilde C}}&{\ba \tilde C^2}&{\bb \tilde Cx\tilde C}&{\ba C^2}&2 \\
\hline
%&12&11&10
%&9
%&8
&7
&\tilde{\tilde{\tilde{\tilde{\tilde{\tilde C}}}}}
&\tilde{\tilde{\tilde{\tilde{\tilde C}}}}
&\tilde{\tilde{\tilde{\tilde C}}}&\tilde{\tilde{\tilde C}}&\tilde{\tilde C}&{\bb \tilde C }&{\bb C }&1 \\
\hline
%&&&
%{\ba 1/x^{10}}&
%{\ba 1/x^9}
%&{\ba 1/x^8}
&{\ba 1/x^7}&{\ba 1/x^6}&{\ba 1/x^5}&{\ba 1/x^4}&{\ba 1/x^3}&{\ba 1/x^2}&1/x& 1&0 \\
\hline\hline
        \ldots
        %& -12 & -11 & -10
        %& -9
        %& -8
        & -7 & -6 & -5 & -4 & -3 & -2 & -1 & 0 &  -n \backslash m \\
 \end{array}
\nn
\ee
}

\bigskip

\bigskip

Boldfaced are the entries $(-n,m)$ with $m$ and $n$ not coprime, then the entry is that
of the coprime element, raised to the power of maximal common divisor.
These {\bf boldfaced entries form the rays of apparently commuting operators}.
The commutativity of the ``Hamiltonians'' $\sum_i A_i^k$ here follows
from commutativity of the ``root'' operators: $[A_i,A_j] = 0$ for $i,j=1..N$.

In other quadrants, one needs to change the definition of the sequence,
from $abc\ldots := C^{[a]} x C^{[b]} x C^{[c]} \ldots$, where $[a]$ denotes the number of tildes, $C^{[a]}=(\hat t)^a C$
to  $abc\ldots :=  C^{[a]} \frac{1}{x} C^{[b]} \frac{1}{x} C^{[c]} \ldots  $, in accordance with symmetry properties (see the comment after formula (\ref{eb})).

Clearly, parameter $m$ in the table counts the number of $C$ operators, which is spin,
while $n=$ number of tildes\ -\ number of $x$  = dimension:
\be
m \ = \ \#(C), \ \ \ \ \ \
n \ = \ \#({\rm tilde}) - \#(x)\nn
\ee
It is a question whether these tables (consisting of linear words of numbers) can be somehow lifted to an abstract description of the algebra {\bf E} (so far consisting of {\it trees} of numbers (\ref{01})).
For example, in the 1-body representation with $C = q^{\hat D}$ and $(\hat t)^A C = x^{-A} C$
we get in a given box in the table:
\be
e_{\Box} = \hat t^{\#_\Box({\rm tilde})}  C^{\#_\Box(C)} x^{\#_\Box(x)} \sim  x^{\#_\Box(x)-\#_\Box({\rm tilde})} C^{\#_\Box(C)}
= x^{-n}q^{m\hat D} = e_{-n,m}\nn
\ee
i.e. the same formulas work in this representation, which is of course not a big surprise.
More interesting would be to understand if there is a generalization to the Fock representation: whether there are simple operators
which substitute $C$ and $x$. Tildes are provided by action of the ${\cal O}_v$ automorphism.

    \subsection{Additional automorphisms and symmetries of the $N$-body representation}

    In addition to the powerful $\mathcal{O}_h$ and $\mathcal{O}_v$ symmetries,
    $N$-body representation, as we already mentioned, also celebrates additional
    \textit{reflection} symmetries: $e_{(n,m)}(x;q,t)\sim e_{(-n,m)}(x^{-1};q^{-1},t^{-1})$ and $e_{(n,m)}(q,t)=-e_{(n,-m)}(q^{-1},t^{-1})$. The first symmetry is clearly specific for the $N$-body representation, while the second symmetry is more generally formulated. One may think it is related to vanishing the central charges. However, this is not the case: as we shall see in the next section, the Fock representation with non-zero one of the central charges still enjoys this symmetry. \textbf{These symmetries allow us to write Hamiltonians
      for all quadrants, starting from \eqref{eq:ham-through-cher}.
    }

{\subsection{Many-body integrable systems: generalization of the trigonometric Ruijsenaars system}
    \label{sec:new-integrable-systems}

In the case of $W_{1+\infty}$ algebra and its deformation to the affine Yangian algebra \cite{MMMP1}, we learned that the commutative family associated with (-1,1) ray gave rise to Hamiltonians of the rational Calogero model, while the commutative families associated with (-1,r) ray, to Hamiltonians of generalizations of that model \cite{MMCal}: the $W_{1+\infty}$ algebra described the free fermion point of the rational Calogero model and its generalizations, while the affine Yangian algebra described these same systems with an arbitrary coupling. At the same time, generalization to other rays is known only in the $W_{1+\infty}$ case, i.e. one generates this way only integrable systems at the free fermion point (which corresponds to Hamiltonians describing non-interacting particles, but in external potentials).

At last, the vertical commutative ray in the case of $W_{1+\infty}$ /affine Yangian algebras corresponded to the trigonometric Calogero-Sutherland model \cite{MMCal,MMMP1}.

The situation in the elliptic Hall/DIM algebra is a little bit different. The vertical ray still corresponds to the trigonometric system, to the trigonometric Ruijsenaars model. However, (-1,1) ray is associated with the same system, but after a trivial canonical transformation. Generally, the rays obtained after the action of automorphism ${\cal O}_v$ are equivalent in this sense. Hence, the new many-body systems are generated only by ${\cal O}_h$. The simplest example is given by rays $(-1,r)$, $r=2, 3, \dots$. Manifestly, the first new non-trivial Hamiltonian is given by
    \begin{align}
      H^{(-1,2)}_1 = & \ \sum_{i=1}^N \widetilde{C}_i x_i \widetilde{C}_i\Big|_{symmetric\ functions}
      \\ \notag
      = & \ \sum_{i=1}^N \frac{(q-1)^2}{q^2} \frac{1}{x_i}
    \left (
    \prod_{j \neq i} \frac{(t x_i - x_j)}{(x_i - x_j)}
    \ttop{i}
    \right )^2\nn \\ \notag
    & \ + \frac{(q-1)(t - q)(t - 1)}{q}
    \sum_{i \neq j} \prod_{k \neq i,j}
    \frac{(t x_i - x_k)(t x_j - x_k)}{(x_i - x_k)(x_j - x_k)}
    \frac{1}{(q x_i - x_j)} \ttop{i} \ttop{j},
    \end{align}
This Hamiltonian earlier emerged in \cite[Eqs.(5.10)-(5.12)]{CF} as a Hamiltonian obtained by action of an automorphism
which is a kind of square root of ${\cal O}_v^{-1}$ (\ref{NbOv}) (it makes a replace $q^{\hat D_i}\to x_i^{-1/2}q^{\hat D_i}$)
from the Hamiltonian of the twisted Macdonald-Ruijsenaars
system \cite{CE}. % associated with ray (1,2).
One may expect\footnote{We are grateful to O. Chalykh for this comment.} that ray (-1,r) Hamiltonians are closely related to integrable systems coming from cyclic quivers and cyclotomic DAHAs \cite{CE}.

    Higher Hamiltonians can also be straightforwardly written explicitly
    in terms of $x_i$'s and $\ttop{i}$'s, however, these answers are non-factorizable
    and, therefore, not really illuminating. Hence, we find the Cherednik operator
    description much more compact, convenient and suitable for both computerized
    and symbolic calculations.
  }

\section{Fock representation}

\subsection{{Elements of the representation}}

\subsubsection{Generating elements}

In this section, we introduce the Fock representation with the central charges $c=(1,0)$. Let
$\mathcal{F}_{q_1,q_2}^{(-1,0)}=\mathbb{C}[p_1,p_2,\cdots]$ be the vector space of (graded) polynomials in the variables $p_k$. Denote
$(q_1,q_2,q_3)=(q, t^{-1},q^{-1}t)$ and use the Macdonald polynomials as the basis of polynomials in this vector space. Then, there is a map
\begin{equation}
	\pi:U_{q_1,q_2,q_3}(\hat{\hat{\mathfrak{gl}}}_1)\rightarrow \mathrm{Aut}(\mathcal{F}_{q,t^{-1}}^{(-1,0)})\nn
\end{equation}
defined by manifest action of the generating elements of the algebra $U_{q_1,q_2,q_3}(\hat{\hat{\mathfrak{gl}_1}})$
on the Macdonald polynomials (see, e.g.,\footnote{We use the picture of generators rotated by 90$^\circ$ by the Miki automorphism w.r.t. that used in these references.} \cite{FHHSY}, \cite[sec.2.4 at N=0]{AFS}, or \cite[Eqs.(37)-(45)]{Zenk}) that gives a representation of $U_{q_1,q_2,q_3}(\hat{\hat{\mathfrak{gl}}}_1)$ on $\mathcal{F}_{q,t^{-1}}^{(-1,0)}$, where the superscript $(-1,0)$ refers to the values of the two central charges of this algebra, and we choose $\mathfrak{q}=q_3^{1\over 2}=\sqrt{t/q}$. This representation is manifestly given by the formulas ($n\in\mathbb{Z}_{>0}$)
\be\label{Fock1}
e_{(n,0)}&=&{1\over q^{n/2}-q^{-n/2}}p_n\nn\\
e_{(-n,0)}&=&{n\over t^{n/2}-t^{-n/2}}{\p\over\p p_n}
\ee
and ($n\in\mathbb{Z}_>$)
\be\label{Fock2}
e_{(n,1)}&=&-{q^{n/2}\over(1-q)(1-t^{-1})}\oint_0{dz\over z^{n+1}}\exp\left(\sum_k{1-t^{-k}\over k}z^kp_k\right)\exp\left(-\sum_k(1-q^k)z^{-k}{\p\over\p p_k}\right)\nn\\
e_{(n,-1)}&=&{t^{n/2}\over(1-q^{-1})(1-t)}\oint_0{dz\over z^{n+1}}\exp\left(-\sum_k{1-t^{-k}\over k}z^k\Big({t\over q}\Big)^{k/2}p_k\right)\exp\left(\sum_k(1-q^k)\Big({t\over q}\Big)^{k/2}z^{-k}{\p\over\p p_k}\right)=\nn\\
&=&{q^{-n/2}\over(1-q^{-1})(1-t)}\oint_0{dz\over z^{n+1}}\exp\left(\sum_k{1-t^{k}\over k}z^kp_k\right)\exp\left(-\sum_k(1-q^{-k})z^{-k}{\p\over\p p_k}\right)
\ee
These are Fock representation counterparts of formulas (\ref{eb})-(\ref{ee}) in the $N$-body representation.
The Fock representation also celebrates one of the additional symmetries: $e_{(n,m)}(q,t)=-e_{(n,-m)}(q^{-1},t^{-1})$. Moreover, the symmetry $e_{(n,0)}(q,t)=-e_{(n,0)}(q^{-1},t^{-1})$ is also present in the both representations.

\subsubsection{Vertical ray (0,1)\label{5.1.2}}

Now we construct Hamiltonians of the trigonometric Ruijsenaars system in the Fock representation \cite{AK,Z,MMgenM}, which are associated with ray (0,1).

The first element of this ray is already contained in (\ref{Fock2}) (and similarly for ray (0,-1)):
\be
e_{(0,1)}&=&-{1\over(1-q)(1-t^{-1})}\oint_0{dz\over z}\exp\left(\sum_k{1-t^{-k}\over k}z^kp_k\right)\exp\left(-\sum_k(1-q^k)z^{-k}{\p\over\p p_k}\right):=\nn\\
&:=&-{1\over(1-q)(1-t^{-1})}\oint_0{dz\over z}V(z)\nn
\ee
The Macdonald polynomials $M_\lambda$ are eigenfunctions of this operator with the eigenvalue
\be
e_{(0,1)}M_\lambda=\left(-{1\over (1-q)(1-t^{-1})}+\sum_{(i,j)\in\lambda}q^{j-1}t^{1-i}\right)M_\lambda\nn
\ee
In fact, the Macdonald polynomials are eigenfunctions of the whole commutative set $e_{(0,n)}$ with $n>0$,
the eigenvalues of the other operators at this ray being
\be\label{e0nef}
e_{(0,n)}M_\lambda=\left(-{1\over (1-q^n)(1-t^{-n})}+\sum_{(i,j)\in\lambda}q^{n(j-1)}t^{n(1-i)}\right)M_\lambda
\ee
Manifestly, we first construct (see (\ref{1,r}) and (\ref{0,r}))
\be\label{e12}
e_{(1,2)}&=&[e_{(0,1)},e_{(1,1)}]={q^{1\over 2}\over(1-q)^2(1-t^{-1})^2}\left(\oint_{z=0}\oint_{w=0} {dwdz\over w^2z}
{(z-w)(tz-qw)\over(z-qw)(tz-w)}:V(w)V(z):-\right.\nn\\
&-&\left.\oint_{w=0}\oint_{z=0} {dwdz\over w^2z}
{(w-z)(tw-qz)\over(w-qz)(tw-z)}:V(w)V(z):\right)=\nn\\
&=&{q^{1\over 2}\over(1-q)^2(1-t^{-1})^2}\oint_{z=0}\oint_{w=0} {dwdz\over w^2z^2}
(z-w)\Phi(z,w):V(w)V(z):
\ee
where
\be
\Phi(z,w):={(z-w)(tz-qw)\over(z-qw)(tz-w)}\nn
\ee
and the normal ordering means all derivatives w.r.t. $p_k$ placed to the most right. One can see that the factor $\Phi(z,w)$ controls non-commutativity of $e_{(0,1)}$ and $e_{(1,2)}$, the commutator is zero without it. Note that $\Phi(z,w)$ becomes 1 in the limit $q,t\to 1$. However, this limit is singular. Indeed, $\lim_{q\to 1}\Phi(z,w)\sim 1-q$ and $\lim_{t\to 1}\Phi(z,w)\sim 1-t$. Moreover, the integral does not vanish only when, at least, one degree of $z$ or $w$ comes from expanding the exponential, which gives another factor of $1-t$. Hence, in the limit $t\to 1$, $e_{(1,2)}$ is regular. At the same time, in the limit $q\to 1$, $e_{(1,2)}$ has a pole $(1-q)^{-1}$.

Similarly,
\be\label{Fe1n}
\boxed{e_{(1,m)}={(-1)^mq^{1\over 2}\over(1-q)^m(1-t^{-1})^m}\oint_{z_1=0}...\oint_{z_m=0} \prod_{i=1}^m{dz_i\over z_i^2}
{\cal P}_m(z_1,\ldots,z_{m})\prod_{i>j}\Phi(z_i,z_j)\prod_{i=1}^m:V(z_i):}
\ee
where the polynomial ${\cal P}_m(z_1,\ldots,z_{m})$ of $m$ variables $z_i$ is defined by the recurrent relation
\be
{\cal P}_{m+1}(z_1,\ldots,z_{m+1})=z_{m+1}{\cal P}_m(z_1,\ldots,z_{m})-z_{1}{\cal P}_m(z_2,\ldots,z_{m+1})\nn
\ee
with the first term of the recursion ${\cal P}_1(z_1)=1$, i.e.
\be
{\cal P}_1(z_1)=1\nn\\
{\cal P}_2(z_1,z_2)=z_2-z_1\nn\\
{\cal P}_3(z_1,z_2,z_3)=-z_1z_2+2z_1z_3-z_2z_3\nn\\
{\cal P}_m(z_1,\ldots,z_{m})=\prod_{k=1}^mz_k\cdot
\sum_{k=1}^{m}(-1)^{k+1} \binom{m-1}{k-1}{1\over z_k}\nn
\ee
where $\displaystyle{\binom{m}{k}}$ denotes the binomial coefficient.

In order to generate $\mathfrak{h}_{(0,m)}$, one now has to commute these $e_{(1,m)}$ with $e_{(-1,0)}$ obtaining
\be
\boxed{
\mathfrak{h}_{(0,m)}={(-1)^{m+1}\kappa_1\over(1-q)^m(1-t^{-1})^m}\oint_{z_1=0}...\oint_{z_m=0}\prod_{i=1}^m{dz_i\over z_i^2}
\left(\sum_{i=1}^mz_i\right){\cal P}_m(z_1,\ldots,z_{m})\prod_{i>j}\Phi(z_i,z_j)\prod_{i=1}^m:V(z_i):}\nn
\ee
In particular,
\be
\mathfrak{h}_{(0,2)}&=&-{\kappa_1\over(1-q)^2(1-t^{-1})^2}\oint_{z=0}\oint_{w=0} {dzdw\over w^2z^2}
(z^2-w^2)\Phi(z,w):V(w)V(z):\nn
\ee
The elements $e_{(0,m)}$ are more involved. Parameterizing them\footnote{One can compare these formulas with more involved expressions for the Hamiltonians in another basis \cite{MMgenM}:
$$
\hat H_k=(-1)^{k+1}\cdot k\cdot (t^k-t^{-k})\cdot\sum_\Delta {\psi_{[1^k]}(\Delta)\over z_\Delta}:\left(\prod_{i=1}^{l_{_\Delta}}
\oint_0{dz_i\over z_i}{\hat V_{\delta_i}(z_i)\over t^{2(i-1)}(t^{\delta_i}-t^{-\delta_i})}\right):
\prod_{i<j}{(z_i-z_j)(t^{-2\delta_i+2}z_i-t^{-2\delta_j+2}z_j)\over (z_i-t^{-2\delta_j}z_j)(t^{-2\delta_i}z_i-z_j)}
$$
where $\Delta=(\delta_1,\ldots,\delta_{l_{_\Delta}})$ is a partition of length $l_{_\Delta}$, $\psi_{[1^k]}(\Delta)$ is the character of the symmetric group $S_k$, the sum runs over all partitions, and
$$
\hat V_m(z):=\exp\left(\sum_{k>0}{(1-t^{-2mk})p_kz^k\over k}\right)\cdot\exp\left(\sum_{k>0}{t^{2mk}-1\over
t^{2k}-1}{q^{2k}-1\over z^k}
{\partial\over\partial p_k}\right)
$$
}
\be\boxed{
e_{(0,m)}={(-1)^m\kappa_1\over\kappa_m}{1\over(1-q)^m(1-t^{-1})^m}\oint_{z_1=0}...\oint_{z_m=0}\prod_{i=1}^m{dz_i\over z_i^2}{\cal E}_m(z_1,\ldots,z_{m})\prod_{i>j}\Phi(z_i,z_j)\prod_{i=1}^m:V(z_i):}\nn
\ee
one obtains
\be
{\cal E}_1(z_1)&=&z_1\nn\\
{\cal E}_2(z_1,z_2)&=&\kappa_1z_1z_2-2z_1^2+2z_2^2=\kappa_1z_1z_2+2(z_1+z_2){\cal P}_2(z_1,z_2)\nn\\
{\cal E}_3(z_1,z_2,z_3)&=&\kappa_1^2z_1z_2z_3-3\kappa_1z_3(z_1^2-z_2^2)+3(z_1+z_2+z_3)(z_1z_2-2z_1z_3+z_2z_3)=\nn\\
&=&\kappa_1^2z_1z_2z_3+3\kappa_1(z_1+z_2){\cal P}_2(z_1,z_2)z_3+3(z_1+z_2+z_3){\cal P}_2(z_1,z_2,z_3)\nn\\
\ldots\nn
\ee
and generally, after some routine drill with the complete homogeneous symmetric polynomials, one obtains
\be\label{Ee0n}
{\cal E}_m(z_1,\ldots,z_m)&=&\sum_{k=1}^{m}\kappa_1^{m-k} \binom{m}{k-1}\left(\sum_{i=1}^{k}z_i\right)
{\cal P}_{k}(z_1,\ldots,z_{k})\prod_{i=k+1}^mz_i
\ee

\subsubsection{Generic elements}

In fact, this is the general structure of any spin $m$ element ($m\ne 0$) of the algebra in the Fock representation:
\be
e_{(n,m)}=\oint_{z_1=0}...\oint_{z_m=0}\prod_{i=1}^mdz_i
{\cal F}_m^{(n)}(z_1,\ldots,z_{m})\prod_{i>j}\Phi(z_i,z_j)\prod_{i=1}^m:V(z_i):\nn
\ee
where ${\cal F}_m^{(n)}(z_1,\ldots,z_{m})$ is some rational function (in fact, a Laurent polynomial) of only $z_i$, $q$ and $t$. Hence, to make the notation compact, we introduce ``an average"
\be\label{ave}
\Big<\ldots\Big>_m:={1\over(1-q)^m(1-t^{-1})^m}\oint_{z_1=0}...\oint_{z_m=0}\prod_{i=1}^m{dz_i\over z_i^2}\ldots\prod_{i>j}\Phi(z_i,z_j)\prod_{i=1}^m:V(z_i):
\ee
so that
\be
e_{(1,m)}&=&(-1)^mq^{1\over 2}\Big<{\cal P}_m(z_1,\ldots,z_{m})\Big>_m\nn\\
\mathfrak{h}_{(0,m)}&=&(-1)^{m+1}\kappa_1\left<{\cal P}_m(z_1,\ldots,z_{m})\cdot \sum_{i=1}^mz_i\right>_m\nn\\
e_{(0,m)}&=&{(-1)^m\kappa_1\over\kappa_m}\Big<{\cal E}_m(z_1,\ldots,z_{m})\Big>_m\nn
\ee
with
\be\label{P1}
{\cal P}_m(z_1,\ldots,z_{m})&=&\prod_{k=1}^mz_k\cdot
\sum_{k=1}^{m}(-1)^{k+1} \binom{m-1}{k-1}{1\over z_k}\nn\\
{\cal E}_m(z_1,\ldots,z_m)&=&\sum_{k=1}^{m}\kappa_1^{m-k} \binom{m}{k-1}\left(\sum_{i=1}^{k}z_i\right)
{\cal P}_{k}(z_1,\ldots,z_{k})\prod_{i=k+1}^mz_i
\ee
Now evaluating commutators reduces to manipulating with these polynomials. For instance,
\be
e_{(2,5)}=[e_{(1,3)},e_{(1,2)}]=-q\Big<P_3(z_1,z_2,z_3)P_2(z_4,z_5)-P_2(z_1,z_2)P_3(z_3,z_4,z_5)\Big>_5=\nn\\
=-q\Big<-z_1z_2z_4+z_1z_2z_5+3z_1z_3z_4-4z_1z_3z_5+z_1z_4z_5-2z_2z_3z_4+3z_2z_3z_5-z_2z_4z_5\Big>_5\nn
\ee
One can similarly construct
\be
e_{(2,2n+1)}=-q\Big<{\cal P}^{(2)}_n\Big>\nn
\ee
where
\be\label{P2}
{\cal P}^{(2)}_n=[{\cal P}_{n+1},{\cal P}_n]=\prod_{i=1}^{2n+1}z_i
\sum_{i<j}^{2n+1}(-1)^{i+j}{j-i\over n}\binom{n}{i-1}\binom{n}{2n+1-j}{1\over z_iz_j}
\ee
etc.

An arbitrary element $e_{(n,m)}$ with $m\ne 0$ can be presented in the form
\be\label{avee}
e_{(n,m)}=\Big<{\bf P}_k\Big>_m
\ee
with $k=m-n$, i.e. $e_{(n,m)}$ can be associated with a homogeneous Laurent polynomial ${\bf P}_k$, but this polynomial is not unique: there are polynomials with vanishing averages, see examples in the next subsection.

Though the elements $e_{(0,m)}$ can not be presented as the average (\ref{ave}), their commutator with any can: for any element (\ref{avee}) and $k>0$
\be
\phantom{.}[e_{(0,k)},e_{(n,m)}]=-t^k\Big<{\bf P}_k\sum_iz_i^k\Big>_m\nn\\
\phantom{.}[e_{(0,-k)},e_{(n,m)}]=q^{-k}\Big<{\bf P}_k\sum_{i=1}^nz_i^{-k}\Big>_m\nn
\ee
It is interesting to understand if this correspondence between the elliptic Hall algebra and polynomials has its origin in the isomorphism between the former and the shuffle algebra.

Let us define a product of polynomials of $n$ variables ${\bf P}_n$ and of $m$ variables ${\bf Q}_m$ as
\be
{\bf P}_n*{\bf Q}_m:={\bf P}_n(z_1,\ldots,z_n){\bf Q}_m(z_{n+1},\ldots,z_{n+m})\nn
\ee
and the commutator of two polynomials as
\be
\phantom{.}[{\bf P}_n,{\bf Q}_m]:={\bf P}_n*{\bf Q}_m-{\bf Q}_m*{\bf P}_n\nn
\ee
Then, any commutator of two elements of the algebra is reduced to the commutator of two corresponding polynomials: for $e_{(n,m)}=\Big<{\bf P}_{m-n}\Big>_{m}$ and $e_{(n',m')}=\Big<{\bf P}'_{m'-n'}\Big>_{m'}$, one obtains
\be
\phantom{.}[e_{(n,m)},e_{(n',m')}=\Big<\Big[{\bf P}_{m-n},{\bf P}'_{m-n}\Big]\Big>_{m+m'}\nn
\ee

In the forthcoming subsections, we are going to describe elements $e_{(n,m)}$ in the Fock representation. Technically, we do this  describing the commutative rays, every such a ray is associated with a set of polynomials, and the whole problem of description is reduced to manipulations with polynomials. Thus, we manifestly describe these polynomials.

\subsubsection{An alternative description of polynomials ${\cal P}$ }

There is another way to write the polynomials like ${\cal P}$ and ${\cal P}^{(2)}$. For coprime $n$  and $m=kn+a$ ($a<n$), let us define following (\ref{ena3}), the polynomials ${\cal P}^{(n,a)}_k$ through
\be
e_{n,kn+a} = q^{n/2} \left< {\cal P}^{(n,a)}_{k}(z_1,\ldots,z_{kn+a})\right>_{kn+a}\nn
\ee
The Miki automorphism ${\cal O}_v$ acts simply (see sec.\ref{MikiF}) because it does not change the number of arguments.
The second Miki automorphism
\be
{\cal O}_h \Big\{ {\cal P}_{k}^{(n,a)} (z_1,\ldots,z_{kn+a})\Big\}={\cal P}_{k+1}^{(n,a)} (z_1,\ldots,z_{(k+1)n+a})\nn
\ee
may look more sophisticated, since the number of arguments changes.
However, this can be described in a rather universal way, by conversion with a  simple generating function,
as we can see from the example that follow from (\ref{P1}):
\be
{\cal P}_k^{(1)} (z_1,\ldots,z_k)={\cal P}_k (z_1,\ldots,z_k)  =
{\cal O}_h^{k-1} \Big\{{\cal P}_1^{(1)}\Big\}
\sim z_1\ldots z_k \oint_{d\xi}  (1-\xi)^{m-1} \sum_{i=1}^\infty \frac{1}{z_i\xi^i}
\label{genf}
\ee

Despite ${\cal P}_{k}^{(n,a)}$ depends on $kn+a$ variables $z_i$,
{\bf it is $n$-linear in $z^{-1}$}, i.e. in the generating function
$g(\xi):=\sum_{i=1}^\infty \frac{1}{z_i\xi^i}$.
What we list now are the weights $\pi^{(n,a)}_k$ in
\be
{\cal P}_k^{(n,a)}\sim \oint \ldots \oint \pi_k^{n,a}(\xi_1,\ldots,\xi_n) \, g(\xi_1)d\xi_1\ldots g(\xi_n) d\xi_n\nn
\ee
and {\bf $\pi_k^{n,a}$ has minimal degree $0$ in the first of its $n$ variables ($\xi_1$) and maximal degree $kn+a-1$ in
the last one ($\xi_n$)}, which explains why ${\cal P}_k^{(n,a)}$ depends on many more $z$-variables $z_1,\ldots,z_{kn+a}$.
The number of variables $z$ is converted into degree of the weight in $\xi$'s.
The Miki automorphism ${\cal O}_h$ acts as ${\cal O}_h: \ k\longrightarrow k+1$ on $\pi_k^{n,a}$.

As we already know from (\ref{genf}),
\be
\pi_k^{(1)} = (1-\xi)^{k-1}\nn
\ee
Similarly,
\be
\pi_k^{(2)} = \overbrace{(1-\xi_1)^{k}}^{\pi_{k+1}^{(1)}}\cdot \overbrace{\xi_2^{k+1}}^{\rm shift} \overbrace{(1-\xi_2)^{k-1}}^{\pi_{k}^{(1)}}
- \overbrace{(1-\xi_1)^{k-1}}^{\pi_{k}^{(1)}}\cdot \overbrace{\xi_2^k}^{\rm shift} \overbrace{(1-\xi_2)^k}^{\pi_{k+1}^{(1)}}
= \xi_2^k (1-\xi_1)^{k-1}(1-\xi_2)^{k-1}\underline{\big(-1+2\xi_2 -\xi_1\xi_2\big)}\nn
\ee
To obtain this formula we need to take ${\cal P}_{k+1}  \longrightarrow (1-\xi_1)^k$, multiply it by
${\cal P}_k \longrightarrow \xi_2^{k+1}(1-\xi_2)^k$, where the factor $\xi_2^{k+1}$ shifts the arguments
from $z_1,\ldots z_k$ to $z_{k+2},\ldots,z_{2k+1}$, and subtract  ${\cal P}_{k}  \longrightarrow (1-\xi_1)^{k-1}$
times ${\cal P}_{k+1} \longrightarrow \xi_2^{k+1}(1-\xi_2)^k$, where now the shift is to $z_{k+1},\ldots,z_{2k+1}$,
i.e. smaller by one.

Further, from ${\cal P}^{(3,1)}_k={\cal P}^{(2)}_k\circ {\cal P}_k$ we read off:
{\footnotesize
\be
\!\!\!\!\!\!\!\!\!\!\!\!
\pi_k^{(3,1)} = \overbrace{\xi_2^k (1-\xi_1)^{k-1}(1-\xi_2)^{k-1}\big(-1+2\xi_2 -\xi_1\xi_2\big)}^{\pi^{(2)}_k}
\cdot \overbrace{\xi_3^{2k+1}}^{\rm shift}\overbrace{(1-\xi_3)^{k-1}}^{\pi^{(1)}_k}
%- \nn \\
- \overbrace{(1-\xi_1)^{k-1}}^{\pi^{(1)}_k}\cdot \overbrace{\xi_2^{k}}^{\rm shift}
\overbrace{\xi_3^k (1-\xi_2)^{k-1}(1-\xi_3)^{k-1}\big(-1+2\xi_3 -\xi_2\xi_3\big)}^{\pi^{(2)}_k} =
\nn
\ee
}
\be
= \xi_2^k\xi_3^k (1-\xi_1)^{k-1}(1-\xi_2)^{k-1}(1-\xi_3)^{k-1}
\underline{\Big(1-2\xi_3+\xi_2\xi_3 -\xi_3^{k+1}(1-2\xi_2+\xi_1\xi_2)\Big)}
\nn
\ee
Similarly, from   ${\cal P}^{(3,2)}_k=-{\cal P}^{(2)}_k\circ {\cal P}_{k+1}$
{\footnotesize
\be
\!\!\!\!\!\!\!\!\!\!\!\!
\pi_k^{(3,2)} = -\overbrace{\xi_2^k (1-\xi_1)^{k-1}(1-\xi_2)^{k-1}\big(-1+2\xi_2 -\xi_1\xi_2\big)}^{\pi^{(2)}_k}
\cdot \overbrace{\xi_3^{2k+1}}^{\rm shift}\overbrace{(1-\xi_3)^{k}}^{\pi^{(1)}_{k+1}}
%+ \nn \\
+ \overbrace{(1-\xi_1)^{k}}^{\pi^{(1)}_{k+1}}\cdot \overbrace{\xi_2^{k+1}}^{\rm shift}
\overbrace{\xi_3^k (1-\xi_2)^{k-1}(1-\xi_3)^{k-1}\big(-1+2\xi_3 -\xi_2\xi_3\big)}^{\pi^{(2)}_k} =
\nn
\ee
}
\be
= \xi_2^k\xi_3^k (1-\xi_1)^{k-1}(1-\xi_2)^{k-1}(1-\xi_3)^{k-1}
\underline{\Big(-(1-\xi_1)\xi_2(1-2\xi_3+\xi_2\xi_3) +\xi_3^{k+1}(1-\xi_3)(1-2\xi_2+\xi_1\xi_2)\Big)}
\nn
\ee
More examples are
{\footnotesize
\be
\!\!\!\!\!\!\!\!\!
\pi^{(4,1)}_k = (\xi_2 \xi_3 \xi_4)^k \Big((1-\xi_1) (1-\xi_2) (1-\xi_3) (1-\xi_4)\Big)^{k-1}
\cdot \nn \\
\cdot
\underline{\Big(-(1-2\xi_4+\xi_3\xi_4) + (\xi_4^{k+1}+\xi_4^{2k+1})(1-2\xi_3+\xi_2\xi_3)
- \xi_3^{k+1}\xi_4^{2k+1}(1-2\xi_2+\xi_1\xi_2)
\Big)}
\nn \\ \nn \\
\pi^{(4,3)}_k = (\xi_2 \xi_3 \xi_4)^k \Big((1-\xi_1) (1-\xi_2) (1-\xi_3) (1-\xi_4)\Big)^{k-1}\cdot \nn \\
\!\!\!\!\!\!\!\!\!
\cdot \underline{\Big(-\xi_2\xi_3(1-\xi_1)(1-\xi_2)(1-2\xi_4+\xi_3\xi_4)
+ (\xi_4^{k+1}+\xi_4^{2k+2})\xi_2(1-\xi_1)(1-2\xi_3+\xi_2\xi_3)(1-\xi_4)
- \xi_3^{k+1}\xi_4^{2k+2}(1-\xi_3)(1-\xi_4)(1-2\xi_2+\xi_1\xi_2)
\Big)} \nn
\ee
}
Thus we see that ${\cal O}_h$ acts on the generating functions in a rather simple way.
Of interest are the underlined polynomials, because they have somewhat non-trivial dependence on $k$.

\subsection{Commutative subalgebras}

\subsubsection{Ray (1,1)}

In the $N$-body representation, this ray is obtained from ray (0,1) with a trivial transform. In the Fock representation, the Hamiltonians of the ray look as follows. We already know the first element of the ray, it is
\be
e_{(1,1)}&=&-{q^{1/2}\over(1-q)(1-t^{-1})}\oint_0{dz\over z^{2}}\exp\left(\sum_k{1-t^{-k}\over k}z^kp_k\right)\exp\left(-\sum_k(1-q^k)z^{-k}{\p\over\p p_k}\right)\nn
\ee
According to  (\ref{cr}) and (\ref{1,1}), the higher ray Hamiltonians are generated from elements $e_{(n,n+1})$. We already know $e_{(1,2)}$ from (\ref{e12}). Generally,
\be
\boxed{e_{(n,n+1)}=q^{n\over 2}\left<\sum_{k=1}^{n+1}(-1)^{k}\binom{n}{k-1}z_k\right>_{n+1}
}\nn
\ee
and the linear polynomial which is averaged is proportional to ${\cal P}_{n+1}(z_1^{-1},\ldots,z_{n+1}^{-1})$.

Now we can evaluate the Hamiltonians. The second one is
\be
\mathfrak{h}_{(2,2)}&=&-{q\kappa_1\over(1-q)^2(1-t^{-1})^2}\oint_{z=0}\oint_{w=0} {dwdz\over w^2z^2}\Big({1\over z}+{1\over w}\Big)
(z-w)\Phi(z,w):V(w)V(z):\nn
\ee
and generally
\be
\boxed{\mathfrak{h}_{(n,n)}=-q^{n\over 2}\kappa_1\left<\left(\sum_{k=1}^n{1\over z_k}\right)\left(
\sum_{i=1}^{n}(-1)^{k}\binom{n-1}{k-1}z_k\right)\right>_n
}\nn
\ee
At last,
\be\label{enn}
\boxed{
e_{(n,n)}=(-1)^nq^{n\over 2}{\kappa_1\over\kappa_n}\Big<{\cal E}_n(z_1,\ldots,z_{n})\prod_{i=1}^n{1\over z_i}\Big>_n
}
\ee
Hence, we observe in the Fock representation too that the Hamiltonians of ray (1,1) are obtained from those of ray (0,1) by a trivial replace
\be
\exp\left(-\sum_k(1-q^k)z^{-k}{\p\over\p p_k}\right)\longrightarrow {q^{1\over 2}\over z}\exp\left(-\sum_k(1-q^k)z^{-k}{\p\over\p p_k}\right)\nn
\ee

\subsubsection{Ray (p,1)\label{MikiF}}

Similarly to the $N$-body representation, Hamiltonians of rays (p,1) can be also obtained by a trivial replace
\be
\exp\left(-\sum_k(1-q^k)z^{-k}{\p\over\p p_k}\right)\longrightarrow {q^{p\over 2}\over z^p}\exp\left(-\sum_k(1-q^k)z^{-k}{\p\over\p p_k}\right)\nn
\ee
For instance, at $p=2$, the first two Hamiltonians are (see (\ref{Fock2}))
\be
e_{(2,1)}&=&-{q\over(1-q)(1-t^{-1})}\oint_0{dz\over z^{3}}\exp\left(\sum_k{1-t^{-k}\over k}z^kp_k\right)\exp\left(-\sum_k(1-q^k)z^{-k}{\p\over\p p_k}\right)\nn\\
e_{(4,2)}&=&{\kappa_1\over\kappa_2}
{q^2\over(1-q)^2(1-t^{-1})^2}\oint_{z=0}\oint_{w=0} {dwdz\over w^4z^4}
(\kappa_1z_1z_2-2z_1^2+2z_2^2)\Phi(z,w):V(w)V(z):\nn
\ee
and generally
\be
e_{(2n,n)}&=&(-1)^nq^{n}{\kappa_1\over\kappa_n}\Big<{\cal E}_n(z_1,\ldots,z_{n})\prod_{i=1}^n{1\over z_i^2}\Big>_n\nn
\ee
etc.

Hence, we finally obtain for the (p,1) ray Hamiltonians
\be
\boxed{
e_{(pn,n)}=(-1)^nq^{pn\over 2}{\kappa_1\over\kappa_n}\Big<{\cal E}_n(z_1,\ldots,z_{n})\prod_{i=1}^n{1\over z_i^p}\Big>_n
}\nn
\ee

\subsubsection{Ray (1,2)}

Now we study the simplest ray (1,2), which, as we demonstrated in the $N$-body representation, is not obtained from the vertical ray (0,1) by a trivial transform. In fact, we already know the first Hamiltonian of the ray, (\ref{e12}):
\be
\mathfrak{h}_{(1,2)}&=&-\kappa_1q^{1\over 2}\Big<z_2-z_1\Big>_2=q^{1\over 2}\kappa_1\Big<{\cal H}_2(z_1,z_2)\Big>_2\nn\\
e_{(1,2)}&=&q^{1\over 2}\Big<{\cal H}_2(z_1,z_2)\Big>_2\nn
\ee
The second one are
\be
\mathfrak{h}_{(2,4)}&=&q\kappa_1\Big<z_1z_2-2z_1z_3+2z_2z_4-z_3z_4\Big>_4=q\kappa_1\Big<
{\cal H}_4(z_1,z_2,z_3,z_4)\Big>_4\nn\\
e_{(2,4)}&=&q{\kappa_1\over\kappa_2}\Big<\kappa_1(z_1-z_2)(z_3-z_4)-2(z_1z_2-2z_1z_3+2z_2z_4-z_3z_4)\Big>_4=\nn\\
&=&q{\kappa_1\over\kappa_2}\Big<\kappa_1{\cal H}_2(z_1,z_2){\cal H}_2(z_3,z_4)-2{\cal H}_4(z_1,z_2,z_3,z_4)\Big>_4\nn
\ee
In the generic case, the answer is
\be
\boxed{
\mathfrak{h}_{(n,2n)}=(-1)^nq^{n\over 2}\kappa_1\Big<{\cal H}_{n}(z_1,\ldots,z_{2n})\Big>_{2n}
}\nn
\ee
where the polynomials
\be
{\cal H}_1(z_1,z_2)=z_2-z_1={\cal P}_2(z_1,z_2)\nn\\
{\cal H}_2(z_1,z_2,z_3,z_4)=z_1z_2-2z_1z_3+2z_2z_4-z_3z_4=[{\cal P}_3,{\cal P}_1]\nn\\
{\cal H}_3(z_1,z_2,z_3,z_4,z_5,z_6)=[[{\cal P}_3,{\cal P}_2],{\cal P}_1]\nn\\
\ldots\nn\\
{\cal H}_{n}(z_1,\ldots,z_{2n})=[\ldots[{\cal P}_3,\underbrace{{\cal P}_2],\ldots,{\cal P}_2]}_{n-2},{\cal P}_1]\nn
\ee
do not depend on $q$ and $t$.

Similarly,
\be\boxed{
e_{(n,2n)}=q^{n\over 2}{\kappa_1\over\kappa_n}\Big<{\cal E}^{(2)}_{n}(z_1,\ldots,z_{2n})\Big>_{2n}
}\nn
\ee
where the polynomials
\be
{\cal E}^{(2)}_{1}(z_1,z_2)&=&{\cal H}_1\nn\\
{\cal E}^{(2)}_2(z_1,\ldots,z_4)&=&\kappa_1{\cal H}_1^2-2{\cal H}_2\nn\\
{\cal E}^{(2)}_3(z_1,\ldots,z_6)&=&\kappa_1^2{\cal H}_1^3-3\kappa_1{\cal H}_2*{\cal H}_1+3{\cal H}_3\nn\\
{\cal E}^{(2)}_4(z_1,\ldots,z_8)&=&\kappa_1^3{\cal H}_1^4-4\kappa_1^2{\cal H}_2*{\cal H}_1^2+2\kappa_1\Big({\cal H}_2^2+2{\cal H}_3*{\cal H}_1\Big)-4{\cal H}_4\nn\\
{\cal E}^{(2)}_{5}(z_1,\ldots,z_{10})&=&\kappa_1^4{\cal H}_1^5-5\kappa_1^3{\cal H}_2*{\cal H}_1^3+
5\kappa_1^2\Big({\cal H}_3*{\cal H}_1^2+{\cal H}_2^2*{\cal H}_1\Big)-
5\kappa_1\Big({\cal H}_4*{\cal H}_1+{\cal H}_3*{\cal H}_2\Big)+5{\cal H}_{5}\nn\\
\ldots\nn
\ee
are solutions to the equations
\be
(-1)^n\kappa_1{\cal H}_{n}=h_n\Big\{-\kappa_1{\cal E}^{(2)}_{k}\Big\}\nn
\ee
Although the multiplication $*$ is not commutative, in these formulas, one can deal with polynomials ${\cal H}_k$ as with $c$-numbers, since they commute after taking the average (\ref{ave}): they satisfy relations that are implied by the commutativity of distinct $\mathfrak{h}_{(n,m)}$
\be
\Big<\alpha{\cal H}_2*{\cal H}_1+\beta{\cal H}_1*{\cal H}_2\Big>_6&=&0\ \ \ \ \ \ \ \ \hbox{if }\alpha+\beta=0\nn\\
\Big<\alpha{\cal H}_2*{\cal H}_1^2+\beta{\cal H}_1*{\cal H}_2*{\cal H}_1+\gamma{\cal H}_1^2*{\cal H}_2\Big>_8&=&0
\ \ \ \ \ \ \ \ \hbox{if }\alpha+\beta+\gamma=0\nn\\
\Big<\alpha{\cal H}_3*{\cal H}_1+\beta{\cal H}_1*{\cal H}_3\Big>_8&=&0\ \ \ \ \ \ \ \ \hbox{if }\alpha+\beta=0\nn
\ee
etc.

In more detail,
\be
\Big<{\cal H}_2*{\cal H}_1-{\cal H}_1*{\cal H}_2\Big>_6&=&
\Big<z_1z_2z_5-z_1z_2z_6-z_1z_3z_4+2z_1z_3z_6-2z_1z_4z_6+z_1z_5z_6+\nn\\
&+&z_2z_3z_4-2z_2z_3z_5+2z_2z_4z_5-z_2z_5z_6
-z_3z_4z_5+z_3z_4z_6\Big>_6=0\nn
\ee
Hence, some integrals vanish even when $\Phi(z,w)$ is present (see a comment after formula (\ref{e12}) and in sec.\ref{531} below), since some commutators vanish because of the algebra.

\subsubsection{Ray (1,r)}

The scheme we applied for ray (1,2) is literally the same for ray (1,p). In this case, one has
\be
\mathfrak{h}_{(1,p)}&=&(-1)^{p}\kappa_1q^{1\over 2}\Big<{\cal H}^{(p)}_{p-1}(z_1,\ldots,z_p)\Big>_p\nn
\ee
In the generic case, the answer is
\be
\boxed{
\mathfrak{h}_{(n,pn)}=(-1)^{n+p}q^{n\over 2}\kappa_1\Big<{\cal H}^{(p)}_{n(p-1)}(z_1,\ldots,z_{np})\Big>_{np}
}\nn
\ee
where the polynomials
\be
{\cal H}^{(p)}_{p-1}(z_1,\ldots,z_p)={\cal P}_p(z_1,\ldots,z_p)\nn\\
{\cal H}^{(p)}_{2(p-1)}(z_1,\ldots,z_{2p})=[{\cal P}_{p+1},{\cal P}_{p-1}]\nn\\
{\cal H}^{(p)}_{3(p-1)}(z_1,\ldots,z_{3p})=[[{\cal P}_{p+1},{\cal P}_p],{\cal P}_{p-1}]\nn\\
\ldots\nn\\
{\cal H}^{(p)}_{n(p-1)}(z_1,\ldots,z_{np})=[\ldots[{\cal P}_{p+1},\underbrace{{\cal P}_p],\ldots,{\cal P}_p]}_{n-2},{\cal P}_{p-1}]\nn
\ee
do not depend on $q$ and $t$.  This corresponds to the general rule of sec.\ref{sec:commutative-subalgebra-pr}, since ${\cal P}_{p+1}$ correspond to $e_{(1,n+1)}$ and can be obtained as the commutator of $e_{(0,1)}$ with $e_{(1,n)}$ associated with ${\cal P}_{p}$.

Similarly,
\be\boxed{
e_{(n,np)}=q^{n\over 2}{\kappa_1\over\kappa_n}\Big<{\cal E}^{(p)}_{n(p-1)}(z_1,\ldots,z_{np})\Big>_{np}
}\nn
\ee
where the polynomials
\be
{\cal E}^{(p)}_{p-1}(z_1,\ldots,z_p)&=&{\cal H}^{(p)}_{p-1}\nn\\
{\cal E}^{(p)}_{2(p-1)}(z_1,\ldots,z_{2p})&=&\kappa_1\Big({\cal H}^{(p)}_{p-1}\Big)^2-2{\cal H}^{(p)}_{2(p-1)}\nn\\
{\cal E}^{(p)}_{3(p-1)}(z_1,\ldots,z_{3p})&=&\kappa_1^2\Big({\cal H}^{(p)}_{p-1}\Big)^3-3\kappa_1{\cal H}^{(p)}_{2(p-1)}*{\cal H}^{(p)}_{p-1}+3{\cal H}^{(p)}_{3(p-1)}\nn\\
{\cal E}^{(p)}_{4(p-1)}(z_1,\ldots,z_{4p})&=&\kappa_1^3\Big({\cal H}^{(p)}_{p-1}\Big)^4-4\kappa_1^2{\cal H}^{(p)}_{2(p-1)}*\Big({\cal H}^{(p)}_{p-1}\Big)^2+2\kappa_1\Big(\Big({\cal H}^{(p)}_{2(p-1)}\Big)^2+2{\cal H}^{(p)}_{3(p-1)}*{\cal H}^{(p)}_{p-1}\Big)-4{\cal H}^{(p)}_{4(p-1)}\nn\\
{\cal E}^{(p)}_{5(p-1)}(z_1,\ldots,z_{5p})&=&\kappa_1^4\Big({\cal H}^{(p)}_{p-1}\Big)^5-5\kappa_1^3{\cal H}^{(p)}_{2(p-1)}*\Big({\cal H}^{(p)}_{p-1}\Big)^3+
5\kappa_1^2\Big({\cal H}^{(p)}_{3(p-1)}*\Big({\cal H}^{(p)}_{p-1}\Big)^2+\Big({\cal H}^{(p)}_{2(p-1)}\Big)^2*{\cal H}^{(p)}_{p-1}\Big)-\nn\\
&-&5\kappa_1\Big({\cal H}^{(p)}_{4(p-1)}*{\cal H}^{(p)}_{p-1}+{\cal H}^{(p)}_{3(p-1)}*{\cal H}^{(p)}_{2(p-1)}\Big)+5{\cal H}^{(p)}_{5(p-1)}\nn\\
\ldots\nn
\ee
are solutions to the equations
\be
(-1)^n\kappa_1{\cal H}_{n(p-1)}=h_n\Big\{-\kappa_1{\cal E}^{(p)}_{n(p-1)}\Big\}\nn
\ee
or, in terms of generating functions,
\be
\sum_k{\cal E}^{(p)}_{n(p-1)}{z^k\over k}=-{1\over\kappa_1}\log\left(1+\kappa_1\sum_n(-1)^nz^n{\cal H}_{n(p-1)}\right)\nn
\ee
where the power of ${\cal H}$ is understood as the multiple $*$-product.

Again, in these formulas, one can deal with polynomials ${\cal H}_k$ as with $c$-numbers, since they commute after taking the average (\ref{ave}).

\subsection{General elements of algebra in the Fock representation}

\subsubsection{General structure of elements in one quadrant\label{531}}

\fbox{The general structure of elements $e_{(n,m)}$ with $n\ge 0$, $m>0$ is as follows.}

Assume $gcd(n,m)=1$. Then,
\begin{itemize}
\item Any element $e_{(n,m)}$ has the form
\be\label{gcd1}
e_{(n,m)}=q^{n\over 2}\Big<{\bf P}_{m-n}(z_i)\Big>_m
\ee
where ${\bf P}_{m-n}$ is a homogeneous Laurent polynomial of degree $m-n$, which does not depend\footnote{Note that this property is strongly related to using the commutators of only admissible pairs in the definition of the algebra (which is definitely necessary for self-consistency): for instance, if one calculates the commutator of $e_{(1,0)}$ and $e_{(1,3)}$, one immediately generates the polynomial depending on $q$ and $t$:
$$
{\bf P}\sim z_2(z_1+z_3)\sum_i q_i+z_1z_3\sum_i (q_i+q_i^{-1})
$$
instead of
$$
{\bf P}\sim z_2(z_1+z_3)-2z_1z_3
$$
corresponding to $e_{(2,3)}$.
} on $q$ and $t$.
\item Any element $e_{(kn,km)}$ has the form
\be\label{gcdk}
e_{(kn,km)}=q^{kn\over 2}{\kappa_1\over\kappa_k}\Big<{\bf E}_{km-kn}(z_i;\kappa_1)\Big>_m
\ee
where ${\bf E}_{km-kn}$ is a homogeneous Laurent polynomial of degree $km-kn$, which depends on $q$ and $t$ only through the combination $\kappa_1$ polynomially in $\kappa_1$: it is a polynomial in $\kappa_1$ of degree $k-1$. In fact, the dependence on $\kappa_1$ can be manifestly described:
\be
\sum_k{\bf E}_{km-kn}{z^k\over k}=-{1\over\kappa_1}\log\left(1-\kappa_1\sum_kz^k{\bf H}_{km-kn}\right)\nn
\ee
\end{itemize}
where ${\bf H}_{km-kn}$ is a polynomial of the ${\bf P}_{m-n}$ type (\ref{gcd1}), i.e. a homogeneous Laurent polynomial of degree $km-kn$, which does not depend on $q$ and $t$, and the power of ${\bf H}$ is understood as the multiple $*$-product.

Note that all the polynomials with $m>1$ vanishes at all arguments equal to each other: ${\bf P}_{m-n}(z_i)\Big|_{z_1=z_2=\ldots=z_m}=0$. This is equivalent to vanishing commutators of elements of the algebra without $\Phi(z_i,z_j)$ (as we earlier mentioned), since, in this case, the average is the product of $m$ independent integrals.

\subsubsection{Automorphisms and symmetries of the Fock representation\label{532}}

\begin{itemize}
\item[1.] {\bf Miki automorphism.} In sec.\ref{MikiF}, we described the automorphism
\be\label{Mikiv}
{\cal O}_v:\ \ \exp\left(-\sum_k(1-q^k)z^{-k}{\p\over\p p_k}\right)\longrightarrow {q^{1\over 2}\over z}\exp\left(-\sum_k(1-q^k)z^{-k}{\p\over\p p_k}\right)
\ee
that maps $e_{(n,m)}\to e_{(n+m,m)}$. This is a particular realization of one generator of the Miki automorphism, ${\cal O}_v$ (see sec.\ref{Miki}). The second generator, ${\cal O}_h$ is more involved, and we do not know its simple action on the polynomials associated with $e_{(n,m)}$.

\item[2.]{\bf Reflection of gradings.} Another automorphism maps $e_{(n,m)}\to e_{(-n,m)}$, it manifestly given as follows. Assume $gcd(n,m)=1$. Then, representation of (\ref{gcd1}) and (\ref{gcdk}) changes as follows:
\begin{itemize}
\item[$\bullet$] Any element $e_{(-n,m)}$ has the form
\be\label{gcd1m}
e_{(-n,m)}=(-1)^{n+1}t^{-{n\over 2}}\Big<{\bf P}_{m-n}(-z_i^{-1})\prod_{i=1}^mz_i^2\Big>_m
\ee
where ${\bf P}_{m-n}$ is the same Laurent polynomial as in (\ref{gcd1}).
\item[$\bullet$] Any element $e_{(kn,km)}$ has the form
\be\label{gcdkm}
e_{(kn,km)}=(-1)^{kn+1}t^{-{kn\over 2}}{\kappa_1\over\kappa_k}
\Big<{\bf E}_{km-kn}(-z_i^{-1};-\kappa_1)\prod_{i=1}^mz_i^2\Big>_m
\ee
where ${\bf E}_{km-kn}$ is  the same Laurent polynomial as in (\ref{gcdk}).
\end{itemize}
Note that one can generate some $e_{(n,m)}$ with $n<0$ from those with $n>0$ using the inverse of Miki automorphism (\ref{Mikiv}). For instance,
\be
{\cal O}_v^{-1}(e_{(1,2)})=\sqrt{t\over q}e_{(-1,2)}\nn\\
{\cal O}_v^{-1}(e_{(3,5)})={t\over q}e_{(-2,5)}\nn
\ee
which was expected from formula (\ref{MikiST}) (the integer $p_{g,g(\vec\alpha)}$ is equal to 1 in this case).

\item[3.]{\bf Reflection of spins.} At last, there is a symmetry peculiar for the representation: $e_{(n,m)}(q,t)=-e_{(n,-m)}(q^{-1},t^{-1})$. This symmetry implies that one has to use for the below half-plane the average (\ref{ave}) with the replaced $q,t\to q^{-1},t^{-1}$.

\framebox{\parbox{15cm}{Because of this symmetry and of the automorphism $e_{(n,m)}\to e_{(-n,m)}$, and since elements $e_{(n,0)}$ are explicitly described by (\ref{Fock1}), we could restrict ourselves with constructing elements $e_{(n,m)}$ with $n,m\ge 0$ only.}}
\end{itemize}

\subsubsection{Application: ray (p,r)}

Many of (p,r) rays can be constructed from the rays (1,r) using these automorphisms. For instance, ${\cal O}_v$ maps $e_{(1,2)}\to e_{(3,2)}$ and further
$e_{(3,2)}\to e_{(5,2)}$. Similarly, one obtains:
\be
e_{(1,1)}\to e_{(2,1)}\to e_{(3,1)}\to e_{(4,1)}\to e_{(5,1)}\to e_{(6,1)}\to e_{(7,1)}\nn\\
e_{(1,2)}\to e_{(3,2)}\to e_{(5,2)}\to e_{(7,2)}
e_{(1,3)}\to e_{(4,3)}\to e_{(7,3)}\nn\\
e_{(-1,3)}\to e_{(2,3)}\to e_{(5,3)}\nn\\
e_{(-1,4)}\to e_{(3,4)}\to e_{(7,4)}\nn\\
e_{(1,4)}\to e_{(5,4)}\nn\\
e_{(1,5)}\to e_{(6,5)}\nn\\
e_{(1,6)}\to e_{(7,6)}\nn
\ee
Hence, one can produce all $e_{(n,m)}$ with mutually coprime $n$ and $m$, $n>m-2$ up to $n=7$ for exception of ray (7,5). In this case, one applies the $ABC$-procedure and constructs ray (2,5) from the polynomials $A={\cal P}_2$ and $C={\cal P}_3$ associated with $e_{(1,2)}$ and $e_{(1,3)}$ correspondingly. Then, applying ${\cal O}_v$, one obtains ray (7,2).

Generally, having the polynomials ${\cal P}$ describing the elements $e_{(1,n)}$ (\ref{Fe1n}), one can construct the polynomials ${\bf P}$, ${\bf H}$ and ${\bf E}$ of this section following sec.\ref{334}. Indeed, one can just repeat the procedure applying it to the commutators of the corresponding polynomials. Note that the element $e_{(k,0)}$ is not associated with a polynomial since it does not have the forms (\ref{gcd1}) or (\ref{gcdk}). However, its commutator with any $e_{(n,m)}$ associated with a polynomial just multiplies this polynomial with ${1-t^{-k}\over k}\sum_i z_i^k$. Another possibility is to generate following sec.\ref{334} only the elements $E_k^{(n,a)}$ with $k>0$, which does not involve elements $e_{(k,0)}$. This generates the elements $e_{(n,m)}$ with $m>n$ lying above the diagonal of the quadrant. The elements lying under the diagonal are all then constructed with using the operator ${\cal O}_v$.

\section{Other representations}

$N$-body and Fock are the simplest of a vast variety of DIM and Yangian representations.
The next examples are the matrix and MacMahon representations.

The $N$-body representation can be considered as a reduction of a matrix representation to the eigenvalue one,
but it is unclear whether such lifting exists at all for the Yangian and DIM algebras, at least in the naive form.
The main difficulty is to get rid of permutation operator in the Dunkl operators, which can be easily done for the $W_{1+\infty}$ algebra \cite{MMMP1}, but is far more problematic for non-Lie algebras.
The problem is well known in matrix model theory: though superintegrable eigenvalue models are always available \cite{SIrev,SIU,MMNek},
their realization in terms of full-fledged matrix integrals is straightforward only for the Schur polynomials case,
and not for the Jack \cite{OKJ,MMP}, Uglov \cite{MiakMish,GMT}, Macdonald \cite{Max,MPSh,MMell,Ch3,LWYZ} and elliptic Macdonald function \cite{MMell,MMell2} cases, i.e. disappears already under the $\beta$-deformation \cite{beta1,beta2,beta3,beta4}.

In the MacMahon representation, the ordinary partitions enumerating the basis in the Fock representation space are substituted by the plane partitions, and the ordinary Young diagrams by their 3-dimensional counterparts.
We do not consider these examples in the present text, for partial consideration
of their various aspects see \cite{Fei,Fei1,Proc,Zenkevich,Galakhov}.

Another problem is enumerating representations of a given kind, which would be a counterpart of
the classification of finite-dimensional representations of the Lie algebras $A_n$ by the Young diagrams.
This is also a question to be analyzed elsewhere.

\section{$q,t$ matrix (eigenvalue) models}

\subsection{Generating functions of the Hamiltonians}

There are various generating functions of commutative subalgebras in the elliptic Hall algebra. For instance, a natural one is
\be
T_{\vec\gamma}(z)=\exp\left(-\sum_n e_{n\vec\gamma}{z^n\kappa_n\over n}\right)=1+\sum_k\mathfrak{h}_{k\vec\gamma}z^k\nn
\ee
Another natural generating function is
\be
\mathfrak{T}_{\vec\gamma}(z)=\exp\left(-\sum_n e_{n\vec\gamma}{(-z)^n\over n}\right)\nn
\ee
and $T_{\vec\gamma}(z)$ is a product of six $\mathfrak{T}_{\vec\gamma}(z)$ at points $(t/q,t^{-1},q,-q/t,-t,-q^{-1})$.

We used $\mathfrak{T}_{\vec\gamma}(z)$ in constructing a $(q,t)$-extension \cite{Ch3} of the WLZZ matrix models \cite{China1,China2, Ch1,Ch2}. This generating function conjecturally satisfies the pentagon identity \cite{Zenk}
\be\label{Pent1}
\mathfrak{T}_{(0,1)}(v)\mathfrak{T}_{(1,0)}(u)=\mathfrak{T}_{(1,0)}(u)\mathfrak{T}_{(1,1)}(uv)\mathfrak{T}_{(0,1)}(v)
\ee
In fact, one can now generate a lot of similar pentagon identities just making an $SL(2,\mathbb{Z})$ transform of (\ref{Pent1}). In particular,
\be\label{Pent2}
\mathfrak{T}_{(-1,0)}(v)\mathfrak{T}_{(0,1)}(u)=\mathfrak{T}_{(0,1)}(u)\mathfrak{T}_{(-1,1)}(uv)\mathfrak{T}_{(-1,0)}(v)
\ee

Note also that the generating function of the trigonometric Ruijsenaars Hamiltonians in the $N$-body representation is given in these terms as (see (\ref{De}))
\be
{\mathfrak D}(z)=\mathfrak{T}_{(0,1)}(q^{1\over 2}z)\cdot\mathfrak{T}_{(0,1)}^{-1}(q^{-{1\over 2}}z)\nn
\ee
When constructing the matrix model partition functions, we typically deal with the generating functions of the form
\be\label{U}
U_{\vec\gamma}\{\bar p_k\}=\exp\left(\sum_n e_{n\vec\gamma}{\bar p_n\over n}\right)
\ee
which means that
\be
U_{\vec\gamma}\{\bar p_k\}=\prod_i\mathfrak{T}_{\vec\gamma}(z_i)\nn
\ee
with the Miwa parameterized variables $p_k=-\sum_i (-z_i)^k$. The partition function of matrix model is a vector given by action of  $U_{\vec\gamma}\{\bar p_k\}$ on a special vector in the representation.

\subsection{Operator $\hat{\cal O}(u)$ and the pentagon identity\label{pent}}

Let us note that the operator $\hat{\cal O}(u)$ (\ref{O}) is nothing but ${\cal T}_{(0,1)}(-u)$ entering the pentagon identity (\ref{Pent1}). First of all, it explicitly gives the result of action of $\hat{\cal O}(u)$ that we described in sec.\ref{cones}. Indeed, an evident corollary of the Pentagon identity is that the generating function of the Hamiltonians (\ref{OH}) is equal to
\be
\exp\left(-\sum_n\hat{\mathfrak{H}}_{(n,n)}{v^n\over n}\right)=\hat{\cal O}(u)\cdot{\cal T}_{(1,0)}(-v)\cdot\hat{\cal{O}}(u)^{-1}
={\cal T}_{(1,0)}(-v)\cdot{\cal T}_{(1,1)}(uv)\nn
\ee
or, more explicitly,
\be\label{PentO}
\exp\left(-\sum_n\hat{\mathfrak{H}}_{(n,n)}{v^n\over n}\right)=\hat{\cal O}(u)\cdot
\exp\left(-\sum_ne_{(n,0)}{v^n\over n}\right)\cdot\hat{\cal{O}}(u)^{-1}
=\exp\left(-\sum_ne_{(n,0)}{v^n\over n}\right)\exp\left(-\sum_ne_{(n,n)}{(-uv)^n\over n}\right)\nn\\
\ee
One can read off the action of $\hat{\cal O}(u)$ on concrete $e_{(n,0)}$ expanding this formula in the $v$-series. Moreover, one can deal with more generic elements making a proper $SL(2,\mathbb{Z})$ transform of this formula. For instance,
\be
\hat {\cal O}_h\cdot e_{(n,0)}\cdot\hat{\cal O}_h^{-1}=e_{(n,n)}\nn
\ee
and $\hat{\cal O}(u)$ commutes with ${\cal O}_h$ since $\hat{\cal O}(u)$ is made of elements of the vertical ray (0,1), which are invariant under the action of the automorphism ${\cal O}_h$. Hence, one gets from (\ref{PentO})
\be\label{81}
\hat{\cal O}(u)\cdot\exp\left(-\sum_ne_{(n,n)}{v^n\over n}\right)\cdot\hat{\cal{O}}(u)^{-1}
=\exp\left(-\sum_ne_{(n,n)}{v^n\over n}\right)\exp\left(-\sum_ne_{(n,2n)}{(-uv)^n\over n}\right)
\ee
etc.

Similarly, one can rewrite (\ref{Pent2}) in the form
\be
\exp\left(-\sum_n\hat{\mathfrak{H}}_{(-n,n)}{v^n\over n}\right)=\hat{\cal O}(u)^{-1}\cdot{\cal T}_{(-1,0)}(-v)\cdot\hat{\cal{O}}(u)
=\cdot{\cal T}_{(-1,1)}(uv){\cal T}_{(-1,0)}(-v)\nn
\ee
or, more explicitly,
\be
\exp\left(-\sum_n\hat{\mathfrak{H}}_{(-n,n)}{v^n\over n}\right)=\hat{\cal O}(u)^{-1}\cdot
\exp\left(-\sum_ne_{(-n,0)}{v^n\over n}\right)\cdot\hat{\cal{O}}(u)
=\exp\left(-\sum_ne_{(-n,n)}{(-uv)^n\over n}\right)\exp\left(-\sum_ne_{(-n,0)}{v^n\over n}\right)\nn
\ee
and then generate more formulas similar to (\ref{81}) using the Miki automorphism ${\cal O}_h$.

\subsection{Automorphisms in Macdonald polynomial basis\label{Oh}}

\subsubsection{Operator $\hat{\cal O}(u)$}

There are many important operators that are diagonal in the basis of the Macdonald polynomials. For instance, the operator $\hat{\cal O}(u)$ considered in sec.\ref{cones}, (\ref{O}). Here we use slightly different normalization of this operator following \cite{Ch3}:
\begin{equation}
\hat{\cal{O}}(u)={\cal N}_{q,t}(u)(1-q)^{-\hat{E}}\mathfrak{T}_{(0,1)}(-u)=
{\cal N}_{q,t}(u)(1-q)^{-\hat{E}}\exp\left(-\sum_n e_{(0,n)}{u^n\over n}\right)\nn
\end{equation}
where $\hat{E}$ is a grading operator, which is not an element of the elliptic Hall algebra, but in concrete representations it is equal to $\hat{E}=\sum_{n\ge1 }np_n{\partial\over\partial p_n}$ (Fock representation), or to $\hat{E}=\sum_{i=1}^Nx_i{\p\over\p x_i}$ ($N$-body representation). It is introduced in order to have a better Yangian limit $t=q^\beta$, $q\to 1$.  The normalization constant here is
\be
{\cal N}_{q,t}(u)=\prod_{m,k>0}(1-uq^mt^{-k})\nn
\ee
The operator $\hat{\cal O}(u)$ simply acts on the Macdonald polynomials. Indeed, as we explained in sec.\ref{5.1.2}, the Macdonald polynomials are the eigenfunctions of the Hamiltonians from the vertical ray (0,1):
\be\label{e0nef-alt}
e_{(0,n)}M_\lambda=\left(-{1\over (1-q^n)(1-t^{-n})}+\sum_{(i,j)\in\lambda}q^{n(j-1)}t^{n(1-i)}\right)M_\lambda:=
\Lambda_\lambda^{(n)} M_\lambda
\ee
Hence,
\be\label{evOH}
\boxed{
\begin{array}{rcl}
\hat{\cal O}(u)M_\lambda\{p_k\}&=&{\cal N}_{q,t}(u)(1-q)^{-\hat{E}}\exp\left(-\sum_n e_{(0,n)}{(-t^N)^{n}\over n}\right)M_\lambda\{p_k\}=
\xi_\lambda(u) M_\lambda\{p_k\}\cr
&&\cr
\xi_\lambda(u)&=&\prod_{i,j\in\lambda}{1-uq^{j-1}t^{1-i}\over 1-q}=(1-q)^{-|\lambda|}{M_\lambda\{p^*_k(u)\}\over M_\lambda\{p^*_k(0)\}}
\end{array}
}
\ee
where
\be
p^*_k(u):={1-u^k\over 1-t^k}\nn
\ee

\subsubsection{Operator $\hat{\cal O}_h$\label{OhF}}

Another operator diagonal in the Macdonald polynomial basis is $\hat{\cal O}_h$. Indeed, all elements of the vertical ray (0,1) are invariant under the action of the automorphism $\hat{\cal O}_h$:
\be
\hat {\cal O}_h^{-1}\cdot e_{(0,n)}\cdot\hat{\cal O}_h=e_{(0,n)}\nn
\ee
Hence, the Macdonald polynomials, being the eigenfunctions of $e_{(0,n)}$ are also the eigenfunctions of $\hat{\cal O}_h$ with some eigenvalues $f_\lambda$. Now we obtain these eigenvalues. To this end, note that
\be
\hat {\cal O}_h^{-1}\cdot e_{(1,1)}\cdot\hat{\cal O}_h=e_{(1,0)}\nn
\ee
i.e.
\be
f_\lambda e_{(1,1)}M_\lambda\{p_k\}=\hat{\cal O}_h\cdot e_{(1,0)}M_\lambda\{p_k\}=
{1\over q^{1/2}-q^{-1/2}}\hat{\cal O}_hp_1M_\lambda\{p_k\}=\nn\\
={1\over q^{1/2}-q^{-1/2}}\hat{\cal O}_h
\sum_i C_{\lambda,i} M_{\lambda+1_i}\{p_k\}={1\over q^{1/2}-q^{-1/2}}
\sum_i C_{\lambda,i} f_{\lambda+1_i}M_{\lambda+1_i}\{p_k\}\nn
\ee
where we used the Pieri formula \cite{Mac}
\be
p_1M_\lambda\{p_k\}=\sum_i C_{\lambda,i} M_{\lambda+1_i}\{p_k\}\nn
\ee
with
\be
C_{\lambda,i}=\prod_{j=1}^{i-1}{(1-q^{\lambda_i-\lambda_j+1}t^{j-i-1})(1-q^{\lambda_i-\lambda_j}t^{j-i+1})\over
(1-q^{\lambda_i-\lambda_j+1}t^{j-i})(1-q^{\lambda_i-\lambda_j}t^{j-i})}\nn
\ee
$\lambda+1_i$ here denotes the diagram obtained from the diagram $\lambda$ by adding one box in line $i$.  If $\lambda+1_i$ is not a
Young diagram, the coefficient $C_i$ vanishes automatically.

It remains to notice that (see, e.g., \cite[Eqs.(2.11),(2.16)]{AFS}\footnote{Notice a slightly different normalization used in this paper.})
\be\label{e11F}
e_{(1,1)}M_\lambda\{p_k\}={1\over q^{1/2}-q^{-1/2}}\sum_i  q^{\lambda_i}t^{1-i} C_{\lambda,i}M_{\lambda+1_i}\{p_k\}
\ee
in order to obtain that
\be
{f_{\lambda+1_i}\over f_\lambda}=q^{\lambda_i}t^{1-i}\nn
\ee
and, finally,
\be\label{OMF}
\boxed{
\begin{array}{rcl}
\hat{\cal{O}}_hM_\lambda\{p_k\}&=&f_\lambda M_\lambda\{p_k\}\cr
&&\cr
f_\lambda&=&q^{\nu'_\lambda/2}t^{-\nu_\lambda/2}
\end{array}
}
\ee
where $\nu_\lambda:=2\sum_i(i-1)\lambda_i$, $\nu'_\lambda:=\nu_{\lambda^\vee}$. The quantity $f_\lambda$ is the standard Taki's framing factor \cite{Taki}.

In the $N$-body representation, one can repeat the same consideration and obtain a similar formula just making the replace of variables in the Macdonald polynomials $p_k=\sum_i^Nx_i^k$: since
\be
e_{(1,1)}M_\lambda(x_i)={qt\over q-1}\sum_i  {q^{\lambda_i}\over t^{i-1}} C_{\lambda,i}M_{\lambda+1_i}(x_i)\nn
\ee
one obtains
\be\label{OMN}
\hat{\cal{O}}_hM_\lambda(x_i)=q^{1/2}tf_\lambda M_\lambda(x_i)
\ee
However, one should understand that the Macdonald polynomials do not form a basis in the $N$-body representation.

\subsubsection{Operator $\hat{\cal O}_v$}

Note that there is no a similar construction for the operator $\hat{\cal O}_v$. Indeed, it is not diagonal in the Macdonald basis. At the same time, it satisfies the equations
\be
e_{(1,0)}\cdot\hat{\cal O}_v\ M_\lambda\{p_k\}=\hat{\cal O}_v\cdot e_{(1,0)}\ M_\lambda\{p_k\}\nn
\ee
and
\be
e_{(0,k)}\cdot\hat{\cal O}_v\ M_\lambda\{p_k\}=\hat{\cal O}_v\cdot e_{(k,k)}\ M_\lambda\{p_k\}\nn
\ee
One can immediately check that these equations do not admit a non-zero solution in the form
\be
\hat{\cal O}_v\ M_\lambda\{p_k\}=\sum_\mu{\cal O}_v^{\lambda\mu}M_\mu\{p_k\}\nn
\ee
The reason is simple: the first of these equations implies that $\hat{\cal O}_v$ is a function of $p_k$ only. In order to realize the map (\ref{Mikiv}) with the adjoint action of the operator $\hat{\cal O}_v$, one has to choose
\be
\hat{\cal O}_v=\exp\left(\sum_kc_kp_k\right)\nn
\ee
where $c_k$ are some coefficients, and to solve the condition
\be
\exp\left(-\sum_k(1-q^k)z^{-k}c_k\right)=q^{1/2}z^{-1}\nn
\ee
which is singular. This should not come as a surprise, since the operators $\hat{\cal O}_v$ in formulas (\ref{O1}) and  (\ref{NbOv}) are also singular, and should be treated as hints rather than real operators.

\subsection{Matrix model partition functions}

\subsubsection{$W$ representation in matrix models}

As was explained in a series of papers, \cite{Ch1,Ch2,MMMP1} any commutative family of $W_{1+\infty}$ algebra in the Fock representation naturally generates a $W$ representation of partition function of a matrix model. It is definitely no longer a model with integrals over matrices, the integrals (if such a representation exists at all) are deformations of eigenvalue integrals in matrix models. However, we use the traditional term ``matrix model" in this case.

If the commutative family consists of elements $\hat H_k$ of grading $k$, the $W$ representation is given by
\be
\left\{\begin{array}{ll}
Z_+(p_k;\bar p_k,N)=\exp\left(\sum_k{\hat H_k\bar p_k\over k}\right)\cdot 1&\hbox{at}\ k>0\cr
&\cr
Z_-(p_k;\bar p_k,g_k,N)=\exp\left(\sum_k{\hat H_k\bar p_k\over k}\right)\cdot \exp\left(\sum_k{g_kp_k\over k}\right)
&\hbox{at}\ k<0
\end{array}
\right.\nn
\ee
where $\bar p_k$ and $p_k$ are arbitrary parameters. In particular, the WLZZ models \cite{China1,China2} get into this class of matrix models. Note that $Z_+$ is a hypergeometric type $\tau$-function \cite{GKM2,OS} of the Toda lattice hierarchy \cite{Ch1,Ch2}, while $Z_-$ is a skew hypergeometric $\tau$-function \cite{Ch1,Ch2}. Hence, these partition functions are naturally expanded into the sums over partitions of bilinear combinations of the Schur and skew Schur functions.

\subsubsection{Partition functions for the family (\ref{OH})}

We can now consider a $q,t$-deformation of these partition functions. One such a deformation was considered in \cite{Ch3}: using the commutative family of Hamiltonians $\hat{\mathfrak{H}}_{(k,km)}$ (\ref{OH}), one can construct partition functions of the $q,t$-deformed matrix models\footnote{In \cite{Ch3}, there was used the operator $\prod_{i=1}^m\hat {\cal O}(u_i)$ depending on $m$ arbitrary parameters $u_i$ in order to generate partition functions. For the sake of simplicity, we put here all these parameters equal to a single $u$.} (see particular cases in \cite{Max,MPSh,MMell}) using the Cauchy identity
\be
\exp\left(\sum_k{1-t^k\over 1-q^k}{p_k\bar p_k\over k}\right)=\sum_\lambda{M_\lambda\{p_k\}M_\lambda\{\bar p_k\}\over ||\lambda||^2}\nn
\ee
and formula (\ref{evOH}):
\be\label{Z+O}
Z_+^{(m)}(p_k;\bar p_k,u;q,t)&=&\exp\left(\sum_k(t^k-1){\bar p_k\over k}\hat {\mathfrak{H}}_{(k,km)}\right)\cdot 1=
\hat{\cal O}(u)^{m}\cdot \exp\left(\sum_k(t^k-1){e_{(k,0)}\bar p_k\over k}\right)\cdot\hat{\cal{O}}(u)^{-m}\cdot 1=\nn\\
&=&\hat{\cal O}(u)^{m}\cdot \exp\left(\sum_k{1-t^k\over 1-q^k}{q^{k/2}p_k\bar p_k\over k}\right)\cdot\hat{\cal{O}}(u)^{-m}\cdot 1=
\sum_\lambda\hat{\cal O}(u)^{m}\cdot {q^{|\lambda|/2}M_\lambda\{p_k\}M_\lambda\{\bar p_k\}\over ||M_\lambda||}=\nn\\
&=&\boxed{\sum_\lambda q^{|\lambda|/2}\xi_\lambda(u)^m{M_\lambda\{p_k\}M_\lambda\{\bar p_k\}\over ||\lambda||^2}}
\ee
and
\be\label{Z-O}
Z_-^{(m)}(p_k;\bar p_k,g_k,u;q,t)=\exp\left(\sum_k(t^k-1){\bar p_k\over k}\hat {\mathfrak{H}}_{(-k,km)}\right)
\cdot \exp\left(\sum_k{1-t^k\over 1-q^k}{g_kp_k\over k}\right)=\nn\\
=\hat{\cal O}(u)^{-m}\cdot
\exp\left(-\sum_k(t^k-1){\bar p_k\over k}e_{(-k,0)}\right)\cdot\hat{\cal O}(u)^{m}\cdot\sum_\lambda{M_\lambda\{p_k\}M_\lambda\{g_k\}\over ||\lambda||^2}=\nn\\
=\hat{\cal O}(u)^{-m}\cdot
\exp\left(\sum_kt^{k/2}\bar p_k{\p\over\p p_k}\right)\cdot\sum_\lambda\xi_\lambda(u)^m{M_\lambda\{p_k\}M_\lambda\{g_k\}\over ||\lambda||^2}=\hat{\cal O}(u)^{-m}\cdot\sum_\lambda\xi_\lambda(u)^m{M_\lambda\{p_k+t^{k/2}\bar p_k\}M_\lambda\{g_k\}\over ||\lambda||^2}=\nn\\
=\hat{\cal O}(u)^{-m}\cdot\sum_\lambda\xi_\lambda(u)^m{M_{\lambda/\mu}\{t^{k/2}\bar p_k\}M_\mu\{p_k\}
M_\lambda\{g_k\}\over ||\lambda||^2}=\boxed{\sum_\lambda t^{(|\lambda|-|\mu|)/2}\left({\xi_\lambda(u)\over\xi_\mu(u)}\right)^m{M_{\lambda/\mu}\{\bar p_k\}M_\mu\{p_k\}
M_\lambda\{g_k\}\over ||\lambda||^2}}\nn\\
\ee
These formulas coincide with those in \cite{Ch3}\footnote{We use slightly different normalized $\bar p_k$.}.

\subsubsection{Partition functions for ($\pm 1,r$) rays}

In complete analogy with the previous paragraph, one can generate the partition function by ($\pm 1,p$) rays, with the operator $\hat{\cal O}_h$ replacing $\hat{\cal O}$. In this case, the $W$ representations are generated by the operators $U_{(\pm 1,p)}$ (\ref{U}), and one obtains:
\be\label{Z+}
Z_+^{(1,r)}(p_k;\bar p_k;q,t)=U_{(1,r)}\{(t^k-1)\bar p_k\}\cdot 1=\exp\left(\sum_k(t^k-1){\bar p_k\over k}e_{(k,kr)}\right)\cdot 1=\nn\\
=\hat{\cal O}_h(u)^{r}\cdot \exp\left(\sum_k(t^k-1){e_{(k,0)}\bar p_k\over k}\right)\cdot\hat{\cal{O}}_h(u)^{-r}\cdot 1=
\boxed{
\sum_\lambda q^{|\lambda|/2}f_\lambda^r{M_\lambda\{p_k\}M_\lambda\{\bar p_k\}\over ||\lambda||^2}}\nn\\
\ee
and
\be\label{Z-}
Z_-^{(-1,r)}(p_k;\bar p_k,g_k;q,t)=U_{(-1,r)}\{(t^k-1)\bar p_k\}\cdot\exp\left(\sum_k{1-t^k\over 1-q^k}{g_kp_k\over k}\right)=
\nn\\
=\exp\left(\sum_k(t^k-1){\bar p_k\over k}e_{(-k,kr)}\right)
\cdot \exp\left(\sum_k{1-t^k\over 1-q^k}{g_kp_k\over k}\right)=\nn\\
=\hat{\cal O}_h(u)^{-r}\cdot
\exp\left(-\sum_k(t^k-1){\bar p_k\over k}e_{(-k,0)}\right)\cdot\hat{\cal O}_h(u)^{r}\cdot\sum_\lambda{M_\lambda\{p_k\}M_\lambda\{g_k\}\over ||\lambda||^2}=\nn\\
=\boxed{\sum_\lambda t^{(|\lambda|-|\mu|)/2}\left({f_\lambda\over f_\mu}\right)^r{M_{\lambda/\mu}\{\bar p_k\}M_\mu\{p_k\}
M_\lambda\{g_k\}\over ||\lambda||^2}}
\ee

\section{Common eigenfunctions of commuting Hamiltonians}

Let us return to the Fock representation and construct eigenfunctions of the Hamiltonians much similar to what we did in \cite[sec.12]{MMMP1}.

\subsection{Eigenfunctions for ray (-1,p)}

The key role is played by the automorphism operator $\hat{\cal O}_h$, since it simply acts on the Macdonald polynomials.
With this operator, one can immediately construct the eigenfunctions for the rays (-1,m): it is sufficient to note that
\be
{\cal O}_h^k(e_{(-n,0)})=\hat{\cal O}_h^{-k}\cdot e_{(-n,0)}\cdot\hat{\cal O}_h^{k}=e_{(-n,nk)}\nn
\ee
i.e.
\be
e_{(-n,nk)}=\hat{\cal O}_h^{-k}\cdot {n\over t^{n/2}-t^{-n/2}}{\p\over\p p_n}\cdot\hat{\cal O}_h^{k}\nn
\ee
Now, using this formula, formula (\ref{OMF}) and
\be
n{\p\over\p p_n}\exp\left(\sum_{k=1}(-1)^{k+1}{p_k\bar p_k\over k}\right)=(-1)^{n+1}\bar p_n\exp\left(\sum_{k=1}(-1)^{k+1}{p_k\bar p_k\over k}\right)\nn
\ee
along with the Cauchy identity,
\be
\exp\left(\sum_{n=1}(-1)^{n+1}{p_n\bar p_n\over n}\right)=\sum_\lambda M_\lambda\{p_k\}\overline{M}_{\lambda^\vee}\{\bar p_k\}\nn
\ee
one immediately obtains
\be\label{hr0}
\boxed{
\begin{array}{rcl}
e_{(-k,mk)}\{p\} \Psi^{(m)}\{p,\bar p\}& =&{(-1)^{k+1}\bar p_k\over t^{k/2}-t^{-k/2}} \Psi^{(m)}\{p,\bar p\}\cr
&&\cr
\Psi^{(m)}\{p,\bar p\}& =& \sum_\lambda f_\lambda^{-m}M_\lambda\{p\}\overline{M}_{\lambda^\vee}\{\bar p\}
\end{array}
}
\ee
where $\bar p_k$ are just arbitrary parameters.

\subsection{Eigenfunctions for $\mathfrak{H}_{(k,mk)}$}

One can construct the eigenfunctions for the $\mathfrak{H}_{(-k,mk)}$ family\footnote{Again, one can use the product $\prod_{i=1}^m\hat{\cal O}(u_i)$ with $m$ different $u_i$.} (\ref{OH}),
\be\label{OHm}
\mathfrak{H}_{(-k,mk)}=\hat{\cal O}(u)^{-m}\cdot e_{(-k,0)}\cdot\hat{\cal O}(u)^{m}
\ee
in a similar way using (\ref{evOH}):
\be\label{hr1}
\boxed{
\begin{array}{rcl}
\mathfrak{H}_{(-k,mk)}\{p\} \Psi_H^{(m)}\{p,\bar p\}& =&{(-1)^{k+1}\bar p_k\over t^{k/2}-t^{-k/2}} \Psi_H^{(m)}\{p,\bar p\}\cr
&&\cr
\Psi^{(m)}_H\{p,\bar p\}& =& \sum_\lambda \xi_\lambda(u)^{-m}M_\lambda\{p\}\overline{M}_{\lambda^\vee}\{\bar p\}
\end{array}
}
\ee

Note that now one can consider two natural generalizations of {\bf the Harish-Chandra-Itzykson-Zuber integral} \cite{MMS,MMMP1,MOP1,Mor}: one generalization,
\be\label{IZ1}
\boxed{
\hbox{IZ}_1:=\Psi^{(1)}\{p,\bar p\}=\sum_\lambda {M_\lambda\{p\}\overline{M}_{\lambda^\vee}\{\bar p\}\over f_\lambda}
=\sum_\lambda q^{-{1\over 2}\nu'_\lambda}t^{{1\over 2}\nu_\lambda}M_\lambda\{p\}\overline{M}_{\lambda^\vee}\{\bar p\}
}
\ee
is an eigenfunction of the $e_{(-1,1)}$ ray Hamiltonians: these are (\ref{HpR}) at $p=1$
after making the reflection of grading,
\be
e_{(-k,k)}=-t^{-k\over 2}{\kappa_1\over\kappa_k}\Big<{\cal E}_k(-z_i^{-1};-\kappa_1)\prod_{i=1}^kz_i^3\Big>_k\nn
\ee
Another generalization,
\be\label{IZ2}
\boxed{
\hbox{IZ}_2:=\Psi_O^{(1)}\{p,\bar p\}=\sum_\lambda {M_\lambda\{p\}\overline{M}_{\lambda^\vee}\{\bar p\}\over \xi_\lambda(u)}=
\sum_\lambda (1-q)^{|\lambda|}{M_\lambda\{p^*_k(0)\}\over M_\lambda\{p^*_k(u)\}}M_\lambda\{p\}\overline{M}_{\lambda^\vee}\{\bar p\}
}
\ee
has a simpler limit to the $\beta$-deformed \cite{Brezin,Ey,MMS,MOP1,MOP2} and further to usual \cite{HC,IZ} Harish-Chandra-Itzykson-Zuber integrals, but instead satisfies more involved eigenvalue equations with the Hamiltonians $\mathfrak{H}_{(-k,k)}$ (\ref{OHm}). For instance, the first Hamiltonian of this series (the only one which is relatively simple, all other are very involved) is
\be
\mathfrak{H}_{(-1,1)}=(1-q)^{-1}\Big(e_{(-1,0)}-ue_{(-1,1)}\Big)=
{1\over (1-q)( t^{1/2}-t^{-1/2})}{\p\over\p p_1}
+u{t^{-1/2}\over (1-q)^2(1-t^{-1})}\oint_0dzV(z)\nn
\ee

\subsection{Eigenfunctions for ray (1,p)}

For positive grading Hamiltonians, one similarly obtains that
\be\label{r2}
e_{(k,mk)}=\hat {\cal O}_h(u)^{m}\cdot {1\over q^{k/2}-q^{-k/2}}p_k\cdot\hat {\cal O}_h(u)^{-m}
\ee
Hence, from
\be
p_n\exp\left(\sum_{k=1}(-1)^{k+1}{p_k\bar p_k\over k}\right)=n{\p\over\p \bar p_n}\exp\left(\sum_{k=1}(-1)^{k+1}{p_k\bar p_k\over k}\right)\nn
\ee
it immediately follows that formula \eqref{hr0} is dual to a similar relation for the positive ray:
\begin{equation}\label{dH}
	e_{(k,mk)}\Phi^{(m)} \left\{ p, \bar{p} \right\} ={(-1)^{k+1}k\over q^{k/2}-q^{-k/2}}\dfrac{\partial}{\partial \bar{p}_k} \Phi^{(m)} \left\{ p, \bar{p} \right\}
\end{equation}
with
\begin{equation}\label{def}
	\Phi^{(m)} \left\{p,\bar{p} \right\} = \sum_\lambda f_\lambda^mM_\lambda\{p\}\overline{M}_{\lambda^\vee}\{\bar p\}
\end{equation}
which is nothing but formula,
\be
\Phi^{(m)} \left\{p,\bar{p} \right\}=\exp\left( \sum_{k} \dfrac{\bar{p}_k e_{(k,mk)}}{k} \right) \cdot 1\nn
\ee
In order to generate the eigenvalue equation instead of (\ref{dH}), one can make the reflection grading.

\section{Yangian and further $W_{1+\infty}$ limits}

\subsection{Yangian limit}

The main advantage of the DIM algebra as compared with the affine Yangian or even with $W_{1+\infty}$ algebra is that commutators preserve gradings and spins:  $[e_{(n_1,m_1)},e_{(n_2,m_2)}] \sim e_{(n_1+n_2,m_1+m_2)}$ for admissible pairs in the DIM algebra instead of $[W_{(n_1,m_1)},W_{(n_2,m_2)}] \sim \sum_{k=2}W_{(n_1+n_2,m_1+m_2-k)}$ in the Yangian/$W_{1+\infty}$ algebra.
This is already clear from the one-body representation: in the DIM algebra, the generators are given by $e_{(n,m)}\sim x^nq^{m\hat D}$  so that $[ x^{n_1}q^{m_1\hat D}, x^{n_2}q^{m_2\hat D}]\sim  x^{n_1+n_2}q^{(m_1+m_2)\hat D}$, while, in the $W_{1+\infty}$ algebra, the generators are given by $\mathfrak{e}_{(n,m+1)}\sim x^n\hat D^m$ so that
$[ x^{n_1}\hat D^{m_1}, x^{n_2}\hat D^{m_2}]\sim  x^{n_1+n_2}\Big((\hat D+n_2)^{m_1}\hat D^{m_2}
-\hat D^{m_1}(\hat D+n_1)^{m_2}\Big)$ so that there is no term $x^{n_1+n_2}\hat D^{m_1+m_2}$ in the commutator.

In the Yangian limit, every DIM generator from the upper half-plane maps to an infinite sum of generators of different spins (but with the same grading). It results in two properties (see Figs.\ref{YWfig}, \ref{Dimfig}): the Yangian generators are enumerated by points of the integer upper half-lattice (instead of the whole integer lattice), and commutators do not respect spins. As a result, the DIM algebra turns out to have simpler properties. In particular, there are natural $SL(2,\mathbb{Z})$ automorphisms that act on this algebra.

In fact, the limit of the generating elements $e_{(\pm 1,m)}$ and $e_{(0,m)}$ to the affine Yangian algebra generated by $\{e_i,f_i,\psi_i$ \cite{Ts,Proc,MMMP2} is defined in the following way \cite{Ts,Nawata}:
\be\label{Yanl}
q_i&=&\exp\Big(\hbar \epsilon_i\Big),\ \ \ \ \ \ \ c_2=0,\ \ \ \ \ \ \ c_1=\sigma_3\psi_0\nn\\
e_{(1,m)}&=&{1\over\hbar}\sum_{j=0}{(m\hbar)^j\over j!}e_j,\ \ \ \ \ \ \
e_{(-1,m)}={1\over\hbar}\sum_{j=0}{(m\hbar)^j\over j!}f_j,\ \ \ \ \ \ \
e_{(0,m)}=-{1\over\sigma_3}\sum_{j=0}{(m\hbar)^{j-2}\over j!}k_j
\ee
where $m\in\mathbb{Z}_{>0}$, $\sigma_3=\epsilon_1\epsilon_2\epsilon_3$, and
\be
\sum_{j=0}k_jz^{j+1}=\log\Big(1+\sigma_3\sum_{i=0}\psi_iz^{i+1}\Big)\nn
\ee
One can see that the DIM element of an arbitrary spin maps to the sum of Yangian elements of all spins starting from the zero one. Hence, one of the problems with the limit is that, in order to generate Hamiltonians of higher spins in the Yangian limit, one has to start with linear combinations of DIM Hamiltonians of different spins. Such combinations called cones are not simple to construct in the DIM algebra. A technical trick that allows one to deal with them is to introduce a special operator $\hat{\cal O}(u)$, (\ref{O}). This operator depends on an arbitrary parameter $u$, which, in the Yangian or $W_{1+\infty}$ limit is a free parameter of the construction, sometimes associated also with $N$ \cite{MMMP1}, and it is completely absent in the DIM algebra.

The result of action  on the algebra elements of the operator $\hat{\cal O}(u)$ giving an inner automorphism of the DIM algebra is rather involved (see secs.\ref{cones},\ref{pent}), however, action of this automorphism (generally, a few times and with different parameters $u_i$) provides an immediate Yangian limit. This illustrates once more a non-triviality of the limiting transition from the DIM algebra to the affine Yangian algebra. We will return to this issue elsewhere.

The limit to the $W_{1+\infty}$ algebra is much simpler, since this algebra is a Lie algebra, and the limit to it can be taken in terms of the $qW_{1+\infty}$ algebra, which provides a proper Lie algebra limit of the DIM algebra. In the meanwhile, the cones are easily realized in the $qW_{1+\infty}$ algebra: the degree of generators in the DIM algebra become just the algebra elements in the $qW_{1+\infty}$ case, i.e. it is sufficient to deal with the Lie algebra, not with its universal enveloping algebra.

Below we provide a few examples of the limit in $N$-body and Fock representations in order to illustrate a phenomenon. We discuss these issues in detail elsewhere.

\subsection{$N$-body representation}

\subsubsection{Limit to Yangian algebra}

      Whether the word construction of sec.\ref{Chpq} in the $N$-body representation survives in the peculiar
      Yangian limit, is a very good question that we postpone for future research,
      listing here only immediate promises and complications.

      First of all, since the left hand (Yangian) side of
      \eqref{eq:ray-1r-hamilts} is quite parallel to the right hand (DIM) side,
      one may hope that the $\mathcal{O}_h$ operation \eqref{eq:oh-oper-words} literally survives.

      The challenge is, however, to obtain the $\mathcal{O}_v$ operation (which,
      on the DIM side, ironically, looks much simpler)
      and to associate, instead of $C^{(\alpha)}_i$, some operators
      to the numbers $\alpha >1$ in a word $W_{\vec\alpha}$.

      At first glance, the Yangian limit of $\mathcal{O}_v$
      is given by anticommutator with $x_i$
      \begin{align}
        \mathcal{O}_v \Big|_{\text{Yangian}} (\bullet) =
        \{x_i,\bullet\} := x_i \bullet + \bullet x_i\nn
      \end{align}
      Indeed, while the first Hamiltonian of ray (-1,1), $f_1 = H^{(-1,1)}_1$ is equal to the sum of Dunkl operators \cite{MMCal,MMMP1}
      \begin{align}
        f_1 = \sum_{i=1}^N \tilde{\mathcal{D}}_i\nn
      \end{align}
      while the whole commutative family is formed by
      \be
      H_k^{(-1,1)}=\sum_{i=1}^N \tilde{\mathcal{D}}_i^k\nn
      \ee
      It is commutative since
      \be
      \phantom{.}[ \tilde{\mathcal{D}}_i, \tilde{\mathcal{D}}_j]=0\nn
      \ee

      Moving further, the operator of zero grading and the same spin is
      \be
      \mathcal{O}_v \Big|_{\text{Yangian}} (f_1) =\psi_2= \sum_{i=1}^N \left(x_i \tilde{\mathcal{D}}_i
        + \tilde{\mathcal{D}}_i x_i
        \right)\nn
      \ee
      while the cut-and-join operator $\psi_3$ ($=6\hat W_0$ in the notation of \cite{MMMP1}), is, modulo zero-modes, a sum of squares:
      \begin{align}
        \psi_3 = \frac{3}{4} \sum_{i=1}^N \left(x_i \tilde{\mathcal{D}}_i
        + \tilde{\mathcal{D}}_i x_i
        \right)^2
        - \frac{3}{4} N \left(1 + \beta(N-1)\right)^2\nn
      \end{align}
      and commutes with $\psi_2$, and higher (completed cycles) cut-and-join operators forming the whole commutative family are similarly made \cite{DMP} from power sums of root operators
      $A_i = x_i \tilde{\mathcal{D}}_i + \tilde{\mathcal{D}}_i x_i $. The family is commutative since
      \be
      \phantom{.}[A_i,A_j]=0\nn
      \ee

      \noindent The next commutative family is generated by
            \begin{align}
        e_1 = \sum_{i=1}^N \{x_i,\{x_i, \tilde{\mathcal{D}}_i\}\}\nn
      \end{align}
      and consists of Hamiltonians that do not have form \cite{MMMP1}
      \be
      H_k^{(1,1)} = \sum_{i=1}^N \Big(\{x_i,\{x_i, \tilde{\mathcal{D}}_i\}\}\Big)^k\nn
      \ee
since
      \be
      \phantom{.}[\{x_i,\{x_i, \tilde{\mathcal{D}}_i\}\},\{x_j,\{x_j, \tilde{\mathcal{D}}_j\}\}]\ne 0\nn
      \ee
Similarly, the generating element of grading $p$ is given by
      \begin{align}
        e^{(p)}_1 = \sum_{i=1}^N \underbrace{\{x_i,\{x_i,\ldots\{x_i,}_{p+1} \tilde{\mathcal{D}}_i\}\ldots\}\nn
      \end{align}
Again, in this case, in contrast with the DIM case, the operators in the summand do not
      commute with each other
      \begin{align}
        \underbrace{\{x_i,\{x_i,\ldots\{x_i,}_{p+1} \tilde{\mathcal{D}}_i\}\ldots\},
          \underbrace{\{x_j,\{x_j,\ldots\{x_j,}_{p+1} \tilde{\mathcal{D}}_j\}\ldots\}] \neq 0\nn
      \end{align}
      unless $p = 0$ and, hence, do not give rise to commutative families.

This means that either the Yangian limit of the automorphism ${\cal O}_v$ is more tricky, or that it does not act on the affine Yangian algebra transitively.

  \subsubsection{Limit to the Calogero type models}

As we already explained in sec.\ref{sec:new-integrable-systems}, ray (1,1) do not produce a new integrable system in the $N$-body representation of the DIM algebra, while, in the Yangian limit, an analogous ray gives rise to the rational Calogero model. The reason is, again, in the non-trivial Yangian limit: the action of operator $\hat {\cal O}$ makes a non-trivial integrable system. At the same time, since $\hat {\cal O}$ does not deform the vertical ray associated with the trigonometric Ruijsenaars system in the DIM case, its limit to the trigonometric Calogero-Sutherland system in the Yangian case is quite immediate.

\subsection{Fock representation}

We already discussed the behaviour of $e_{(1,2)}$ in the limits when $t\to 1$ and $q\to 1$ after formula (\ref{e12}). First of all, let us note that it is sufficient to consider only $e_{(n,m)}$ in the first quadrant since other three quadrants are related with the first one by symmetries of sec.\ref{532}. Now, one can check that the behaviour of $e_{(n,m)}$ with arbitrary $n\ne 0$ have the same asymptotic behaviour: when acting on the Macdonald polynomials, they are regular at $t\to 1$ and have a simple pole at $q=1$, which is consistent with formula (\ref{Yanl}). At $n=0$, i.e. $e_{(0,m)}$ has simple poles both at $q=1$ {\it and} at $t=1$. This is also consistent with the behaviour (\ref{Yanl}).

Note that, in principle, the elements of algebra $e_{(n,m)}$ admit consistent changing the normalization with non-singular factors $\alpha_{1,2}(q,t)$ so that $e_{(n,m)}\to \alpha_{1}(q,t)^n\alpha_{2}^m(q,t)e_{(n,m)}$. This does not change the commutation relations of the algebra (\ref{1}), (\ref{2}) and the asymptotic behaviour (\ref{Yanl}), but adds to the non-singular terms. This is the origin of emergency of additional parameters like $u$ and $N$ \cite{MMMP1,MMMP2} in the affine Yangian and $W_{1+\infty}$ algebra limits.

As we explained above, action of the inner automorphism ${\cal O}(u)$ makes the Yangian limit immediate. In the very clear way, it can be observed at an example of the eigenfunctions of the (-1,1) ray Hamiltonians and the Itzykson-Zuber formulas (\ref{IZ1}), (\ref{IZ2}). If one looks at the immediate Yangian limit of the Hamiltonians, one obtains from formula (\ref{IZ1}) in the leading order a trivial eigenfunction
\be
\hbox{IZ}_1:=\sum_\lambda q^{-{1\over 2}\nu'_\lambda}t^{{1\over 2}\nu_\lambda}M_\lambda\{p\}\overline{M}_{\lambda^\vee}\{\bar p\}\ {\stackrel{\hbox{\footnotesize Yangian limit}}{\longrightarrow}}\ \exp\left(\sum_{n=1}(-1)^{n+1}{p_n\bar p_n\over n}\right)\nn
\ee
This is not surprising since
\be
e_{(-1,1)}{\stackrel{(\ref{Fock2})}{=}}-{q^{-1/2}\over(1-q)(1-t^{-1})}\oint_0dzV(z)\
{\stackrel{\hbox{\footnotesize Yangian limit}}{\longrightarrow}}\  {1\over\beta\hbar}{\p\over\p p_1}+O(\hbar)\nn
\ee
and, generally, using formulas (\ref{enn}) and (\ref{gcd1m}),
\be
e_{(-n,n)}\
{\stackrel{\hbox{\footnotesize Yangian limit}}{\longrightarrow}}\  {1\over\beta\hbar}{\p\over\p p_n}+O(\hbar)\nn
\ee
However, if one looks at the Yangian limit of the Hamiltonians transformed with use of $\hat{\cal O}(u)$, one obtains from formula (\ref{IZ2}) in the leading order the Itzykson-Zuber formula for the $\beta$-ensemble \cite{MMS,MOP1}:
\be
\hbox{IZ}_2:=\sum_\lambda {M_\lambda\{p\}\overline{M}_{\lambda^\vee}\{\bar p\}\over \xi_\lambda(u)}\
{\stackrel{\hbox{\footnotesize Yangian limit}}{\longrightarrow}}\
\sum_\lambda {J^{(\beta)}_\lambda\{p\}J^{(\beta^{-1})}_{\lambda^\vee}\{\bar p\}\over \xi_\lambda^{(\beta)}(u)}\nn
\ee
since the Hamiltonians in the Yangian limit become less trivial rational Calogero Hamiltonians in $p_k$ variables $\hat H_{-k}^{(1)}$ \cite{MMMP1}. Here $J^{(\beta)}_\lambda\{p_k\}$ are the Jack polynomials \cite{Mac}, and $\xi_\lambda^{(\beta)}(u)=\prod_{i,j\in\lambda}(\log_q u+j-1-\beta(1-i))$.

Another example is discussed in detail in \cite[secs.2.4-2.5]{Ch3}, where the quantity {\bf E}$_1(q,t|N)$ is nothing but ${\cal O}(t^N)^{-1}\cdot e_{(1,0)}\cdot{\cal O}(t^N)$, etc.

\newpage

\section{Conclusion}

This paper contains a detailed (though not quite complete) description of rays of commuting
generators of the elliptic Hall / DIM $U_{q,t}(\widehat{\widehat{\mathfrak{gl}}}_1)$ algebra in three simplest representations:
one-body, $N$-body and Fock representations.
An extension to the MacMahon representation and beyond should involve some new technicalities and is not discussed.

\bigskip

We define the algebra as the elliptic Hall algebra in sec.\ref{ellHall}.
The main heroes there, the generators $e_{(n,m)}$ at all points of integer $2d$ lattice (we call the first index $n$ grading, and the second one, $m$, spin)
and their commutation relations are introduced,
however, the algebra is generated by just four elements $e_{(0,\pm 1)}$ and $e_{(\pm 1,0)}$.
In particular, their normalization prescribes normalizations of all others.
Generic commutation relations are, however, somewhat complicated and  highly non-linear,
they involve a hierarchy of non-admissible pairs
(whose commutators \textit{can} be evaluated, but this requires
%at the moment, non-algorithmic re-expression in terms of
repeated commutators of admissible pairs)
and are ultimately expressed through symmetric functions of the generators.
Our goal in this paper was to decipher these definitions,
and find maximally explicit expressions for $e_{(n,m)}$ in a few simplest representations of the DIM algebra, which naturally arise in physical applications:
\begin{itemize}
  \item The one- and $N$-body representations are used to construct integrable systems (in the DIM case, variations of the trigonometric Ruijsenaars system),
  \item The Fock representation is related to deformed integrable hierarchies,
    and related matrix/eigenvalue models
    (the $(q,t)$-deformations of KP/Toda hierarchies and WLZZ matrix models, respectively),
  \item The (not considered here) MacMahon representation should be,
    similarly, a proper framework for constructive and operational
    description of topological string theory,
   a very intriguing branch of science, in which only the first steps
   have been made, in part, due to absence of the (generalized) integrability
    framework.
\end{itemize}

An advantage of the elliptic Hall algebra description is that
the main {\bf statement about commutative rays} in this formulation is just one of the {\it axioms}:
it is eq.(\ref{sb}) or eq.(\ref{1}).
Generators along each straight line, which passes through the origin, form a Heisenberg subalgebra,
and, along each of the rays, they just commute (constitute halves of the Heisenberg subalgebra).
Moreover, in the DIM algebra, one can rotate lines and rays and convert them one into another (Miki automorphisms),
see (\ref{MikiST}),
thus {\bf commutativity along a single ray implies that along  any  other}, at least, in principle.
Thus commutativity is promised by the very consistency of the definition, but {\it what} commutes, i.e. the generators $e_{(kn,km)}$, in concrete representations, are still to be
constructed. In practice, their construction amounts to building them
from elementary ones $e_{(\pm 1, 0)}, \ e_{(0, \pm 1)}$, and the abstract blueprint/procedure
to do so, which would work in any representation, is highly desirable.
Additionally, the Miki automorphisms, in each concrete representation imply certain
symmetries/dualities between related integrable systems/hierarchies, therefore
understanding the form that the Miki automorphisms take in concrete representations is also
of importance, in practice.

\bigskip

We manage to realize the above ambitious program to various degrees of success, and \textbf{the main claims} of this paper are:

\begin{itemize}
\item[1)] A universal scheme for constructing $e_{(n,m)}$:
  \subitem{$\bullet$} The elements $e_{(n,m)}$ with coprime $n$ and $m$, i.e. with $gcd(m,n)=1$, are described by subsequent commutators (\ref{subsequentcom}),
  as a {\it tree} generated by the ABC-procedure, i.e. by the action of Miki operators ${\cal O}_v$ and ${\cal O}_h$
  on the basic pair $e_{(1,0)}$ and $e_{(0,1)}$.
  \subitem{$\bullet$}
  The commuting families are straightforwardly constructed in the basis of auxiliary operators (Hamiltonians) $\mathfrak{h}_{(kn,km)}$, they are built from corresponding $e_{(n,m)}$ with coprime $n$ and $m$ by a canonical procedure described in sec.\ref{sec:commutative-subalgebra-pr}:
one represents $B:=e_{(n,m)}$ as an admissible pair commutator $[A,C]$ by the ABC-procedure, then realizes
$\mathfrak{h}_{(kn,km)} = [A,{\rm ad}^{k-1}_B C]$.
Commutativity of $\mathfrak{h}_{(kn,km)}$ follows from the theory
of the elliptic Hall algebra exposed in sec.\ref{ellHall}.

\item[2)] Explicit expressions for $e_{n,m}$ in concrete representations:
  \subitem{$\bullet$} In the $N$-body representation, we discover a
  concise description of $e_{(n,m)}$ both with coprime $n$ and $m$, and non-coprime,
  which continues the line of reasoning in \cite{MMMP1}:
  namely,
  $e_{(n,m)} = \Tr {\cal C}_{n,m}$ while $e_{(kn,km)} = \Tr {\cal C}^k_{n,m}$,
and the ``root'' operators $\mathcal{C}_{n,m}$ are provided by
  explicit monomials of suitable transformed Dunkl-difference (Cherednik) operators.
  This construction should prove indispensable in constructing the related
  $(q,t)$-deformed matrix models \cite{Ch3}, particularly
  in the language of Miwa-deformations \cite{MMPS1,MMPS2}.
  \subitem{$\bullet$} In the one-body representation, this picture (almost) trivializes
  (see sec.\ref{s1body}), but, crucially, normalizations of generators
  %(and the arising singularities/zeroes in commutation relations)
  allow one to track/explain
  admissibility/non-admissibility, as well as to narrow the anzatz for
 commutators of non-admissible pairs.
  \subitem{$\bullet$} In the Fock representation, the generators $e_{(n,m)}$ are related
  to certain Laurent polynomials in variables $z_i, i = 1 .. n$, while commuting
  $e_{(n,m)}$'s amount to commuting these polynomials with a simple rule.
\item[3)] A realization of the Miki automorphisms:
  \subitem{$\bullet$} In the $N$-body representation, the basis generators of the Miki automorphisms $\mathcal{O}_h$
  and $\mathcal{O}_v$ are quite explicit: $\mathcal{O}_v$ is simply a canonical
  transformation that multiplies $\ttop{i}$ by $x_i$ (see \eqref{NbOv}),
  while $\mathcal{O}_h$ is more involved and amounts to modifying
  a word encoding the Cherednik operator monomial, in an explicit way \eqref{eq:oh-oper-words}
  that resembles growing branches on a tree.
  \subitem{$\bullet$} In the Fock representation, the $\mathcal{O}_v$ operator
  is similarly straightforward \eqref{Mikiv}, while description of $\mathcal{O}_h$,
  at the level of Laurent polynomials, remains obscure.
  \subitem{$\bullet$} However, in the dual fashion, $\mathcal{O}_h$ acts diagonally,
  with explicit Taki-factor eigenvalues \eqref{OMF} in the Macdonald polynomial basis
  of the Fock representation; for $\mathcal{O}_v$ no such a description is available.
\item[4)] Applications:
  \subitem{$\bullet$} By considering common eigenfunctions of commutative ray Hamiltonians,
  we find \textit{two} natural candidates for the role of $(q,t)$-deformation of the
  Harish-Chandra-Itzykson-Zuber integral (\ref{IZ1}) and (\ref{IZ2}): the one which is more natural
  from the point-of-view of the DIM algebra itself, and another one, which has a better
  Yangian limit, and therefore is more closely related to our previous superintegrability
  studies \cite{Ch1,Ch2,MOP1}.
  \subitem{$\bullet$} Similarly, two natural versions of $(q,t)$-deformed counterparts
  of the WLZZ matrix models are possible (\ref{Z+O})-(\ref{Z-O}) and (\ref{Z+})-(\ref{Z-}), while previously
  only the first one was considered \cite{Ch3} (see also earlier works \cite{Max,MPSh,MMell}).
\end{itemize}

\bigskip

\noindent Some important questions are \textbf{left unanswered} by this paper,
and are natural topics for future research:
\begin{itemize}
\item Is there an algorithm/combinatorial formula to evaluate the commutator
  of any two $e_{(n,m)}$ operators (when they form a non-admissible pair)?
  %Does it involve a triangulation of a non-admissible triangle into admissible ones?
\item Is there a more explicit formula for the \textit{cone} Hamiltonians of sec.\ref{cones} (see also sec.\ref{pent})
  and their more complicated larger cone counterparts (particularly, obtained by successive actions with a few $\hat{\cal O}(u_i)$)?
\item How these cone constructions can be used to describe a peculiar
  limit of the commutative subalgebras of the elliptic Hall/DIM algebra to those of the affine Yangian algebra explicitly?
\end{itemize}
We hope that this paper provides sufficient background to help answer these questions.

\bigskip

An ``ideal" example of what we are aimed at is provided by the one-body representation.
There the generators are just proportional to $x^mq^{nx\frac{d}{dx}}$, and the commutativity along all rays is a trivial exercise,
the Miki rotations being provided by operators (\ref{1bodyrotations}).
In the $N$-body case, commutative rays become (\ref{eq:ray-1r-hamilts}), still simple in an appropriate notation,
while the Miki rotations are already a little more involved, see (\ref{eq:ov-oper-words}), (\ref{eq:oh-oper-words}).
Finally, in the Fock representation, the Hamiltonians are made from polynomials, see sec.\ref{531},
and rotations are given in (\ref{Mikiv}) and sec.\ref{OhF}.
Moreover, in these representations, there are additional symmetries apart from the Miki rotations:
the reflection of gradings, $e_{(n,m)}\to e_{(-n,m)}$,
and the reflection of spin,  $e_{(n,m)}(q,t)=-e_{(n,-m)}(q^{-1},t^{-1})$.
This allows us to restrict consideration to a single quadrant of the whole $2d$ lattice of elements $e_{(n,m)}$.

One of the possibilities to construct answers in the representation is to use that
the coprime pair $(n,m)$ can be always decomposed with the help of the Euclid algorithm
and reduced to some simple pair, say, $(-1,1)$. For instance, $(-7,11)  \stackrel{{\cal O}_h}{\longrightarrow}  (-7,4)
\stackrel{{\cal O}_v}{\longrightarrow}   (-3,4)
\stackrel{{\cal O}_h}{\longrightarrow}   (-3,1)
\stackrel{{\cal O}_v^2}{\longrightarrow}  (-1,1)$  implies that
$e_{(-7,11)} = {\cal O}_h^{-1} {\cal O}_v^{-1} {\cal O}_h^{-1} {\cal O}_v^{-2} \,  e_{(-1,1)}$. This produces a decomposition of $e_{(n,m)}$ into a series of Miki automorphisms (\ref{MikiST}).

What remains is to apply the action of ${\cal O}_{v,h}$ in the given representation, see the representative example in sec.\ref{Nex}.

A peculiar feature of the $N$-body representation inherited
from the one-body representation is that elements of a commutative ray
$e_{(kn,km)}$ are given by sums of powers $k$ of the same monomial product of operators:
tilded Cherednik operators and $x_i$.
This specific property has no clear continuation to the Fock and more complicated
representations.
This is actually a well known fact surveyed also in Appendix B of \cite{MMMP1}:
for example, what in the $N$-body representation is just squaring a differential when coming from the first Hamiltonian to the second one: $\sum_i \frac{\p}{\p x_i} \ \longrightarrow \ \sum_i \left(\frac{\p}{\p x_i}\right)^2$,
in the Fock representation, becomes changing the grading and the spin of operators:
$\sum_k kp_{k-1}\frac{\p}{\p p_k}\longrightarrow
\sum_{k,l}  kl p_{k+l-2}\frac{\p^2}{\p p_k \p p_l} + \sum_k k(k-1)p_{k-2}\frac{\p}{\p p_k}$.

\bigskip

For convenience, we collect these statements in the table,
which can be considered as a summary of our main results:
\be
%\!\!\!\!\!\!\!\!\!\!\!\!\!\!\!
\begin{array}{|c||c|c|c|c|}
\hline
{\rm reps} & \text{ one-body}\ (qW_{1+\infty}) & N\text{-body} & {\rm Fock} & \stackrel{\rm Other}{\rm reps}
\\
\hline\hline
{\rm Hams}\ \  H^{(n,m)}_k = e_{(-kn,km)}
& z^{-kn} q^{km\hat D}
&
%(\ref{eq:table-of-ham-words}):??? \
\sum_{i=1}^N \left(  C_i^{[a]} x_i C_i^{[b]} \ldots \right)^k
&q^{kn\over 2}{\kappa_1\over\kappa_k}\Big<{\bf E}_{km-kn}(z_i;\kappa_1)\Big>_m &  ?
 \\
\hline
{\rm Miki\ ``rotations"} \ (\ref{MikiST}) &(\ref{1bodyrotations}) &  (\ref{eq:ham-through-cher}) & (\ref{Mikiv})\ \hbox{and sec}.\ref{OhF} & ? \\
\hline
\ldots &&&&
\end{array}
\nn
\ee
The table looks simple but it is not obvious {\it a priori}, and justifying calculations
are somewhat tedious, they were briefly represented in the text.

\bigskip

Once we have commutative operator rays, a natural question is what are their common eigenfunctions.
For the vertical ray $e_{(0,k)}$, these are just the Macdonald polynomials, but, for other rays, the answer is
 more interesting and is much similar to the $W_{1+\infty}$/Yangian algebra case \cite{MMMP1}.
Already for the ray $e_{(-k,k)}$ the common eigenfunctions are bilinear combinations (\ref{IZ1}) (and (\ref{IZ2}) for the rotated Hamiltonians) of Macdonald polynomials,
actually defining the $q,t$-deformation of character expansion \cite{Kazak,Bal,Mor2} of the Itzykson-Zuber integral \cite{IZ}.
The variety of eigenfunctions is parameterized by the power sum arguments of the Macdonald polynomials in the bilinear
decompositions.

Also deserves mentioning the commutative rays (a variety of integrable systems)
and especially their interrelations (unitary equivalence) first attracted attention in the case of $W_{1+\infty}$ algebra and
in their Yangian deformation \cite{MMCal}, but there they are considerably more difficult to describe.
At the level of the DIM algebra, things simplify conceptually, but the problem of taking the Yangian limit of the commutative rays
in an elegant way persists.
However, an exact relation between commutative families in the affine Yangian algebra and the elliptic Hall/DIM algebra is somewhat non-trivial,
which is well seen already at the one-body level: the elliptic Hall algebra generators are represented by $z^m q^{n\hat D}$, while those of the affine Yangian algebra, by $z^m\hat D^n$.
In the Fock representation of the elliptic Hall algebra, the generators are made from the vertex operators of a ($q,t$-deformed) type $e^{\phi(z)}$,
while those of the affine Yangian algebra, from powers of $\p\phi(z)$.
A better analysis and clearer presentation is still required at this point.

Other open questions are generalizations to other representations and to the elliptic DIM algebra,
where the role of Macdonald polynomials is played by more general elliptic Macdonald polynomials \cite{AKMM,MMZ}.

\bigskip

To summarize, this paper describes the network of integrable systems of the Calogero-Ruijsenaars type,
originally implied by the WLZZ models
in terms of commutative rays in the DIM algebra.
The full set of Hamiltonians (not all commuting)
is identified with the full set of generators in the elliptic Hall version of the DIM algebra.

What we do want from such a description are explicit formulas for all the Hamiltonians, and commutation relations
for those of them which do not commute (belong to different rays).
From the algebraic point of view, this is equivalent to explicit expressions for all the generators
in particular representations.
Within this context, the generators at integer points of the $2d$ lattice are divided into two sets:
the {\it core} operators $e_{(n,m)}$ with coprime $m$ and $n$, which serve as first Hamiltonians of
various integrable systems, i.e. are the {\it germs} of the rays,
and $e_{(kn,km)}$, defining the $k$-th Hamiltonians in the commutative sets.

\bigskip

What we describe in the text is the current state of affairs about this pattern:

\vspace{2.3cm}

\!\!\!\!\!\!\!\!\!\!\!\!\!\!\!\!\!\!\!\!\!%\!\!\!\!\!\!\!\!\!\!\
%\!\!\!\!\!\!\!\!\!\!\!\!
%\!\!\!\!\!\!\!\!\!\!\!\!
\begin{tabular}{|c|c|}
\hline &\\
Already simple/explicit  &  Still difficult/obscure \\
&\\
\hline \hline
&
\\
All generators $e_{(n,m)}$ (points in the $2d$ lattice of Fig.2)
&
This procedure is  ambiguous, and can involve
\\
for coprime $n$ and $m$
are explicitly represented
&
 splittings (trees). Devising it from top to bottom,
\\
as repeated commutators
of $e_{(0,1)}$ and $e_{(1,0)}$
&
i.e. original decomposition $e_{(n,m)}:=B=[A,C]$,
\\
via the Euclid algorithm and the Miki rotations,
&
 requires some work
\\
see (\ref{subsequentcom})-(\ref{01})&
Commutators of different $e_{(n,m)}$ can be rather non-trivial
\\
\hline
All $e_{(kn,km)}$ with different $k$ (points along rays)  commute
&
Their expressions through  $e_{(0,1)}$ and $e_{(1,0)}$
\\
&
goes through intermediate step with $\mathfrak{h}_{(kn,km)}$
\\
\hline
All $\mathfrak{h}_{(kn,km)}$ with different $k$
&
These $\mathfrak{h}_{(kn,km)}$ are explicit polynomials
\\
 (Ruijsenaars-like Hamiltonians along rays)
&
 of commuting $e_{(k'n,k'm)}$ along the same rays,
 \\
are explicitly expressed by the ABC procedure (\ref{ABC})
&
which still need to be resolved
\\ & \\
\hline\hline &\\
For $W_{\infty}$ everything is explicit and clear &
\\
\hline
For the one-body representation of DIM&
\\
everything remains explicit
&
Still, some care is needed with normalizations
\\
\hline
For $N$-body representation there is an additional bonus: &\\
explicit expressions for {\it all} $e_{(n,m)}$ via Cherednik
&
Commutativity of these operators
\\
(generalized Dunkl) operators $C$, different
&
is (still) non-evident, though true
\\
for coprime ($n, m)$ and points $(kn,km)$ in the rays&
\\
\hline
For the Fock representation, there is a peculiar description
&
\\
in terms of polynomials
of relatively few variables
&
The role and quality of this description is unclear.
\\ &
\\
\hline \hline & \\
Probably the  situation with the Fock representation
&
Still comparable level of explicitness for other
\\
is already generic, and thus will not get
&
representations requires work. A lift from $N$-body
\\
more complicated in other representations
&
to matrix representation seems to be lacking
\\
&\\
\hline
\end{tabular}

\bigskip

\bigskip

It is a purpose of the future research to expose the issues of the second column
and bring it to the level of simplicity and explicitness achieved in the first one.
This is partly done in the mathematical literature,
but with accents put mainly on the formal algebra,
and not so much on the physical meaning of formulas and their connection to
traditional theory of integrable systems.
In this sense, the present paper is a natural development of \cite{MMMP1}, extending analysis
from the $W_{1+\infty}$ algebra to the DIM algebra and explaining potential advantages at this level
like the existence of the {\it pair} of Miki automorphisms, which unify the whole picture.

\section*{Acknowledgements}

 We are grateful to Y. Zenkevich for valuable discussions and explanations. This work was supported by the Russian Science Foundation (Grant No.23-41-00049).

\newpage

\appendix

{\section{Admissible pairs algorithm} \label{sec:admissible-algorithm}
  To check, whether any given pair of elements $e_{\vec a}$ and $e_{\vec b}$
  is an admissible pair, one can use the following considerations.
  Consider the triangle $ABC$, formed by vertices $A = (-b_1, -b_2)$,
  $B = (0,0)$, $C = (a_1, a_2)$
  \begin{itemize}
  \item First, by virtue of celebrated Pick's formula, number of integer points inside
    the triangle and on its boundary are related to its \textit{area}
    \begin{align}
      V_{\text{inside}} + \frac{1}{2} V_{\text{boundary}} - 1
      = S_{ABC} = \frac{1}{2} \Bigg{|} \det \left (
      \begin{array}{cc}
        a_1 & a_2 \\
        b_1 & b_2
      \end{array}
      \right )\Bigg{|}\nn
    \end{align}
  \item Second, a number of integer points (including endpoints) on some vector
    $(v_1, v_2)$ equals
    \begin{align}
      V_{(v1,v2)} = 1 + GCD(v_1, v_2)\nn
    \end{align}
    $GCD$ being greatest common divisor.
    From this one can deduce the number of integer points on the sides of the triangle
    $V_{AB} = V_{(b_1, b_2)}$, $V_{AC} = V_{(a_1 + b_1, a_2 + b_2)}$,
    $V_{BC} = V_{(a_1, a_2)}$
  \item Finally, one notices that the total number of boundary points equals
    \begin{align}
      V_{\text{boundary}} = V_{AB} + V_{BC} + V_{AC} - 3\nn
    \end{align}
    because the vertices are being counted twice.
  \end{itemize}

  Note that determining the ``middle point'' $m_{ab}$ is straightforward
  -- and is very similar to the algorithm of sorting Young diagrams.
}

{\section{Cherednik operators} \label{sec:cherendik-def}

Besides the description of the Cherednik operators in sec.\ref{Cho}, there is an alternative (isomorphic,
  see \cite[lem.1.3.12]{book:Ch-daha}) description of the Cherednik operators via the related affine Hecke algebra,
which is, perhaps, more convenient for checking their
  commutativity.

Namely, define
  \begin{align}
    T_i = R_{i,i+1} \cdot s_{i,i+1}, \text{ and, accordingly } T_i^{-1}
    = s_{i,i+1} \cdot R^{-1}_{i,i+1}\nn
  \end{align}
Then, the Cherednik operators are equal to
  \begin{align}
    C_i = & \ T_i \dots T_{N-1} \sigma_\pi T_1^{-1} \dots T_{i-1}^{-1},
    \text{ with }\nn
    \\ \notag
    & \ \sigma_\pi = s_{N-1, N}  \dots s_{i,i+1} \ttop{i} s_{i-1,i} \dots s_{1,2},
  \end{align}
where the operator $\sigma_\pi$ does not depend on the index $i$
  and acts on a tuple of coordinates $(x_1, \dots x_N)$ as follows
  \begin{align}
    (x_1, x_2, \dots, x_N) \rightarrow (x_2, x_3, \dots, x_N, q x_1)\nn
  \end{align}
In particular, the following commutation relation holds
  \begin{align}
    \frac{1}{x_N} \cdot \sigma_\pi = q \ \sigma_\pi \cdot \frac{1}{x_1}\nn
  \end{align}

  \bigskip

  Now, $T_i$, together with $C_i$ obey the following set of the $GL_N$ affine Hecke algebra
  relations  \cite[Eqs.(1.3.1)-(1.3.6)]{book:Ch-daha} (note that our normalization convention is slightly different
  from that in \cite{book:Ch-daha}, hence, some $t$-factors)
  \begin{align} \label{eq:aha-rels}
    T_i T_{i+1} T_i = & \ T_{i+1} T_i T_{i+1},\ \ \  i = 1 .. N-2
    \\ \notag
    T_i T_j = & \ T_j T_i, \ \ \ \ \ \ \ \ \ \ \ \ \ \text{ provided } |i - j| > 1
    \\ \notag
    T_k - t^{-1} T_k^{-1} & \  + \frac{(1-t)}{t} \cdot 1 = 0
    \ \ \ \ \text{ skein relation}
    \\ \notag
       [C_i, C_j] & \ = 0,\ \ \  i,j = 1 .. N
       \\ \notag
          [C_i, T_j] & \ = 0, \ \ \ j \neq i, i-1
          \\ \notag
          T_i^{-1} C_i T_i^{-1} & \ = t C_{i+1}, \ \ \ i = 1 .. N-1
  \end{align}
  Therefore, in this presentation, one can think of generators $T_i$
  as fundamental $\mathcal{R}-$matrices, where $T_i$ intertwines
  $i$-th and $i+1$-th strands out of $N$ total strands.

  \bigskip

  Note that the operators $1/x_i$ are, in a sense, similar to $C_i$
  in this algebra, in that we have
  \begin{align}
    \Big[\frac{1}{x_i}, T_j\Big] & \ = 0, \ \ \ j \neq i, i-1\nn
    \\ \notag
    T_i^{-1} \frac{1}{x_i} T_i^{-1} & \ = \frac{t}{x_{i+1}},
    \ \ \ i = 1 .. N-1
  \end{align}

  \bigskip

  This allows one to conveniently write the $\widetilde{C}_i$ operators as
  \begin{align}
    \widetilde{C}_i = T_i \dots T_{N-1} \frac{1}{x_N} \sigma_\pi
    T_1^{-1} \dots T_{i-1}^{-1}
    = t^{(-1)(n-i)} \frac{1}{x_i} T_i^{-1} \dots T_{N-1}^{-1} \sigma_\pi
    T_1^{-1} \dots T_{i-1}^{-1}\nn
  \end{align}
  moreover, in this form, the operators corresponding to $(1,r)$ rays
  start to look similar, for instance,
  \begin{align}
    \widetilde{C}_i x_i \widetilde{C}_i
    = t^{(-2)(n-i)}
    \frac{1}{x_i} T_i^{-1} \dots T_{N-1}^{-1}
    \underbrace{\sigma_\pi
      T_1^{-1} \dots T_{N-1}^{-1}
      \sigma_\pi}_{\sigma_{\pi'}}
    T_1^{-1} \dots T_{i-1}^{-1}\nn
  \end{align}
Hence, if one manages to show that $\sigma_{\pi'}$
  commutes with everything else as $\sigma_{\pi}$,
one proves mutual commutativity of the $\widetilde{C}_i x_i \widetilde{C}_i$
  operators.

  \bigskip

  Yet another convenient in working with the $GL_N$ affine Hecke algebra
  is the operator
  \be
    P = & \ T_1 \dots T_{N-1} \sigma_\pi T_1^{-1} \dots T_{N-1}^{-1}\nn
\ee
so that
\be
    C_i = & \ T_{i-1}^{-1} \dots T_1^{-1} P T_{N-1} \dots T_i\nn
\ee
One can reformulate the algebra relations as those from (\ref{eq:aha-rels}) that do not involve $C_i$ plus the two additional relations
  \be\label{eq:aha-p-rels-extra}
    P T_{i-1} & = T_i P \text{ and, hence } P^{-1} T_i = T_{i-1} P^{-1}
    \\ \notag
    P^N &\text{ is a central element}
  \ee

  From this representation, it is really convenient to prove mutual commutativity
  of $C_i$ (if it is not postulated from the very beginning), for instance,

  \begin{itemize}
    \item N = 2
      \begin{align}
        C_1 = & \ P T_1 \ \ C_2 = T_1^{-1} P\nn \\ \notag
        C_1 C_2 = & \ P^2 \\ \notag
        C_2 C_1 = & \ T_1^{-1} P^2 T_1 \mathop{=}_{P^2 \text{ is central}} P^2 = C_1 C_2
      \end{align}
    \item N = 3
      \begin{align}
        C_1 = & \ P T_2 T_1 \ \ C_2 = T_1^{-1} P T_2 \ \ C_3 = T_2^{-1} T_1^{-1} P\nn
        \\ \notag
        C_1 C_2 = & \ P T_2 P T_2 = P^2 T_1 T_2 = P^{-1} P^3 T_1 T_2 = P^{-1} T_1 T_2 P^3
        \\ \notag
        C_2 C_1 = & \ T_1^{-1} P T_2 P T_2 T_1 =
        T_1^{-1} P^2 T_1 T_2 T_1
        \mathop{=}_{\text{Yang-Baxter}} T_1^{-1} P^2 T_2 T_1 T_2
        \mathop{=}_{P^3 \text{ is central}} P^{-1} T_1 T_2 P^3
      \end{align}
  \end{itemize}
}

\end{document}